\newcommand{\be}{\begin{equation}}
        \newcommand{\ee}{\end{equation}}
\newcommand{\bega}{\begin{equation}\begin{gathered}}
            \newcommand{\eega}{\end{gathered}\end{equation}}
\newcommand{\bea}{\begin{equation} \begin{aligned}}
            \newcommand{\eea}{\end{aligned} \end{equation}}
\newcommand{\balign}{\begin{align}}
\newcommand{\ea}{\end{array}}
\newcommand{\Grad}{\mbox{\rm Grad}}

\newcommand{\Ddiv}{\mbox{\rm Div}}
\newcommand{\tr}{\mbox{\rm tr} \,}

\newcommand{\T}{^\top}

\newcommand{\MT}{^{-\top}}
\newcommand{\ME}{^{-1}}

\def\endbox{
\leavevmode\kern-.25em\raise-.15em
\vbox{\hrule height 0.8truept
\hbox{\vrule width  0.8truept
\vbox{\kern 4truept
\hbox{\kern 1.5truept
{}\kern 2.5truept}
\kern 3.0truept}
\vrule width  0.8truept}
\hrule height 0.8truept}
}
\def\pmb#1{\setbox0=\hbox{#1}%
    \kern-.025em\copy0\kern-\wd0
    \kern.05em\copy0\kern-\wd0
    \kern-.025em\raise.0433em\box0 }

\newcommand{\bbbb}{\bm{b}}

\newcommand{\bee}{\bm{e}}
\newcommand{\bff}{\bm{f}}

\newcommand{\bt}{\bm{t}}
\newcommand{\buu}{\bm{u}}
\newcommand{\bv}{\bm{v}}

\newcommand{\bcc}{\bm{c}}
\newcommand{\bn}{\bm{n}}
\newcommand{\bC}{\mathbf{C}}
\newcommand{\bD}{\bm{D}}
\newcommand{\bE}{\mathbf{E}}
\newcommand{\bF}{\mathbf{F}}
\newcommand{\bG}{\bm{G}}

\newcommand{\bK}{\mathbf{K}}

\newcommand{\bP}{\bm{P}}

\newcommand{\bR}{\bm{R}}
\newcommand{\bS}{\mathbf{S}}

\newcommand{\bU}{\bm{U}}
\newcommand{\bV}{\bm{V}}

\newcommand{\bX}{\mathbf{X}}
\newcommand{\bY}{\mathbf{Y}}
\newcommand{\bZ}{\mathbf{Z}}

\newcommand{\bzero}{\mbox{$\bf 0$}}

\newcommand{\bvarepsilon}{\boldsymbol{{\varepsilon}}}

\newcommand{\bsigma}{\boldsymbol{\sigma}}

\newcommand{\bomega}{\mathbf{\omega}}
\newcommand{\bChi}{\bm{\mathcal{X}}}

\newcommand{\bSigma}{\mathbf{\Sigma}}

\newcommand{\bPhi}{\mathbf{\Phi}}
\newcommand{\bvarPhi}{\boldsymbol{\varPhi}}

\newcommand{\bOmega}{\mathbf{\Omega}}

\newcommand{\D}{\Delta}
\newcommand{\lam}{\lambda}
\newcommand{\Rint}{\bm{R}}
\newcommand{\Ri}{\bm{R}^{(i)}}
\newcommand{\Rred}{\bm{R}_{red}}

\newcommand{\KT}{\mathbf{K}_{T}}
\newcommand{\KTi}{\mathbf{K}_{T}^{(i)}}
\newcommand{\KTnl}{\mathbf{K}_{T \: nl}}
\newcommand{\ZT}{\mathbf{Z}^\top}
\newcommand{\Pred}{\bm{P}_{red}}
\newcommand{\Pzred}{\bm{P}_{0 \: red}}
\newcommand{\Ured}{\bm{U}_{red}}
\newcommand{\Gred}{\bm{G}_{red}}

\newcommand{\Klinred}{\mathbf{K}_{lin \: red}}
\newcommand{\MDEIM}{\mathbf{M}_{DEIM}}
\newcommand{\fdam}{f_\textrm{dam}}
\newcommand{\dd}{\textrm{d}}
\newcommand{\Dbar}{\bar{D}}
\newcommand{\Ubar}{\bar{\bm{U}}}
\newcommand{\bDbar}{\bar{\bm{D}}}

\newcommand{\KTred}{\mathbf{K}_{T \: red}}
\newcommand{\Phim}{\mathbf{\Phi}_m}
\newcommand{\PhimT}{\mathbf{\Phi}_m^{\top}}
\newcommand{\Omk}{\mathbf{\Omega}_k}

\newcommand{\Rnlred}{\bm{R}_{nl \: red}}
\newcommand{\Rnl}{\bm{R}_{nl}}

\newcommand{\lt}[1]{{}^{(#1)}}
\newcommand{\lb}[1]{{}_{(#1)}}

\newcommand{\ltz}{{}^{(0)}}

\newcommand{\lti}{{}^{(i)}}
\newcommand{\ltipo}{{}^{(i+1)}}

\newcommand{\norm}[1]{\left \| #1 \right \|}

\documentclass[12pt,a4paper,notitlepage]{article}

\usepackage{amsfonts}
\usepackage{amsmath}
\usepackage{amssymb}
\usepackage{amsthm}
\usepackage{color}
\usepackage[T1]{fontenc}
\usepackage[utf8]{inputenc}
\usepackage{graphicx}
\usepackage{hyperref}
\usepackage{mathrsfs}
\usepackage{multirow}
\usepackage[square, numbers, comma, sort&compress]{natbib}
\usepackage[format=hang]{subcaption}
\usepackage{times}
\usepackage{upgreek}
\usepackage{bm}
\usepackage{bbm}
\usepackage{mathtools}
\usepackage{cleveref}
\usepackage{subcaption}
\usepackage{import}
\usepackage{placeins}
\usepackage{algpseudocode}
\usepackage{svg}
\usepackage{setspace}
\usepackage[]{algorithm2e}
\usepackage[style=iso]{datetime2}
\AtBeginDocument{} 

\usepackage{pgf}
\usepackage{tikz}
\usepackage{pgfplots}
\pgfplotsset{compat=newest}
\usepackage{pgfplotstable}
\usepgfplotslibrary{groupplots}
\usepackage{color}
\usetikzlibrary{calc,backgrounds} 

\definecolor{rwth1}{RGB}{0,84,159}      
\definecolor{rwth2}{RGB}{142,186,229}   
\definecolor{rwth3}{RGB}{0,97,101}      
\definecolor{rwth4}{RGB}{0,152,161}     
\definecolor{rwth5}{RGB}{87,171,39}     
\definecolor{rwth6}{RGB}{189,205,0}     
\definecolor{rwth7}{RGB}{255,237,0}     
\definecolor{rwth8}{RGB}{246,168,0}     
\definecolor{rwth9}{RGB}{227,0,102}     
\definecolor{rwth10}{RGB}{204,7,30}     
\definecolor{rwth11}{RGB}{161,16,53}    
\definecolor{rwth12}{RGB}{97,33,88}     
\definecolor{rwth13}{RGB}{122,111,172}  

\date{}

\graphicspath{{./images/}}

\begin{document}

\author{\large{Steffen Kastian,  Jannick Kehls,  Tim Brepols and Stefanie Reese}\\[0.5cm]
  \hspace*{-0.1cm}
  \normalsize{\em Institute of Applied Mechanics, RWTH Aachen
    University,}\\
  \normalsize{\em Mies-van-der-Rohe-Str.\ 1, 52074 Aachen, Germany}\\[0.25cm]
}

\title{\LARGE Discrete Empirical Interpolation Method for nonlinear softening problems involving damage and plasticity}
\maketitle

\small
{\bf Abstract.}
Accurate simulations are essential for engineering applications, and intricate continuum mechanical material models are constructed to achieve this goal.
However, the increasing complexity of the material models and geometrical properties lead to a significant increase in computational effort.
Model order reduction aims to implement efficient methods for accelerating the simulation process while preserving a high degree of accuracy.
Numerous studies have already demonstrated the benefits of this method for linear elastic material modeling.
However, in the present work, we investigate a two-surface gradient-extended damage-plasticity model.
We conducted complex simulations with this model, demonstrating both damage behavior and softening.
The POD-based discrete empirical interpolation method (DEIM) is introduced and implemented.
To accomplish simulations with DEIM and softening behaviour, we propose the implementation of a reduced form of the arc-length method.
Existing research on calculating models with both damage and softening behavior using the DEIM and arc-length method is limited.
To validate the methods, two numerical examples are thoroughly investigated in this study: a plate with a hole and an asymmetrically notched specimen.
The results show that the proposed methods can create a reduced order model with high accuracy and a significant speedup of the simulation.
For both examples, the analysis is conducted in three steps: first, plasticity without damage is examined, followed by damage without plasticity, and finally, the combination of plasticity and damage is investigated.

\vspace*{0.3cm}

\normalsize


\section{Introduction}
In the last few decades, the research field of computational mechanics has been majorly advanced to the point where large-scale simulations are performed throughout all engineering industries. These large-scale simulations deal with a wide variety of phenomena such as damage, thermodynamics, fluids, gasses, electromagnetism, or structural analysis. The possibility to run such resource-intensive simulations became feasible when hardware improved to the point where numerical methods could be employed to solve the aforementioned phenomena. For structural analysis, further hardware improvements were used to calculate bigger and more complex structures with increasingly complex underlying material models. This in turn meant that the simulation time also increased more and more.

To be able to increase the complexity even further, different methods are being taken to reduce the computational effort of the simulations. First of all, high-performance clusters are being used by most of the industry to speed up the simulation times of the huge models that need to be evaluated. Secondly, researchers are trying to increase the processor power of the processing units inside these clusters. Gordon Moore, the namesake of Moore's Law, postulated in the 1960s that the number of transistors per fixed cost, power, and area would double every two years and therefore roughly double the computing power of the processing units every two years. For many decades Moore's law could prove true, but in recent years, processing unit manufacturers struggled more and more to meet the improvement predictions suggested by Moore's law \cite{mooreslaw}. As a response, researchers are now focusing on new ways to increase processing power. Increasing processing power is of course the most obvious way to reduce the simulation time but there is a third measure being taken which is intensely concerned with reducing the amount of effort to obtain a simulation result.

The topic concerned with reducing the computational complexity of underlying mathematical models is called model order reduction (MOR). Model order reduction is not restricted to structural analysis, but rather a topic in almost all research fields where complex mathematical models are being solved numerically. Model order reduction techniques are therefore often investigated in a wide variety of research fields, but due to the mathematical nature of the different techniques, promising methods are quickly adopted.

Looking at model order reduction in the context of structural problems that are being solved by means of the finite element method (FEM), a problem exists in a high-dimensional space in which it needs to be solved. Through projection based model order reduction techniques, one seeks to reduce the problem from the high-dimensional space to a low-dimensional space, in which the problem can be solved much more quickly. After the solution in the reduced order space has been found, it can be projected back onto the original space. Needless to say that this reduction introduces an approximation error that should be kept as small as possible to maintain the validity of the obtained results.

Model order reduction techniques can be categorized in many ways, but according to Benner et al. \cite{MOR1} the two high-level categories for MOR are, firstly, system-theoretic techniques and, secondly, numerical approaches. The first category is concerned with the derivation of a more compact model with the same geometry but a much smaller number of states, whose response accurately approximates the response of the full model. Approaches of the second category are not intruding on the underlying theory of the model but are rather concerned with the reduction of the complexity of the numerical methods that are used to solve the problem. Techniques from both categories share the overall objective to reduce the computational complexity of a simulation.

This thesis will focus on the second category of MOR techniques with a specific focus on the snapshot-based proper orthogonal decomposition~(POD). In this context, snapshot-based means that to reduce a problem, precomputed solutions of the full problem are used to create the basis matrices that are necessary for the projection from the high-dimensional space to the low-dimensional space and vice versa.

Other than the proper orthogonal decomposition, there are different snapshot-based techniques, which are also well-researched and should not be neglected. 
Such methods are, for example, the proper generalized decomposition, where Chinesta et al. explore the capabilities of the Proper Generalized Decomposition (PGD) in tackling highly multidimensional models, offering an efficient alternative to traditional mesh-based discretization approaches \cite{PGD1}, while Nouy studys different formulations of the PDG and introduces a new one called Minimax PGD \cite{PGD2}. 
A second class is the reduced basis method which Rozza et al. used for Stokes equations in domains with affine parametric dependence \cite{RB1} and Fares et al. for the electric field integral equation \cite{RB2}. 
Further we would like to mention the load dependent Ritz method. Noor et al. used this approach already 1980 for nonlinear analysis of structures \cite{ritz3}. Additionaly it is used for predicting the nonlinear dynamic response \cite{ritz1} and dynamic substructuring \cite{ritz2}. These are just a few model reduction methods among many others.
Some methods are used primarily in the field of dynamics, whereas others are mostly used in the field of static problems. Hence, constructing direct comparisons between these methods proves challenging; however, a study comparing the load-dependent Ritz method, modal basis method, and proper orthogonal decomposition for structural quasi-static problems revealed that the proper orthogonal decomposition achieved the highest accuracy while significantly reducing CPU time \cite{radermacher2013}. This finding contributes to the rationale behind selecting the proper orthogonal decomposition as the fundamental basis for this thesis.


Another significant aspect is that one aim of the thesis is the reduction of highly nonlinear problems where iterative solution schemes have to be used. As a result, the aforementioned methods do not lead to speed-ups of the simulation that are desirable. Why this is the case will be explained in section \ref{sec:methods}. There are, however, methods to speed up the simulations of nonlinear problems, the so-called hyperreduction techniques \cite{hyper}. A few promising hyperreduction techniques are based on the POD, such as the gappy POD approach \cite{GPOD1, GPOD2}, the discrete empirical interpolation method (DEIM) \cite{chatsor, DEIM2, DEIM3}, and the empirical cubature \cite{EC1, EC2}. Non-intrusive model order reduction techniques, which do not rely on knowledge of the governing equations, are also well recognized to be present. \cite{NI1, NI2, NI3}. Such methods often rely on modern techniques from computer science, namely machine learning, deep learning, and neural networks in general.

The POD technique has gained significant attention and has been widely applied in various fields of engineering and science. It offers an effective approach to reduce the computational complexity of high-dimensional problems while preserving the essential characteristics of the system's behavior. By extracting dominant modes or patterns from a set of precomputed solutions, the POD provides a low-dimensional representation that captures the system's essential dynamics. This reduced-order representation can then be used for efficient simulations, optimization, and uncertainty quantification.

In summary, this work will briefly describe the material model in section \ref{sec:mat} and the model order reduction techniques in section \ref{sec:methods}.
Results of the accuracy and efficiency of the method are shown in section \ref{sec:results}. Through this research, valuable insights and techniques can be gained to enhance the computational efficiency and accuracy of large-scale simulations in engineering and scientific domains.

\subsection{Notational convention}
The notational convention used throughout the work follows standard tensor notation. Scalars and scalar-valued functions are denoted by regular italic characters in both upper and lower case, e.g., $A$ or $a$. Vectors and vector-valued functions are represented by bold italic letters in upper and lower case, e.g., $\bm{B}$ or $\bm{b}$.
Matrices, tensors of the second order, and tensor-valued functions are written in bold characters of either case, e.g., $\mathbf{C}$ or $\mathbf{c}$. 
Greek symbols used to denote second-order tensors are written in bold italic, e.g., $\bvarepsilon$ or $\bsigma$. Fourth-order tensors and tensor-valued functions are represented by italic uppercase calligraphic letters, e.g., $\mathcal{D}$. The transpose of a vector, matrix, or tensor is displayed with the superscript $\top$ (e.g., $\bm{B}^\top$). 
The operators $\Grad$ and $\Ddiv$ describe the gradient and divergence of a scalar, vector, or tensor with respect to Lagrangian coordinates. The abbreviation $\mathbf{C}' \coloneqq \mathbf{C} - \frac{1}{3}\tr(\mathbf{C})\mathbf{I}$ denotes the deviatoric part of a second-order tensor, where $\mathbf{I}$ is the second-order identity tensor, and $\tr(\mathbf{C})$ represents the trace of $\mathbf{C}$. 
The determinant of a tensor $\mathbf{C}$ is denoted as $\det(\mathbf{C})$, and the single contraction of a first-order tensor $d$ with a second-order tensor $\mathbf{C}$ is represented as $\mathbf{C}[d]$.

\section{Material Model}
\label{sec:mat}

In the following section, the constitutive model used to investigate the proposed model order reduction approach will be briefly introduced. The material model was developed by Brepols et al. \cite{brepols2020} and serves as the finite strain version of the small strain version of the model \cite{brepols2017}. In principle, the material model is based on the micromorphic approach proposed by Forest \cite{forest2009}. The approach of Forest can be seen as a systematic approach to construct high-level gradient material models of already existing 'local' counterparts. The model introduces an additional nodal degree of freedom for the micromorphic damage. The Helmholtz free energy depends on the difference between the global damage and the local damage which can be strongly coupled by a penalty parameter. This approach is chosen to overcome a general problem of conventional continuum mechanical local damage models, where the softening regime is highly dependent on the mesh discretization in a finite element simulation.

Because the model is implemented with finite deformation kinematics, the reference configuration $B_0$ at time $t_0$ and the current configuration $B_t$ at time $t$ of a body are differentiated. For the different configurations, the position of a material point is described by the vectors $\bm{X}$ and $\bm{x}$, respectively. The current position of a material point can be obtained by adding a displacement vector $\buu$ to the position of the material point in reference configuration such that $\bm{x} = \bm{X} + \buu$. Assuming there exists a mapping $\bm{x} = \bChi(\bm{X}, t)$, the deformation gradient can be introduced as $\bF \coloneqq \partial \bChi(\bm{X},t)/\partial \bm{X}$. To model elastoplastic material behavior, the deformation gradient is multiplicatively decomposed into the elastic part $\bF_e$ and the plastic part $\bF_p$. Because the model includes Armstrong-Frederick type non-linear kinematic hardening, the plastic part of the deformation gradient is further decomposed into the recoverable part $\bF_{p_e}$ and the irrecoverable part $\bF_{p_i}$. After these splits, the deformation gradient reads
\be
\bF = \bF_e \bF_p \text{~ with ~} \bF_p = \bF_{p_e} \bF_{p_i}
\ee
and from this, the elastic Cauchy-Green deformation tensors are derived as
\be
\bC_e = \bF_e\T \bF_e, \qquad \bC_{p_e} = \bF_{p_e} \T \bF_{p_e}.
\ee
Having defined these tensors, the Helmholtz free energy is assumed to be
\be
\psi = f_{\textrm{dam}}(D)(\psi_e(\bC_e)+\psi_p(\bC_{p_e}, \xi_p))+\psi_d(\xi_d) + \psi_{\bar{d}}(D-\Dbar, \Grad(\Dbar))
\ee
where $\psi_e$ and $\psi_p$ denote the elastic and plastic part of the free energy, respectively. In addition to the recoverable plastic part of the Cauchy-Green deformation tensor, the plastic part of the free energy also depends on the accumulated plastic strain $\xi_p$. The energy related to damage hardening is stored in $\psi_d$ and the energy related to the micromorphic extension is stored in $\psi_{\bar{d}}$. Furthermore, the damage hardening energy is dependent on the accumulated damage $\xi_d$ and the energy related to the micromorphic extension is dependent on the micromorphic damage variable $\Dbar$. Lastly, the scalar-valued weakening function $f_{\textrm{dam}}(D)$ is introduced. The elastic energy is chosen as a compressible Neo-Hookean type and reads
\be
\psi_e = \dfrac{\mu}{2}\left(\tr \bC_e - 3 - 2\ln\left(\sqrt{\det\bC_e}\right)\right) + \dfrac{\lambda}{4}\left(\det\bC_e -1 -2 \ln\left(\sqrt{\det\bC_e}\right)\right)
\ee
with the two Lamé constants $\mu$ and $\lambda$. The plastic energy is defined as
\be
\psi_p = \dfrac{a}{2}\left(\tr \bC_{p_e} - 3 - 2\ln\left(\sqrt{\det\bC_{p_e}}\right)\right) + e\left(\xi_p + \dfrac{\exp(-f \xi_p)-1}{f}\right).
\ee
The material parameters $a$, $e$ and $f$ are related to kinematic and non-linear Voce isotropic hardening, with another parameter $b$ showing up in the evolution equation of $\dot{\bC}_{p_i}$.
The energy related to Voce damage hardening is defined as
\be
\psi_d = r\left(\xi_d + \dfrac{\exp\left(-s\xi_d\right) - 1}{s} \right)
\ee
where $r$ and $s$ are the two damage hardening material parameters. The energy of the micromorphic extension reads
\be
\psi_{\bar{d}} = \dfrac{H}{2}(D-\Dbar)^2 + \dfrac{A}{2} \Grad(\Dbar)\cdot\Grad(\Dbar)
\ee
with the aforementioned penalty parameter $H$ and the parameter $A$. An internal length scale is introduced into the material as $L \coloneqq \sqrt{A / H}$ which can be understood as a localization limiter, preventing the aforementioned localization of damage into a zone of vanishing volume. Choosing the penalty parameter to be a very high number allows for a strong coupling between the micromorphic and the local damage variable. The constitutive equations with respect to the reference configuration are based on the Clausius-Duhem inequality in order to fulfill the second law of thermodynamics. The derivation of the constitutive equations is carefully described by Brepols et al. \cite{brepols2020} and because it is not the main topic of this thesis, only the resulting equations are given below:

$\bullet$ State laws:
\bea
\bS &= \fdam(D) \, 2 \, \bF_p\ME \dfrac{\partial \psi_e}{\partial \bC_e}\bF_p\MT\\
&= \fdam(D) \left(\mu \, (\bC_p\ME - \bC\ME) + \dfrac{\lambda}{2} \left(\dfrac{\det \bC}{\det \bC_p} - 1 \right) \bC\ME \right)
\eea
\be
a_{0_i} = \dfrac{\partial \psi_{\bar{d}}}{\partial \Dbar} = -H (D-\Dbar)
\label{eq:a}
\ee
\be
\bbbb_{0_i} = \dfrac{\partial \psi_{\bar{d}}}{\partial \Grad(\Dbar)} = A \, \Grad(\Dbar)
\label{eq:b}
\ee
$\bullet$ Thermodynamic conjugate forces - plasticity:
\be
\bX = \fdam(D) \, 2 \, \bF_{p_i}\ME \dfrac{\partial \psi_e}{\partial \bC_{p_e}}\bF_{p_i}\MT = \fdam(D)\,a\,(\bC_{p_i}\ME - \bC\ME)
\ee
\be
q_p = \fdam(D)\,\dfrac{\partial \psi_p}{\partial \xi_p} = \fdam(D)\,e\,(1 - \exp(-f\xi_p))
\ee
$\bullet$ Thermodynamic conjugate forces - damage:
\be
Y = - \dfrac{\dd \fdam(D)}{\dd D}\,(\psi_e+\psi_p)-\dfrac{\partial \psi_{\bar{d}}}{\partial D} = - \dfrac{\dd \fdam(D)}{\dd D}\,(\psi_e+\psi_p) - H(D-\Dbar)
\ee
\be
q_d = \dfrac{\partial \psi_d}{\partial \xi_d} = r\,(1 - \exp(-s \, \xi_d))
\ee
$\bullet$ Auxillary stress tensors:
\be
\bY = \bC \bS - \bC_p \bX, \quad \bY_{\textrm{kin}} = \bC_p \bX
\ee
$\bullet$ Evolution equations are expressed in terms of effective (i.e., 'undamaged') quantities:
\be
\tilde{(\bullet)} \coloneqq \frac{(\bullet)}{\fdam(D)}
\ee
\be
\dot{\bC}_p = 2 \, \dot{\lambda}_p \dfrac{\sqrt{3/2}}{\fdam(D)}\dfrac{\tilde{\bY}^{'} \bC_p}{\sqrt{\tilde{\bY}^{'} \cdot (\tilde{\bY}^{'})\T}}, \quad \dot{\bC}_{p_i} = 2 \, \dot{\lambda}_p \dfrac{b/a}{\fdam(D)} \bY^{'}_{\textrm{kin}} \bC_{p_i}
\ee
\be
\dot{\xi}_p = \dfrac{\dot{\lambda}_p}{\fdam(D)}, \quad \dot{D} = \dot{\lambda}_d, \quad \dot{\xi}_d = \dot{\lambda}_d
\ee
$\bullet$ Yield and damage loading function:
\be
\Phi_p = \sqrt{3/2} \sqrt{\tilde{\bY}^{'} \cdot (\tilde{\bY}^{'})\T} - (\sigma_0 + \tilde{q}_p), \quad \Phi_d = Y - (Y_0 + q_d)
\label{eq:threshold}
\ee
$\bullet$ Loading / unloading conditions
\bea
\dot{\lambda}_p \ge 0, \qquad \Phi_p \le 0, \qquad \dot{\lambda}_p \Phi_p = 0\\
\dot{\lambda}_d \ge 0, \qquad \Phi_d \le 0, \qquad \dot{\lambda}_d \Phi_d = 0
\eea
Adding the initial yield stress $\sigma_0$ and damage threshold $Y_0$ in equation \ref{eq:threshold}, the material model has twelve material parameters in total ($\lambda$, $\mu$, $\sigma_0$, $a$, $b$, $e$, $f$, $Y_0$, $r$, $s$, $A$, $H$) which should be optimally determined from experimental data. Finally, the micromorphic balance equation can be written as
\bea
H(D-\Dbar)+ A \, \Ddiv(\Grad(\Bar{D})) &= 0 \quad \text{in } B_0\\
\Grad(\Dbar) \cdot \bn_0 &= 0 \quad \text{on } \partial B_0
\eea
Having the micromorphic balance equation at hand as well as the balance of linear momentum
\bea
\Ddiv(\bF \bS) + \bff_0 &= 0 \quad &&\text{in } B_0 \\
\bF \bS[\bn_0] &= \bt_0 \quad &&\text{on } \partial B_0\\
\buu &= \hat{\buu} \quad &&\text{on } \partial B_0
\eea
the weak form of the problem can be constructed. This is done via the usual procedure: First, the balance equations are multiplied by the appropriate test functions $\delta \buu$ in case of the linear momentum and $\delta \Dbar$ in case of the micromorphic balance and then integrated over the reference domain $B_0$. After that partial integration and the divergence theorem are applied such that the weak form reads:
\be
g(\buu, \Dbar, \delta \buu) \coloneqq \int_{B_0} \bS \cdot \delta \bE \: \dd V - \int_{B_0} \bff_0 \cdot \delta \buu \: \dd V - \int_{\partial B_{0_t}} \bt_0 \cdot \delta \buu \: \dd A = 0 \quad \forall \delta \buu
\ee
\be
h(\buu, \Dbar, \delta \Dbar) \coloneqq \int_{B_0} (H(D- \Dbar)\delta \Dbar - A \, \Grad(\Dbar) \cdot \Grad(\delta \Dbar))\dd V = 0 \quad \forall \delta \Dbar
\ee
The weak form is generally nonlinear and must be linearized with respect to the unknowns and solved iteratively. Details about the linearization and subsequent finite element discretization can be found in \cite{brepols2020}.

Finally, one ends up with a global system of finite element equations which reads
\be
\underbrace{\left[\begin{array}{l l} \mathbf{K}_{uu} & \mathbf{K}_{u \Dbar}     \\
            \mathbf{K}_{\Dbar u}  & \mathbf{K}_{\Dbar \Dbar}
        \end{array} \right]}_{\mathbf{K}_T}
\underbrace{\left \{\begin{array}{l} \D \bm{u} \\
        \D \bDbar
    \end{array} \right \}}_{\D \bU}
= -
\underbrace{\left \{ \begin{array}{l} \bR_u + \bP_u \\
        \bR_{\Dbar}
    \end{array} \right \}}_{\bG}
\label{eq:soe}
\ee
The abbreviations $\mathbf{K}_T$ for the tangential stiffness matrix, $\D \bU$ for the incremental unknowns vector, and $\bG$ for the residual vector are introduced for later purposes. To get a solution to the problem at hand, the equation system \ref{eq:soe} is solved iteratively until convergence is achieved.
For further details regarding the individual components of the tangential stiffness matrix $\mathbf{K}_{uu}$, $\mathbf{K}_{u \Dbar}$, $\mathbf{K}_{\Dbar u}$, $\mathbf{K}_{\Dbar \Dbar}$ and the residual vector $\bR_u$, $\bP_u$, $\bR_{\Dbar}$ as well as the algorithmic implementation at the Gauss point level, the reader is again kindly referred to the work of Brepols et al. \cite{brepols2020}.
Because of the complexity of the material model, the computation of the stiffness matrix is fairly expensive. It should be noted at this point that in the present work, solely linear interpolation functions with 8 integration points per element are used. For completeness, the work of Barfusz et al. \cite{barfusz2021} should be highlighted, in which a single Gauss point formulation was implemented for the material model at hand and it was validated that the reduced integration approach still yields accurate results and can be used to increase the efficiency.
\section{Methods}
\label{sec:methods}

 In this chapter we delve into the various methods used to tackle the challenges of the proposed examples. Those methods include Proper Orthogonal Decomposition, Discrete Empirical Interpolation Method (DEIM) and the arc-length method. 

 The POD is a mathematical technique used to extract a reduced-order representation of a high-dimensional system. This method has found widespread use in mechanics. Although the primary objective is to decrease the time required for solving the system of equations, it is still necessary to establish the complete system.

 The Discrete Empirical Interpolation Method (DEIM) is another numerical method that has gained popularity in mechanics. It is a technique used to reduce the computational complexity of large-scale nonlinear systems. The DEIM only requires the calculation of a few key elements, as opposed to the entire system. Its combination with the POD technique makes it a time-efficient approach.

The arc-length method is a numerical technique used to solve nonlinear equations in mechanics. It is used to trace the response of a structure or system as it undergoes nonlinear deformation, particularly in relation to softening and snapback behavior.
This method has found wide application in structural engineering, where it is used to analyze the stability of structures under different loads.

\subsection{Proper Orthogonal Decomposition}
\label{sec:POD}
The proper orthogonal decomposition involves the decomposition of a data matrix, $\bD$, into its dominant modes or principal components using singular value decomposition (SVD).
In our case, we first solve the full system to obtain the different solution vectors $\bU_i (i=1,...,l)$, which are then used to construct the snapshot matrix, $\bD=[\bU_1, \bU_2, ..., \bU_l] \in \mathbb{R}^{n \times l}$.
This snapshot matrix is a time series of the state of the system at discrete time intervals with the solution $\bU_i$.
We then apply SVD to the snapshot matrix $\bD=\bPhi \bSigma \bV$, where $\bPhi=[\bvarPhi_1, \bvarPhi_2, ..., \bvarPhi_n]$ $\in \mathbb{R}^{nxn}$ and ${\bV}=[\bv_1, \bv_2, ..., \bv_l]$ $\in \mathbb{R}^{lxl}$ are orthonormal matrices, and $\bSigma \in \mathbb{R}^{nxl}$ is a diagonal matrix containing the singular values. The dominant modes or principal components of the system are then truncated from the left singular vectors of the matrix $\bPhi_m=[\bvarPhi_1, \bvarPhi_2, ..., \bvarPhi_m]$. The matrix $\bPhi_m$ is the so called projection matrix which contains the main characteristics of the system. The galerkin projection projects $\bU_{red}$ to the approximated full solution $\Ubar$
\be
\bU \approx \Ubar = \bPhi_m \bU_{red}
\ee
and the stiffness matrix
\be
\KTred(\Ubar) = \PhimT \KT(\Ubar) \Phim
\ee
into a smaller subspace. The reduced solution scheme for each newton iteration $i$ reads
\bega
\KTred (\Ubar \lti) \, \D \Ured \lti = - \Gred(\Ubar \lti)\\
\Ubar \ltipo = \Ubar \lti + \Phim \D \Ured \lti\\
\Gred \T (\Ubar \lti) \: \D \Ured \lti \leq tolerance\\
i = i + 1.
\label{eq:NRPOD}
\eega

Overall, the POD method provides a powerful tool for reducing the computational cost of solving high-dimensional systems, allowing researchers to gain insights into the most critical features of the data while reducing the computational complexity.

\subsection{Discrete Empirical Interpolation Method}
\label{sec:DEIM}

In this chapter, we explore the Discrete Empirical Interpolation Method (DEIM) and its application in the context of computational mechanics.

The DEIM is a numerical technique that has gained significant attention in recent years due to its ability to reduce the computational complexity of large-scale nonlinear systems. It is particularly useful in the context of computational mechanics, where the behavior of materials is often modeled using partial differential equations (PDEs) and solved numerically. It starts by constructing a set of interpolation points that capture the relevant behavior of the system. These points are typically chosen based on the system's response or other relevant criteria. The system's state variables or parameters are then projected onto these interpolation points, resulting in a reduced-dimensional system.
One of the key advantages of DEIM is its ability to accurately approximate the system's response using only a small number of interpolation points. This allows for significant reduction in computational costs, making it well-suited for large-scale problems commonly encountered in computational mechanics.



\subsubsection{Mathematical formulation}

Similar to the previous chapter on POD, preliminary simulations are necessary in this context as well.
Given the proficiency of POD in handling linear simulations and the complementary nature of DEIM in addressing nonlinear behavior, we partition the internal force vector $\bR(\bU)$ into distinct linear $\bR_{lin}(\bU)$ and nonlinear $\bR_{nl}(\bU)$ components.
This was already proposed by Radermacher and Reese \cite{radermacher2016} to avoid singular matrices.
The linear part is defined as $\bR_{lin}(\bU) = \bK_{lin} \bU$ and therefore the nonlinear part results in
\be  \bR_{nl}(\bU) = \bR(\bU) - \bK_{lin} \bU \ee
where $\bK_{lin}$ is the constant stiffness matrix obtained in the undeformed state.
The residual vector without MOR is given by $\bG = \bR_{lin} + \bR_{nl} - \bP$ where $\bP$ is the external force vector. Since the linear component $\bR_{lin}$ can be well approximated with POD, DEIM is additionally applied to $\bR_{nl}(\bU)$, as otherwise the full stiffness matrix for the nonlinear component would need to be recalculated for each time step.
Applying POD and DEIM to the residual vector results in the following equation:
\be \label{eq:DEIM_Residuum}  \Gred(\Ubar):= \Klinred ~\bU_{red} + \Rnlred(\Ubar) -\PhimT \bP = \bzero \ee

where $\Gred(\Ubar)$ is the reduced residual vector obtained through POD and DEIM, $\bR_{lin,red} =  \Klinred \bU_{red}$ is the approximated linear component using POD,
\be \label{eq:Rnlred} \Rnlred(\Ubar) = \PhimT \Omk (\ZT \Omk)^{-1} \ZT \Rnl(\Ubar) \ee
is the reduced nonlinear component obtained using DEIM and POD, and $\bP_{red} = \PhimT \bP$ is the reduced projection of the load using POD.
This split was proposed by Radermacher et al. \cite{radermacher2016}.
It is important to mention that in equation \ref{eq:Rnlred} most of the quantities are precomputed.
This includes $\Klinred$ and a large part of $\Rnlred(\Ubar)$, which can be combined into the new matrix
\be \MDEIM = \PhimT \Omk (\ZT \Omk)^{-1}. \ee
These matrices are not recalculated in the reduced simulation and are constant throughout the computation.
The matrix $\Phim$ is the matrix we get from the POD approach. New are the matrices $\Omk$ and $\ZT$. The matrix $\Omk$ is a projection matrix for the nonlinear internal force vector, and its interpretation is similar to that of the projection matrix used in POD.
The matrix $\ZT$ is the selection matrix and is composed of individual unit vectors, indicating which degrees of freedom and therefore which elements need to be evaluated.
While the matrix $\Rnl(\Ubar)$ still exists in its full dimension, the multiplication with $\ZT$ determines which entries of $\Rnl(\Ubar)$ are actually needed, and only those entries and their corresponding elements need to be evaluated.

\begin{algorithm}[H]
    \vskip\medskipamount 
    \leaders\vrule width \textwidth\vskip0.4pt 
    \vskip\medskipamount 
    \nointerlineskip
    \KwIn{$\bD_R=[\bR_{nl}(\bU(t_1)) ,\, \bR_{nl}(\bU(t_2)),\, \dots ,\, \bR_{nl}(\bU(t_j))]$, $k$}
        \KwOut{$\bOmega_k$, $\bZ$}
    $\mathbf{\Omega \Sigma V}=\bD_R$\\
    $\bOmega_k=[\bomega_1 ,\, \bomega_2 ,\, \dots ,\, \bomega_k]$\\
    $[\rho,\gamma_1]=\max|\bomega_1|$\\
    $\bZ_1=[\bee_{\gamma_1}]$\\
        \For{i=1 \text{to} k-1}{
    $\bZ_i^T \bOmega_i \bcc_i = \bZ_i^T \bomega_{i+1}$\\
    $\bR_{i+1}=\bomega_{i+1}-\bOmega_i\bcc_i$\\
    $[\rho,\gamma_i]=\max|\bR_{i+1}|$\\
    $\bZ_{i+1}=[\bZ_i \, \bee_{\gamma_i}]$\\
    }
    \leaders\vrule width \textwidth\vskip0.4pt 
    \vskip\smallskipamount 
    \nointerlineskip
    \caption{DEIM algorithm for finding indices}
    \label{algo:DEIM}
\end{algorithm}

To obtain the two matrices, we use the approach proposed by Chaturantabut and Sorensen \cite{chatsor}, or Radermacher and Reese \cite{radermacher2016}.
This is illustrated in Algorithm \ref{algo:DEIM}, where the snapshots of the nonlinear component of the internal force vector $\bD_R$ and the desired number of DEIM modes $k$ are used as input.
By performing a SVD, the matrices $\bD_R = \mathbf{\Omega \Sigma V}$ are obtained. The first $k$ vectors of the matrix $\mathbf{\Omega} = [\bomega_1 , \bomega_2 , \dots , \bomega_j]$ yield the projection matrix $\bOmega_k=[\bomega_1 , \bomega_2 , \dots , \bomega_k]$.
The selection of discrete interpolation points is done in a greedy manner.
In the initial step, the position $\gamma_1$ of the maximum value of the first DEIM mode is found as $[\rho,\gamma_1]=\max|\bomega_1|$. Where $\rho$ is the maximum value and $\gamma_l$ the index which can be interpreted as the position with the corresponding degree of freedom.
This position is chosen as the first interpolation point $\bZ_1=[\bee_{\gamma_1}]$.
In the subsequent stages, the next DEIM mode is considered, and efforts are made to subtract the previously determined modes as efficiently as possible.
This approach takes into account the directions in which significant knowledge already exists, allowing for the selection of interpolation points that provide new information.
The exact formulas for this process are shown in Algorithm \ref{algo:DEIM}.
Finally, the previously introduced residual vector eq. (\ref{eq:DEIM_Residuum}) is solved using the Newton-Raphson method.
A time step of this process is shown in eq.~\ref{eq:NRPOD}.
Using the DEIM in the field of finite element simulations with classically assembled meshes, it should be mentioned that determining the values at individual discrete points in the mesh is not straightforward. In this method, in order to evaluate the values at a single point, the adjacent elements need to be evaluated as well.
Nevertheless, there are strategies available to overcome this issue and maintain the efficiency of the original DEIM method \cite{MDEIM}.
\subsection{Arc-length method}

\label{sec:arcl} 

As discussed in Chapter \ref{sec:mat}, the material model being considered exhibits softening behavior. This means that in a load-controlled simulation, the simulation can only run until the limit load is reached at best. Further loading of the material does not result in equilibrium and no more solutions can be found. Displacement-controlled simulation is the simplest solution to this problem, allowing for simulation of the material beyond the limit load into the softening regime. However, this approach is not suitable for materials that exhibit snap-back behavior.

\begin{figure}[ht]
    \centering
    \def\svgwidth{1\textwidth}
    \begin{subfigure}{.48\textwidth}
        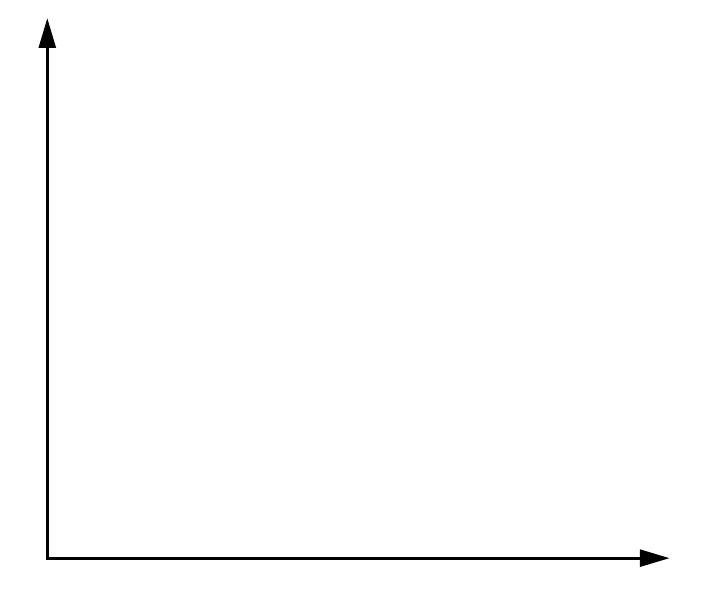
        \caption{Load-controlled simulation.}
        \label{fig:whyarclenload}
    \end{subfigure}
    \def\svgwidth{1\textwidth}
    \begin{subfigure}{.48\textwidth}
        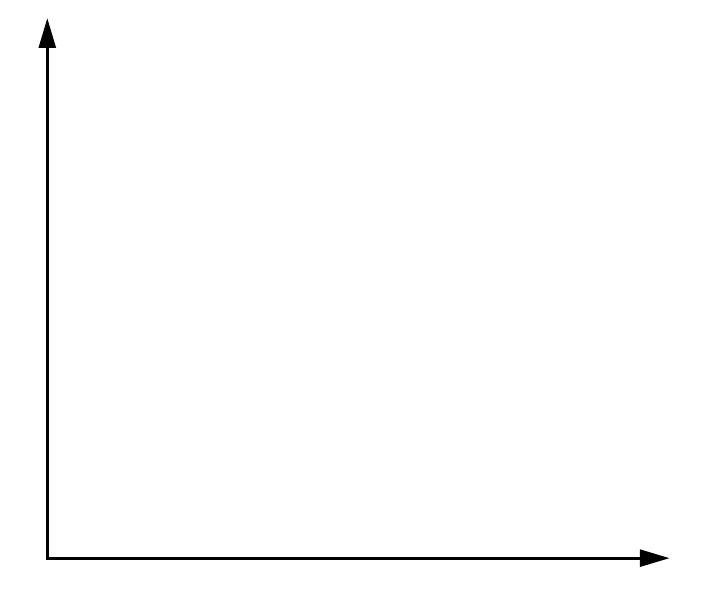 
        \caption{Displacement-controlled simulation.}
        \label{fig:whyarclendisp}
    \end{subfigure}
    \caption{Limitations of standard load-controlled and displacement-controlled techniques.}
    \label{fig:whyarclen}
\end{figure}

Snap-back is a phenomenon where not only the load decreases, but also the displacement, as shown in the exemplary load-displacement curve in Figure \ref{fig:whyarclen}. Figure \ref{fig:whyarclenload} illustrates the problem of load-controlled simulation, where no equilibrium state exists when the prescribed load becomes too large. Figure \ref{fig:whyarclendisp} visualizes the issues with displacement-controlled simulation, where a unique solution does not exist for the prescribed displacement $u_6$.

\begin{figure}[ht]
    \centering
    \def\svgwidth{.8\textwidth}
    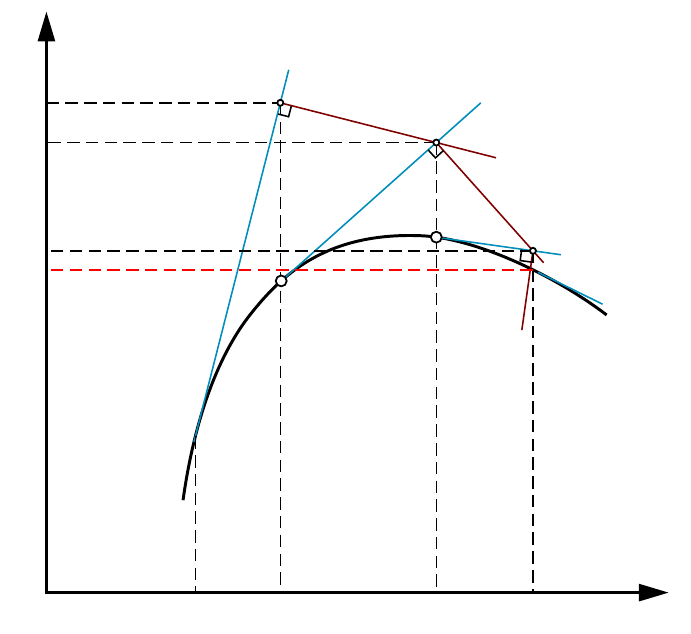
    \caption{Visual representation of the arc-length method by Ramm.}
    \label{fig:arclen}
\end{figure}

To overcome these problems, the constant arc-length method was developed by Riks \cite{riks1972, riks1979} and Wempner \cite{wempner1971}, and made computationally efficient by Ramm \cite{ramm1981} and Crisfield \cite{crisfield1981} in slightly different ways as shown in Fig. \ref{fig:arclen}.

\subsubsection{Modified Riks-Wempner method}

Suppose that a solution exists from a previous pseudo-timestep $t-1$ with $\lt{t-1}\bU = \lt{t}\bU\ltz$ and $\lt{t-1}\lambda = \lt{t}\lambda\ltz$, and we want to obtain the solution for the next pseudo-timestep $\lt{t}\bU$ and $\lt{t}\lambda$. For simplicity, we will omit the left superscript for the pseudo-timestep $t$. We will indicate the iteration counter with a right superscript in brackets. At the beginning of the simulation (or at the first pseudo-timestep), we need to set a few parameters: the arc-length $\Delta s_0$, the reference load $\bP_0$, and a predefined change in the load level $\Delta\lambda_0$. Once these parameters are set, we can initiate a new pseudo-timestep.

First, we increase the load level by the predefined change as follows:

\begin{equation}
    \lambda = \lt{t-1}\lambda + \Delta\lambda_0.
\end{equation}

Next, the tangential stiffness matrix $\bK_T \ltz$ and the vector of internal forces $\Rint \ltz$ can be computed. Once these items have been computed and assembled, the vector of external forces $\bP \ltz$ can be calculated as
\be
\bP \ltz= \bP_0 \lam \ltz.
\ee

It's important to note that the external force vector $\bP \ltz$ is determined by multiplying the reference load $\bP_0$ by the current load factor $\lam \ltz$.

This allows for the computation of the residual vector
\be
\bG \lt0 = \Rint \lt0 - \bP \lt0.
\ee
With this, the linear system of equations can be solved to obtain the incremental displacement $\D \bU \ltz$
\be
\bK_T \ltz \D \bU \lt{1} = \bG \ltz.
\ee
After calculating the current incremental displacement, the arc-length $\D s$ of the pseudo-timestep can be determined
\be
\D s = \norm{\D \bU\lt{1}}.
\label{eq:Dsfull}
\ee

Another important aspect of the algorithm is handling the change of sign of $\D \lam$ at limit points. To provide a comprehensive understanding for interested readers and facilitate the implementation of the method into a FEM code, the conditions for changing the sign will be explained as follows.

First, the stiffness parameter for the current pseudo-timestep needs to be computed

\be
c = \D \bU^{(1) \: \top} \bP_0.
\ee

Next, we evaluate the ratio between the current stiffness parameter and the initial stiffness parameter:

\be
r_c = \dfrac{\ltz c}{c}.
\ee

Assuming a positive stiffness parameter in the first pseudo-timestep, the ratio becomes negative when a limit point is passed, indicating that the current stiffness parameter has become negative. However, to ensure the generality of the implementation, we cannot rely solely on this ratio to determine the sign of $\D \lam$. Instead, starting from the second pseudo-timestep, the following condition is used

\be
r =
\begin{cases}
    \begin{aligned}
         & -\operatorname{sgn}(\lt{t-1} r) \quad \begin{array}{l} \text{ if } (\lt{t-1}r_c > 0 \text{ and } r_c < 0 \text{ and } \lt{t-1}r_c < \lt{t-2}r_c) \\ \text{ or } (\lt{t-1}r_c < 0 \text{ and } r_c > 0 \text{ and } \lt{t-1}r_c > \lt{t-2}r_c) \end{array} \\
         & \lt{t-1} r
    \end{aligned}
\end{cases}
\ee

In the first pseudo-timestep $r$ is initiated as $+1$ and following that, $r$ is either $+1$ or $-1$ depending on the aforementioned conditional statement. Having done all these steps, the correction of the load factor can be calculated as
\be
\D \lam \ltz = \dfrac{\D s_0}{\D s}*r.
\ee
The correction is added to the initial load factor at the beginning of the iteration leading to the load level for the next iteration
\be
\lam \lt{1} = \lam \ltz + \D \lam \ltz - \D \lam_0
\ee
Also, the the incremental change of the displacement vector is updated such that
\be
\D \bU_\lam^{(1)} = \D \bU \lt{1} * \D \lam \ltz = \D \bU_I.
\ee
The expression $\D \bU_I$ for the incremental displacement of the first newton iteration of the pseudo-timestep is introduced for clarity, because it will be used in every Newton iteration after the first one.
Lastly, the vector of unknowns can be updated
\be
\bU \lt{2} = \bU \lt{1} + \D \bU_\lam^{(1)}.
\ee
For there on, the "normal" newton iteration ($i=1,2,\dots$) starts, beginning with the computation and assembly of the tangential stiffness matrix $\KT \lti$  and the vector of internal forces $\Rint \lti$. From there on, the load vector can be calculated in dependency of the previously adjusted load factor
\be
\bP \lti = \bP_0 * \lam \lti.
\label{eq:newtonstart}
\ee
Again, this will be used for the calculation of the residual vector
\be
\bG \lti = \Rint \lti - \bP \lti .
\ee
Following that, two incremental displacement vectors have to be calculated. The first is the result of solving
\be
\bK_T^{(i)} \D \bU \lti = \bG \lti .
\ee
for $\D \bU \lti$ and the second one is the result of solving
\be
\bK_T^{(i)} \D \bU_{II} \lti = \bP_0
\ee
for $\D \lti \bU_{II}$. These two solution vectors can now be used to calculate the new increment for the load factor
\be
\D \lam \lti = - \dfrac{\D \bU_I^\top \D \bU \lti}{\D \bU_I^\top \D \bU_{II}^{(i)}}
\label{eq:Dilamfull}
\ee
The increment is added to the load level from the beginning of the iteration to get the load level for the next iteration
\be
\lam \lt{i+1} = \lam \lti + \D \lambda \lti
\ee
and also to update the increment of the displacement vector
\be
\D \bU_\lam^{(i)} = \D \bU \lti + \D \lam \lti * \D \bU_{II}^{(i)}.
\ee
Lastly, the displacement vector is updated for the next iteration
\be
\bU \lt{i+1} = \bU \lti + \D \bU_\lam^{(i)} .
\ee

This procedure from equation \ref{eq:newtonstart} on is repeated until the desired accuracy or the maximum number of iterations is reached. The method is also presented in algorithm \ref{algo:arclen} for a more compact overview. From the viewpoint of computational complexity, the second system of linear
\begin{algorithm}[!ht]
    \leaders\vrule width \textwidth\vskip0.4pt 
    \vskip\medskipamount 
    \nointerlineskip
    \KwIn{$\lt{t-1}\bU$, $\lt{t-1}\lambda$, $\D s_0$, $\bP_0$, $\D \lambda_0$}
    \KwResult{$\lt{t}\bU$, $\lt{t}\lambda$}
    Compute and assemble $\KT \lt0$ and $\Rint \lt0$\\
    $\lam \ltz = \lt{t-1}\lam + \D \lam_0$\\
    $\bP \ltz = \bP_0 * \lam \ltz$\\
    $\bG \ltz = \Rint \lt0 - \bP \lt0 $\\
    $\bK_T \ltz \D \bU \lt{t-1} = \bG \ltz $ $\rightarrow$ solve for $\D  \bU\lt{1}$\\
    $\D s = \norm{\D \bU \lt{t-1}}$\\
    $c = \D \bU^{(0) \: \top} \bP_0$\\
    $r_c = \dfrac{\ltz c}{c}$\\
    $
        r =
        \begin{cases}
            \begin{aligned}
                 & -\operatorname{sgn}(\lt{t-1}r) \quad \begin{array}{l} \text{ if } (\lt{t-1}r_c > 0 \wedge r_c < 0 \wedge \lt{t-1}r_c < \lt{t-2}r_c) \\ \vee (\lt{t-1}r_c < 0 \wedge r_c > 0 \wedge \lt{t-1}r_c > \lt{t-2}r_c) \end{array} \\
                 & \lt{t-1}r
            \end{aligned}
        \end{cases}
    $\\
    $\D \lam \ltz = \dfrac{\D s_0}{\D s}*r$\\
    $\lam \lt{1} = \lam \ltz + \D \lam \ltz - \D \lam_0$\\
    $\D \bU_\lam^{(1)} = \D \bU \lt{t-1} * \D \lam \ltz = \D \bU_I$\\
    $\bU \lt{2} = \bU \lt{1}+ \D \bU_\lam\ltz$\\
    \For{$i=1$ to $i_{max}$}{
        Compute and assemble $\KTi$ and $\Ri$\\
        $\bP \lti= \bP_0 * \lam \lti$\\
        $\bG \lti = \Ri - \bP \lti$\\
        $\KTi \D \bU \lti = \bG \lti$ $\rightarrow$ solve for $\D \bU \lti$\\
        $\KTi \D \bU_{II}^{(i)} = \bP_0$ $\rightarrow$ solve for $\D \bU_{II}^{(i)}$\\
        $\D \lam \lti = - \dfrac{\D \bU_I^\top \D \bU \lti}{\D \bU_I^\top \D \bU_{II}^{(i)}}$\\
        $\D \bU_\lam^{(i)} = \D \bU \lti + \D \lam \lti * \D \bU_{II}^{(i)}$\\
        $\lam \lt{i+1} = \lam \lti + \D \lambda \lti$\\
        $\bU \lt{i+1} = \bU \lti + \D \bU \lti$\\
        Check for convergence \\
    }
    \leaders\vrule width \textwidth\vskip0.4pt 
    \vskip\smallskipamount 
    \nointerlineskip
    \caption{Algorithmic solution scheme with the arc-length method applied.}
    \label{algo:arclen}
\end{algorithm}
equations that has to be solved in every iteration doubles the CPU-time necessary to solve the system. This is one aspect of why it is of interest to implement model order reduction into the arc-length method. Another aspect is the fact that in a DEIM-reduced simulation it is not possible to calculate some of the values in the full space because for example the stiffness matrix is only computed at interpolation points and can therefore not be used.

\subsubsection{Reduced arc-length method}

Having thoroughly explained the discrete empirical interpolation method and the 'full' Riks-Wempner arc-length method as proposed by Ramm, the implementation of the arc-length method in a reduced vector space is straight forward. It should be mentioned that a hyperreduction was already successfully applied to the the arc-length method proposed by Crisfield \cite{crisfield1981} by Lunay et al. \cite{hrarclen}, but to the knowledge of the author, hyperreduction has never been applied to Ramms arc-length method.

As explained in chapter \ref{sec:DEIM}, the stiffness matrix $\KT$ and the vector of internal forces $\Rint$ are not fully computed, but only for the DOFs that were predefined in the offline step of the simulation. This means, that every value calculated in the arc-length method procedure is calculated from vectors and matrices in the reduced vector space.

Of course, everything stated concerning the offline step of the simulation are still applicable when the reduced arc-length method shall be used. This means that the matrices $\Phim$, $\MDEIM$ and $\ZT$ have been computed already.

Concerning the arc-length method part of the solution scheme, there are a few equations that are majorly influenced. These equations will be explained in the following.

First of all, equation \ref{eq:Dsfull} does actually not change, because the norm of the vector $\D \bU$ is approximately equal to the norm of $\D \Ured$. This is due to the orthogonality of the basis vectors $\Phim$ and $\Omk$. This means that the arc-length for pseudo-timestep $t$ is the same whether it is computed in the reduced vector space or the full vector space. Because the vector $\D \Ured$ is solved in the reduced space, it can only be used to approximate the vector $\D \bU$. Therefore, an approximation error is introduced to the arc-length of the pseudo-timestep. Going further the expression for the stiffness parameter $c$ now reads
\be
c = \D \Ured^{(0)\:\top} \bP_{0 \: red}
\ee
Again, the norm of both vectors is approximately equal to their norm in the full vector space and as a result, the dot product (represented by the matrix multiplication of both vectors, where the first one is transposed) leads to the same result in the reduced space as it would lead to in the original space. The steps after that are not changed in any way and as such, the reduced incremental displacement can be updated and added to the total reduced displacements such that
\be
\Ured^{(1)} = \Ured^{(0)} + \D \bU_{\lam \: red}^{(0)}
\ee
The solution must then be projected back to the full vector space, as it is necessary for the computation of the tangential stiffness matrix and the vector of internal forces. This is done via the conjuncture known from chapter \ref{sec:POD}
\be
\bU \lt{1} = \Phim \Ured^{(1)}
\ee
\begin{algorithm}[!ht]
    \leaders\vrule width \textwidth\vskip0.4pt 
    \vskip\medskipamount 
    \nointerlineskip
    \KwIn{$\lb{t-1}\bU$, $\lb{t-1}\Ured$, $\lb{t-1}\lambda$, $\D s_0$, $\bP_0$, $\D \lambda_0$}
    \KwResult{$\lb{t}\bU$, $\lb{t}\lambda$}
    Compute and assemble $\KTred^{(0)}$ and $\Rred^{(0)}$\\
    $\lam \ltz= \lt{t-1}\lam + \D \lam_0$\\
    $\Pred^{(0)} = \bP_{0 \: red} * \lam \ltz$\\
    $\Gred^{(0)} = \Rred^{(0)} - \Pred^{(0)}$\\
    $\KTred^{(0)} \D \Ured^{(0)} = \Gred^{(0)}$ $\rightarrow$ solve for $\D \Ured^{(0)}$\\
    $\D s = \norm{\D \Ured^{(0)}}$\\
    $c = \D \Ured^{(0) \: \top} \bP_{0 \: red}$\\
    $r_c = \dfrac{\ltz c}{c}$\\
    $
        r =
        \begin{cases}
            \begin{aligned}
                 & -\operatorname{sgn}(r) \quad \begin{array}{l} \text{ if } (\lt{t-1}r_c > 0 \wedge r_c < 0 \wedge \lt{t-1}r_c < \lt{t-2}r_c) \\ \vee (\lt{t-1}r_c < 0 \wedge r_c > 0 \wedge \lt{t-1}r_c > \lt{t-2}r_c) \end{array} \\
                 & \lt{t-1}r
            \end{aligned}
        \end{cases}
    $\\
    $\D \lam \ltz= \dfrac{\D s_0}{\D s}*r$\\
    $\lam \lt{1} = \lam \ltz + \D \lam \ltz - \D \lam_0$\\
    $\D \bU_{\lam \: red}^{(0)} = \D \Ured^{(0)} * \D \lam^{(0)} = \D \bU_{I \: red}$\\
    $\Ured^{(1)} = \Ured^{(0)} + \bU_{\lam \: red}^{(0)}$\\
    $\bU \lt{1} = \Phim \Ured^{(1)}$\\
    \For{$i=1$ to $i_{max}$}{
        Compute and assemble $\KTred^{(i)}$ and $\Rred^{(i)}$\\
        $\Pred^{(i)} = \Pzred * \lam^{(i)}$\\
        $\Gred^{(i)} = \Rred^{(i)} - \Pred^{(i)}$\\
        $\KTred^{(i)} \D \Ured^{(i)} = \Gred^{(i)}$ $\rightarrow$ solve for $\D \Ured^{(i)}$\\
        $\KTred^{(i)} \D \bU_{II \: red}^{(i)} = \Pred^{(i)}$ $\rightarrow$ solve for $\D \bU_{II \: red}^{(i)}$\\
        $\D \lam^{(i)} = - \dfrac{\D \bU_{I \: red}^\top \D \Ured^{(i)}}{\D \bU_{I \: red}^\top \D \bU_{II \: red}^{(i)}}$\\
        $\D \bU_{\lam \: red}^{(i)} = \D \Ured^{(i)} + \D \lam^{(i)} * \D \bU_{II \: red}^{(i)}$\\
        $\lam^{(i+1)} = \lam^{(i)} + \D \lambda^{(i)}$\\
        $\Ured^{(i+1)} = \Ured^{(i)} + \D \bU_{\lam \: red}^{(i)}$\\
        $\bU^{(i+1)} = \Phim \Ured^{(i+1)}$\\
        Check for convergence \\
    }
    \leaders\vrule width \textwidth\vskip0.4pt 
    \vskip\smallskipamount 
    \nointerlineskip
    \caption{Algorithmic solution scheme with the reduced arc-length method applied.}
    \label{algo:arclenred}
\end{algorithm}
After the initial step is finished, the newton iteration starts. Again, the necessary entries of the tangential stiffness matrix and the internal force vector are computed and put into the desired form. From there, two reduced systems of linear equations are solved in order to calculate $\D \Ured^{(i)}$ and $\D \bU_{II \: red}^{(i)}$. The two reduced systems of linear equations read
\be
(\Klinred + \MDEIM \ZT \KTnl^{(i)} \Phim) \D \Ured^{(i)} = \Gred^{(i)}
\ee
and
\be
(\Klinred + \MDEIM \ZT \KTnl^{(i)} \Phim) \D \bU_{II \: red}^{(i)} = \Pzred.
\ee
The obtained vectors are, similarly to equation \ref{eq:Dilamfull}, used to calculate the incremental load factor
\be
\D \lam \lti = - \dfrac{\D \bU_{I \: red}^\top \D \Ured^{(i)}}{\D \bU_{I \: red}^\top \D \bU_{II \: red}^{(i)}}.
\ee
This is then used to update the incremental displacement vector which is added to the vector of reduced unknowns. Similar to the initialization part of the pseudo-timestep, this vector is then projected to the full vector space so that the next iteration can take place. To highlight the differences of the full arc-length method and its reduced counterpart, the algorithmic form is shown in Algorithm \ref{algo:arclenred}, where all the equations and steps necessary for the implementation of a DEIM-reduced arc-length method are layed out.

\FloatBarrier
\section{Numerical examples}
\label{sec:results}
In this section, we present two illustrative examples that demonstrate the possible application and the challenges of the so far proposed methods. These examples have been carefully chosen to showcase the model's capability to accurately predict the behavior of complex geometries and loading conditions. At the same time the limitations of the methods are shown.

The first example focuses on a plate with a hole, a commonly encountered structural component closely related to various engineering applications. The presence of a hole significantly affects the stress distribution and fracture behavior, making it an ideal case to assess the reliability and accuracy of model order reduction methods. By investigating the response of this configuration, we aim to highlight the reduced order models ability to capture strain localization, and potential failure initiation and propagation.

The second example revolves around an asymmetrically notched specimen, which introduces an additional challenge due to the asymmetrical stress concentrations it induces. Notches are frequently encountered in structural components and can lead to significant stress concentration and failure initiation. By considering asymmetric notches, we aim to evaluate the model order reduction techniques in a highly challenging example.

These examples will serve as valuable benchmarks for assessing the accuracy and applicability of our methods in practical scenarios, providing insights into the behavior of complex geometries and loading conditions. Through detailed analysis, we will demonstrate the model's ability to capture key mechanical phenomena, enabling engineers and researchers to make informed design decisions and enhance the reliability of their structural components.

\subsection{Rectangular plate with a hole}

\begin{figure}[!ht]
    \centering
    \begin{subfigure}[t]{.43\textwidth}
        \centering
        \def\svgwidth{\textwidth}
        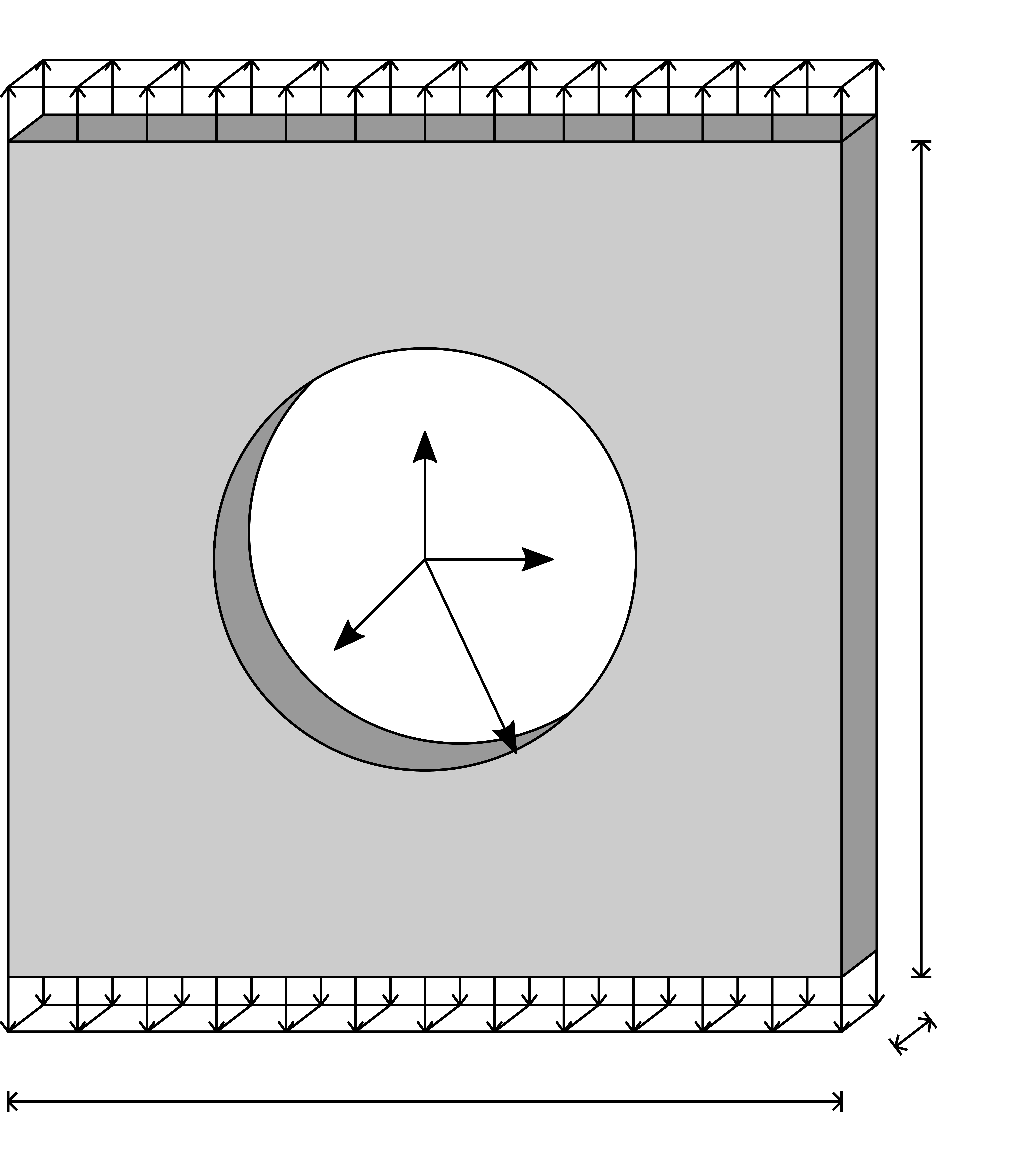
        \caption{Geometric configuration (dimensions in [mm])}
        \label{fig:Plate}
    \end{subfigure}
    \hfill
    \begin{subfigure}[t]{.52\textwidth}
        \centering
        \def\svgwidth{\textwidth}
        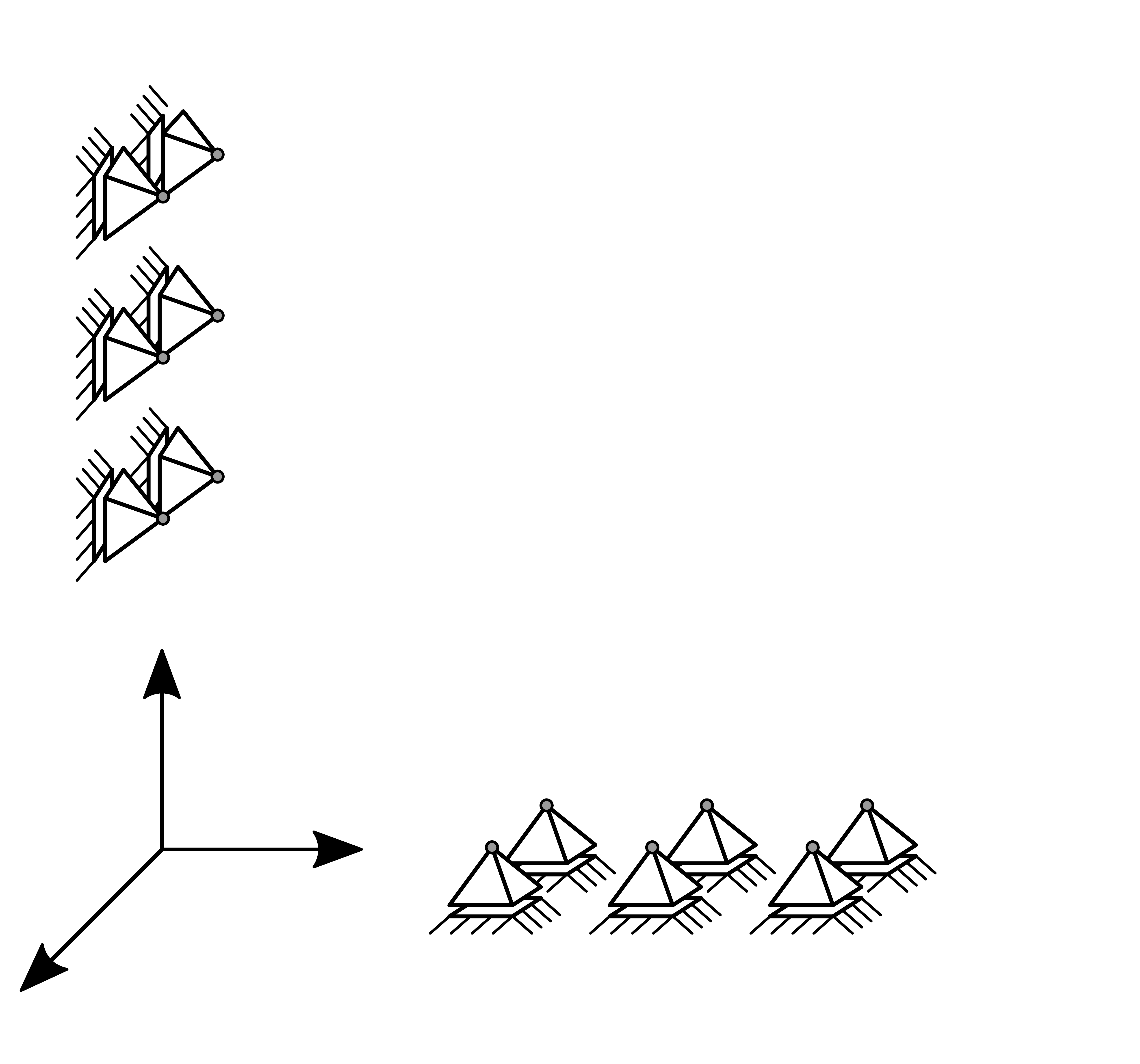
        \caption{Quarter-Symmetry representation with applied boundary conditions and loads}
        \label{fig:Plate_quarter}
    \end{subfigure}
    \caption{Rectangular plate with a circular hole}
    \label{fig:ExamplePlateHole}
\end{figure}

The dimensions of the plate with a hole can be seen in Figure \ref{fig:Plate}. Due to the symmetry of the example, a quarter of the plate can be considered. This quarter is depicted in Figure \ref{fig:Plate_quarter}, along with the corresponding boundary conditions and applied loads.

The material parameters used for the simulations are listed in Table \ref{tab:paraplate}. There parameters are arbitrarily chosen, as the material parameters have not yet been fitted to experimental data. Therefore they have been chosen in such a way that both plasticity and damage develop in an intuitive way.
Furthermore, efforts were made to ensure a complex material behavior, pushing the limits of the methods employed. It is important to note that the objective of this publication is not to investigate a specific material but rather to demonstrate the application of model reduction techniques to complex materialsand more specifically gradient-extended models.

\begin{table}[ht]
    \centering
    \begin{tabular}{l | l | l | l}
        \hline
        Symbol     & Material parameter                   & Value \hspace{1.5cm} & Unit                     \\
        \hline
        $\Lambda$  & first Lamé parameter                 & 75000                & MPa                      \\
        $\mu$      & second Lamé parameter                & 140000               & MPa                      \\
        $\sigma_0$ & yield stress                         & 325                  & MPa                      \\
        $a$        & first kinematic hardening parameter  & 2600                 & MPa                      \\
        $b$        & second kinematic hardening parameter & 12.5                 & [-]                      \\
        $e$        & first isotropic hardening parameter  & 500                  & MPa                      \\
        $f$        & second isotropic hardening parameter & 8                    & [-]                      \\
        $Y_0$      & damage threshold                     & 10                   & MPa                      \\
        $r$        & first damage hardening parameter     & 0.5                  & MPa                      \\
        $s$        & second damage hardening parameter    & 1                    & MPa                      \\
        $A$        & internal length scale parameter      & 50                   & MPa $\text{mm}^\text{2}$ \\
        $H$        & penalty parameter                    & $\text{10}^\text{5}$ & MPa
    \end{tabular}
    \caption{Parameter set for the plate with a hole.}
    \label{tab:paraplate}
\end{table}

\begin{figure}[!ht]
    \centering
    \input{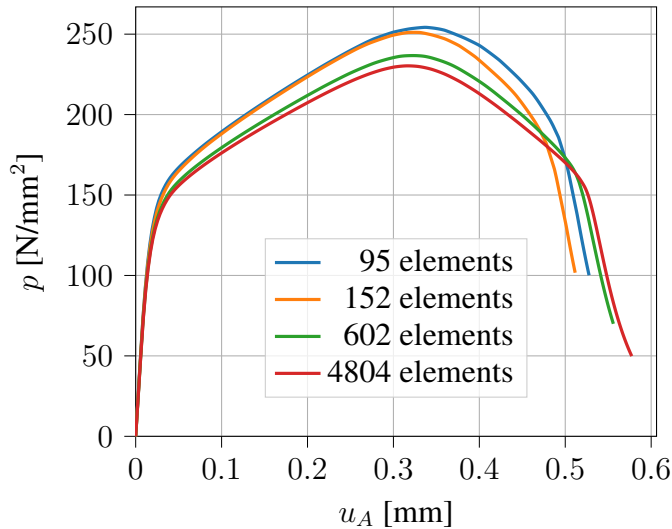}
    \caption{Mesh convergence study of the rectangular plate with a circular hole. $u_a$ is the displacement in Point A shown in Figure \ref{fig:Plate_quarter}}
    \label{fig:mesh_conv_plate}
\end{figure}

In the first step, mesh convergence is investigated to determine the number of elements required for a converged solution. This is done for the full simulation without any MOR. The load-displacement curves are shown in Figure \ref{fig:mesh_conv_plate}. All four calculations were performed using 300 pseudo-timesteps, and the arc-length method was employed to control the load factor. It can be observed that the results converge as the mesh gets finer. Although the solution with 4804 elements has not fully converged, this mesh density is used for subsequent calculations. The limit load, denoted as $p_{max}$, represents the maximum load that can be applied to the structure, beyond which further loading is not possible. To compute the softening regime of the simulation, the arc-length method is used to reduce the load accordingly. For this example it could be calculated displacement-controlled. Due to the desired universal appicability of the methodology, the arc-length method is used nevertheless.

\begin{figure}[!ht]
    \centering
    \begin{subfigure}{.72\textwidth}
        \centering
        \begin{subfigure}{.5\textwidth}
            \centering
            \includegraphics[width=\textwidth]{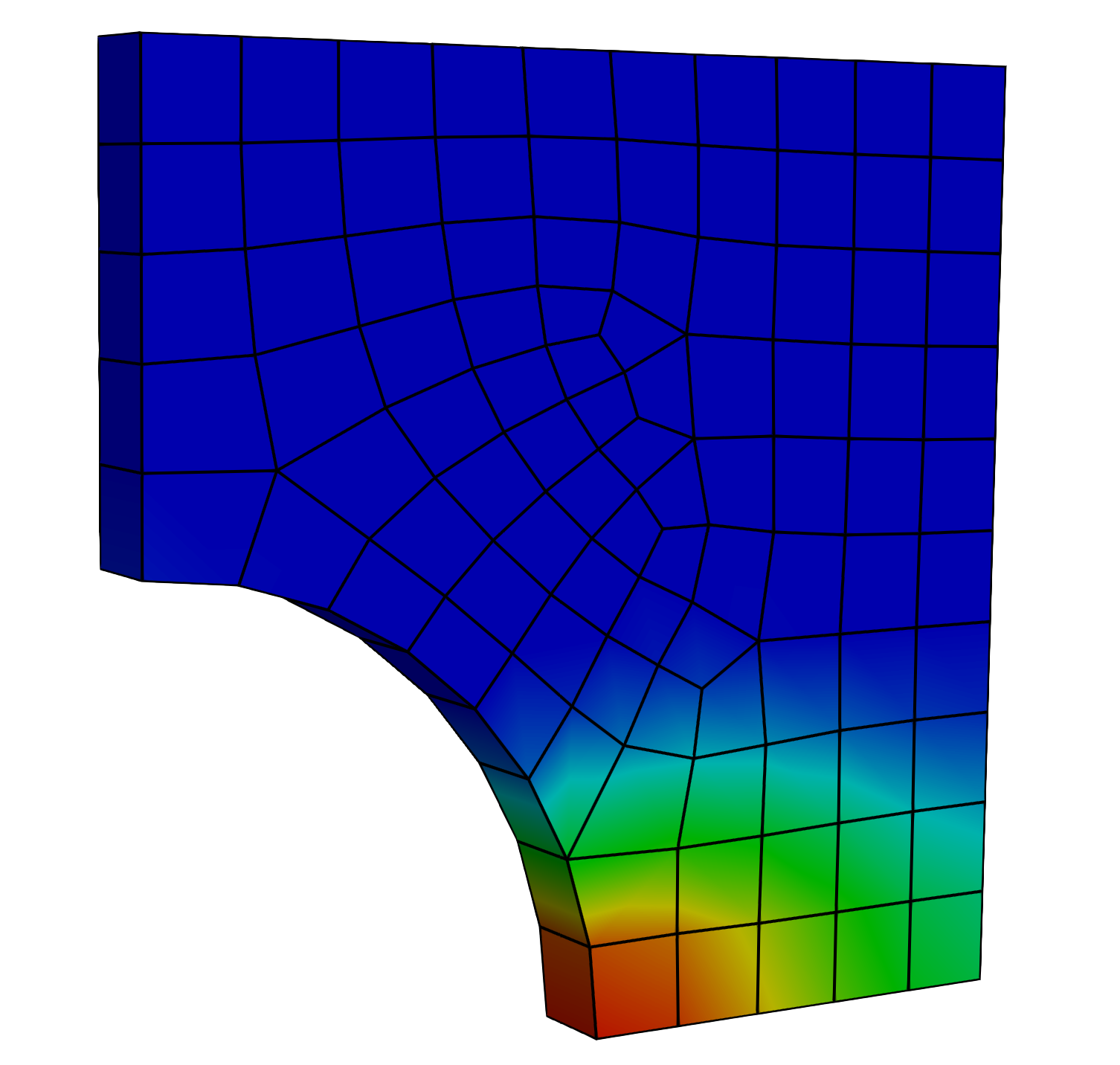}
            \caption{95 elements}
        \end{subfigure}%
        \begin{subfigure}{.5\textwidth}
            \centering
            \includegraphics[width=\textwidth]{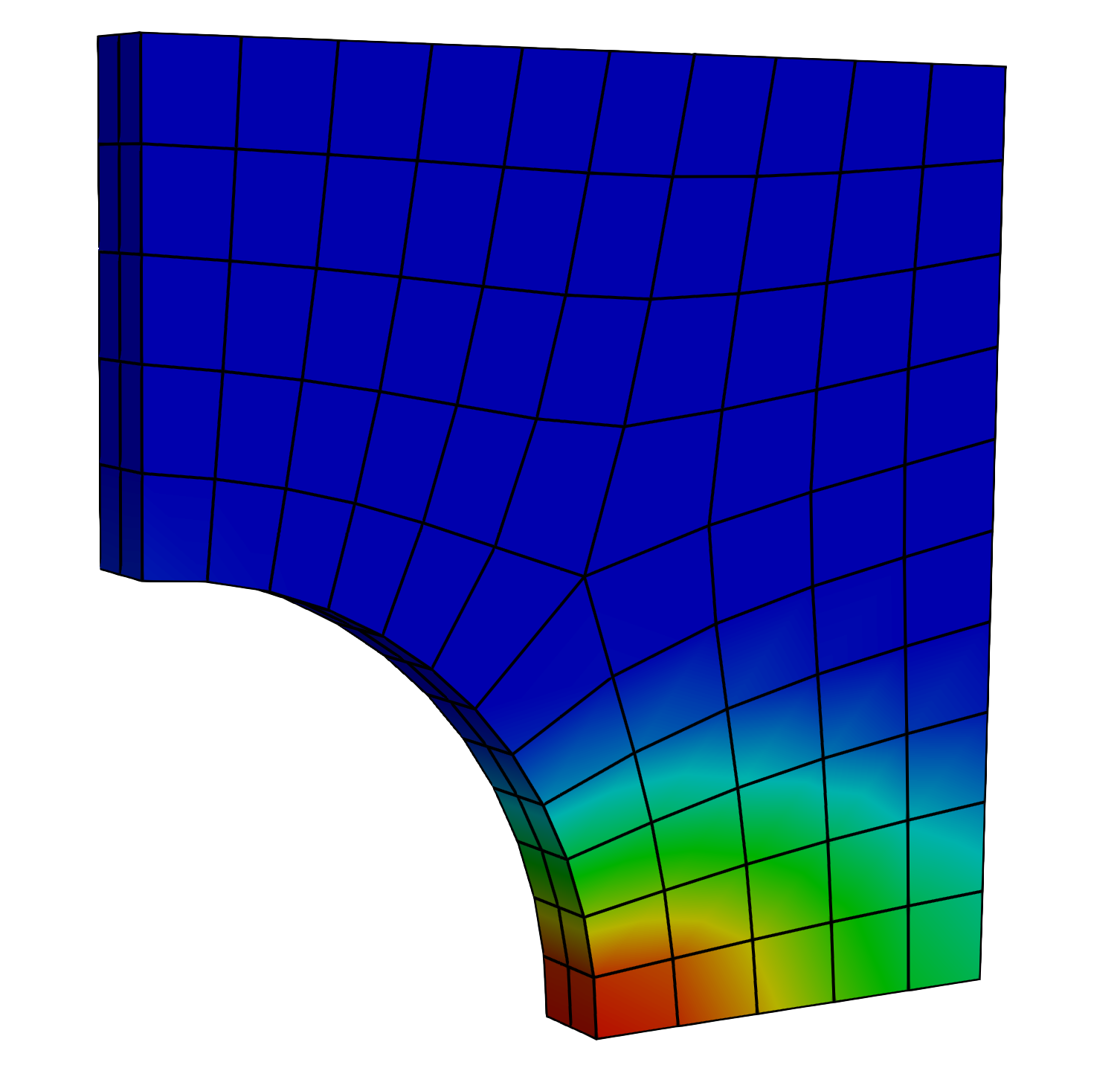}
            \caption{152 elements}
        \end{subfigure}
        \begin{subfigure}{.5\textwidth}
            \centering
            \includegraphics[width=\textwidth]{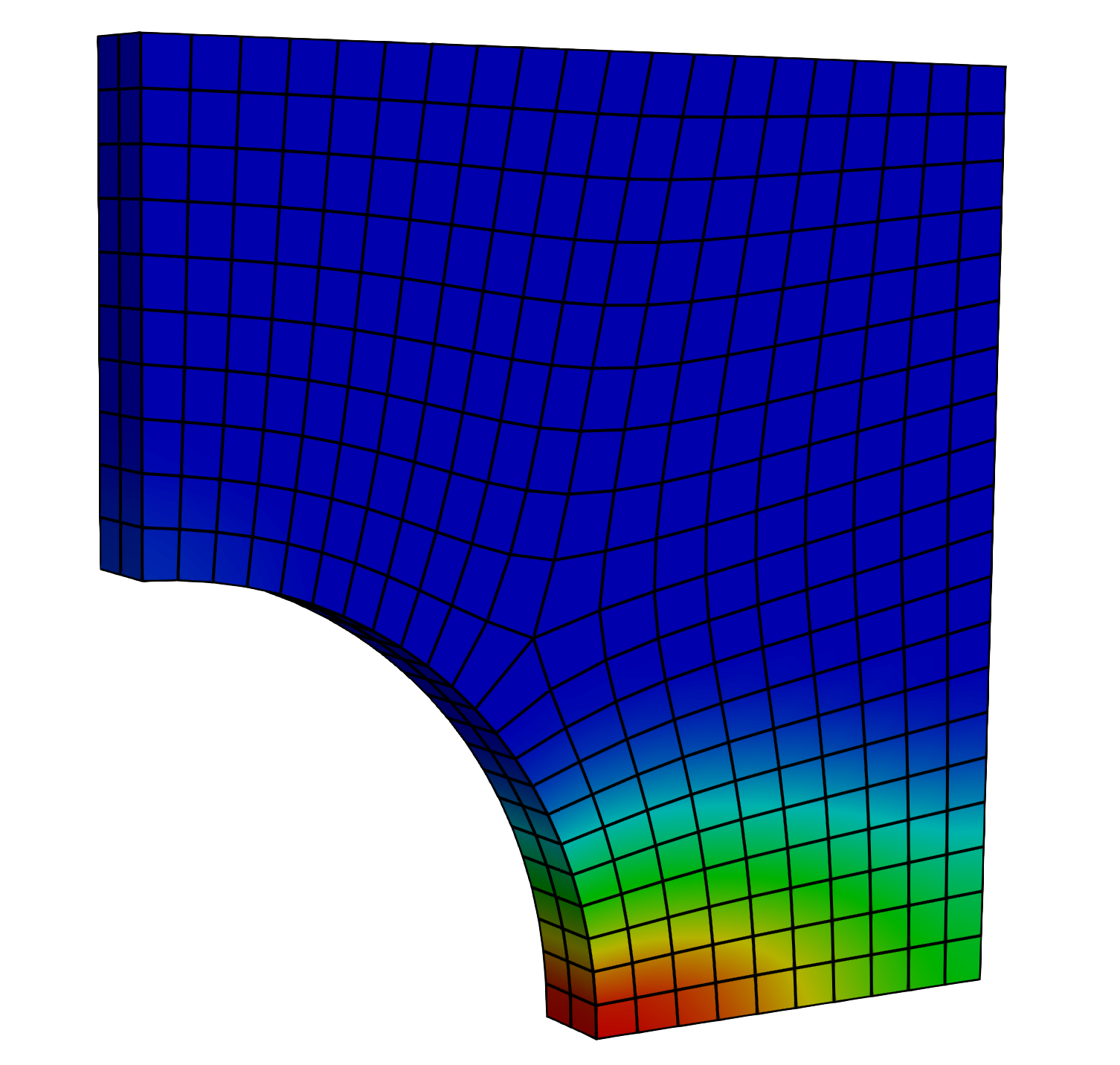}
            \caption{602 elements}
        \end{subfigure}%
        \begin{subfigure}{.5\textwidth}
            \centering
            \includegraphics[width=\textwidth]{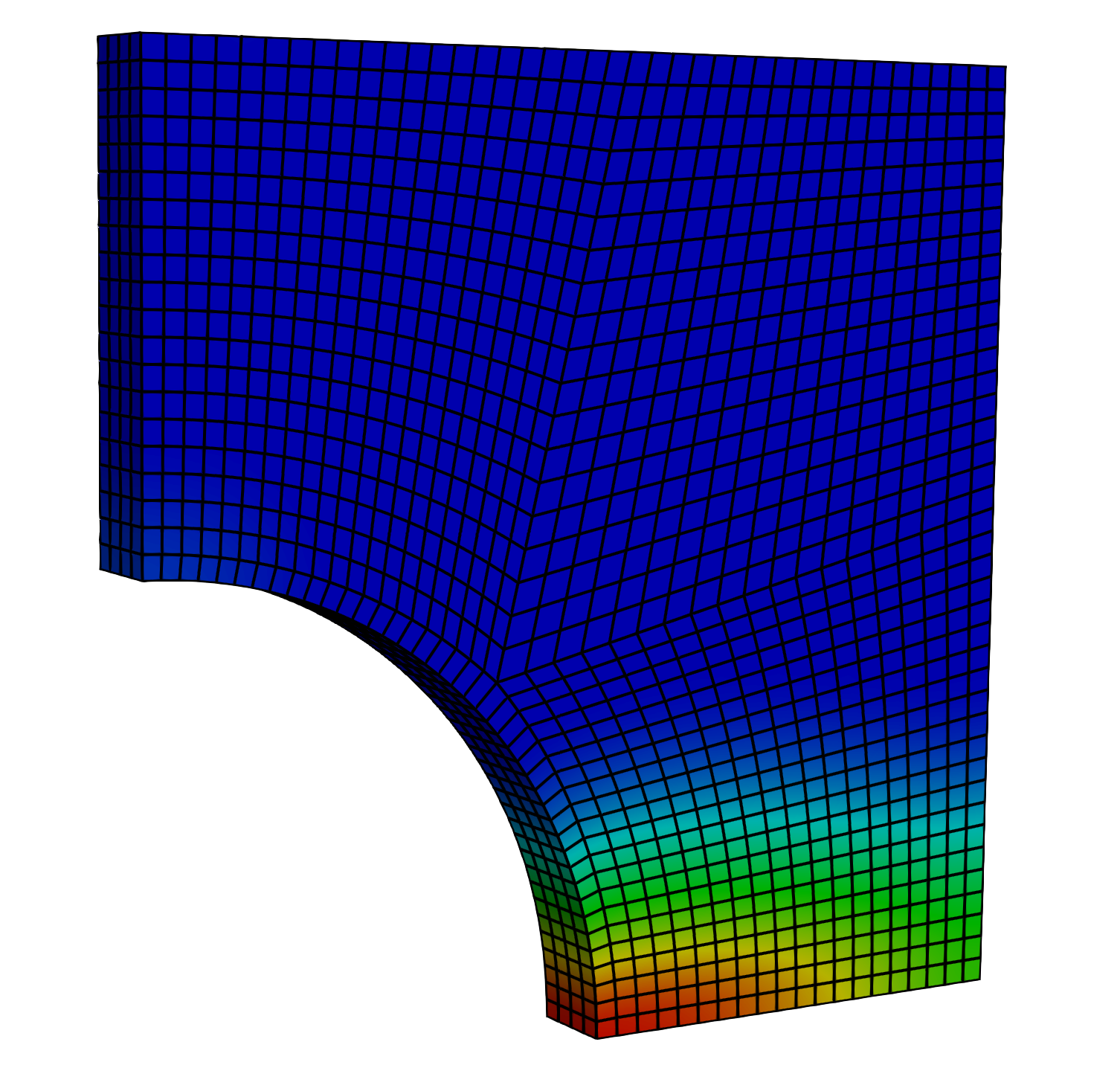}
            \caption{4804 elements}
        \end{subfigure}
    \end{subfigure}
    \begin{subfigure}{.1\textwidth}
        \includegraphics[width=\textwidth]{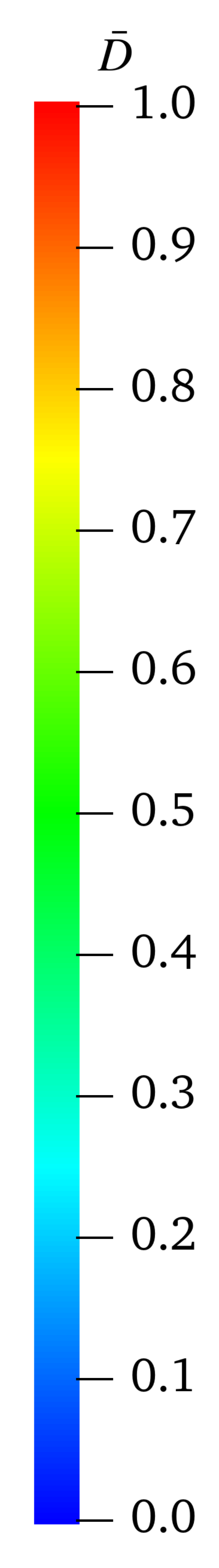}
    \end{subfigure}
    \caption{Damage of the specimen for the different meshes at the end of the simulation.}
    \label{fig:platemeshes}
\end{figure}

The damage evolution for the various mesh densities under investigation is depicted in Figure \ref{fig:platemeshes}. Thanks to the gradient-enhanced two-surface formulation, it can be observed, especially in the finer meshes, that the damage does not only occurs in a single row of elements. In many damage formulations, a common issue is the limited band of damage, where the damage zone becomes narrower as the mesh becomes finer. This is not the case here. Nevertheless, the result seems to be not fully converged. However, since we will be comparing the solutions with the reference solution later on, the lack of convergence does not affect the investigation of the model reduction methods. Nonetheless, it is important to keep in mind that finer meshes would typically be required, leading to better speedups of the methods. To prevent lengthy calculations, we refrained from computing finer meshes.

\subsubsection{Plasticity}

Due to the complexity of the problem, we will investigate the different material behaviours in individual steps. In the first step, we will investigate purely plastic behavior. For this purpose, the damage threshold $Y_0$ will be set sufficiently high to prevent any damage initiation.

\begin{figure}[!ht]
    \centering
    \input{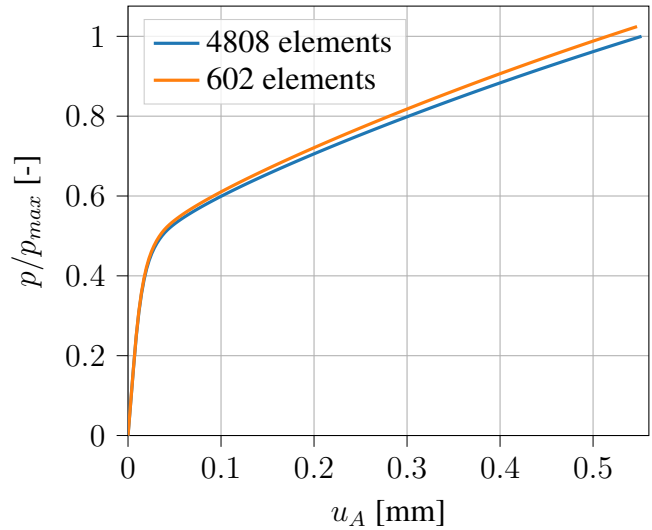}
    \caption{Load-displacement curve for an elastoplastic material behavior for 602 and 4808 elements of the full order model.}
    \label{fig:plate_plasticurve}
\end{figure}

We will examine two different meshes, one with 602 elements and another with 4804 elements, to demonstrate the effectiveness of the methods for both meshes. Additionally, when considering the speedups later on, it will become evident how much more potential these methods hold when applied to larger and more complex models with significantly more elements. The force-displacement curves are presented in Figure \ref{fig:plate_plasticurve}.
In each case, 200 snapshots were taken to construct the projection matrices.

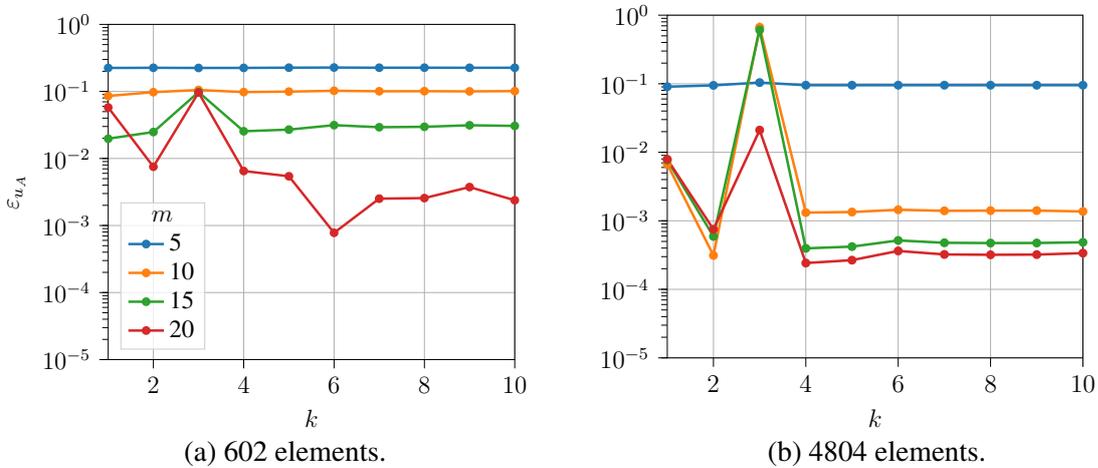
\begin{figure}[!ht]
    \centering
    \begin{subfigure}[t]{.48\textwidth}
        \resizebox{0.95\textwidth}{!}{ 
\begin{tikzpicture}

  \definecolor{crimson2143940}{RGB}{214,39,40}
  \definecolor{darkgray176}{RGB}{176,176,176}
  \definecolor{darkorange25512714}{RGB}{255,127,14}
  \definecolor{forestgreen4416044}{RGB}{44,160,44}
  \definecolor{lightgray204}{RGB}{204,204,204}
  \definecolor{steelblue31119180}{RGB}{31,119,180}

  \begin{axis}[
      legend cell align={left},
      legend style={
          fill opacity=0.8,
          draw opacity=1,
          text opacity=1,
          at={(0.03,0.03)},
          anchor=south west,
          draw=lightgray204
        },
      log basis y={10},
      tick align=outside,
      tick pos=left,
      x grid style={darkgray176},
      xlabel={\(\displaystyle k\)},
      xmajorgrids,
      xmin=1, xmax=10,
      xtick style={color=black},
      y grid style={darkgray176},
      ylabel={\(\displaystyle \varepsilon_{u_{A}}\)},
      ymajorgrids,
      ymin=1e-05, ymax=1,
      ymode=log,
      ytick style={color=black}
    ]
    \addlegendimage{empty legend}
    \addlegendentry{\hspace{-.3cm}$m$}
    \addplot [very thick, steelblue31119180, mark=*, mark size=1.5, mark options={solid}]
    table {%
        1 0.224272486394312
        2 0.225353737056163
        3 0.224166590957005
        4 0.224510991167994
        5 0.226303143210599
        6 0.227363024034014
        7 0.2259475736957
        8 0.226204774154867
        9 0.225519767182192
        10 0.225698707589838
        11 0.225737487992811
        12 0.225453155350387
        13 0.22560946920508
        14 0.225681187113289
        15 0.225619110262091
        16 0.225510816412991
        17 0.225448222455346
        18 0.225772949673425
        19 0.225690810842111
        20 0.225449450575664
      };
    \addlegendentry{5}
    \addplot [very thick, darkorange25512714, mark=*, mark size=1.5, mark options={solid}]
    table {%
        1 0.0856165807351638
        2 0.0975284143043523
        3 0.10522764072977
        4 0.0978428843837166
        5 0.0994250557400799
        6 0.102404612253122
        7 0.100626629136623
        8 0.10089558111284
        9 0.100294465965703
        10 0.101188910129764
        11 0.101478604154934
        12 0.100813118325805
        13 0.100684617292385
        14 0.100529840948176
        15 0.100601272855969
        16 0.100408333727166
        17 0.100397307202375
        18 0.101688167097158
        19 0.101457037756065
        20 0.100649263820588
      };
    \addlegendentry{10}
    \addplot [very thick, forestgreen4416044, mark=*, mark size=1.5, mark options={solid}]
    table {%
        1 0.0197358553733737
        2 0.0248065327960194
        3 0.0978308654909454
        4 0.0254850229030457
        5 0.0269552738910592
        6 0.0314108042635537
        7 0.0292707311934087
        8 0.0296859123422733
        9 0.0312950425682155
        10 0.0306208469716148
        11 0.0313983854547918
        12 0.0298578613605239
        13 0.0304595354758219
        14 0.0304143644024767
        15 0.0300527401711046
        16 0.0292438893871119
        17 0.029281890047646
        18 0.02986548652109
        19 0.0292883364006353
        20 0.0290881367229201
      };
    \addlegendentry{15}
    \addplot [very thick, crimson2143940, mark=*, mark size=1.5, mark options={solid}]
    table {%
        1 0.057727932853576
        2 0.0075325644964263
        3 0.0962419279792656
        4 0.00649769761558516
        5 0.00543025909087832
        6 0.000777932127705835
        7 0.00251619000765443
        8 0.00256354713118013
        9 0.00374682128928542
        10 0.00238079769523677
        11 0.00234736006589161
        12 0.00232428437465514
        13 0.000329083349049298
        14 0.000485087913100721
        15 0.00249343888991128
        16 0.00297700748165268
        17 0.00298402934081399
        18 0.000236083588169269
        19 0.000608450033517141
        20 0.00107778747998949
      };
    \addlegendentry{20}
  \end{axis}
\end{tikzpicture} }
        \setlength{\abovecaptionskip}{0pt}
        \caption{602 elements.}
    \end{subfigure}
    \begin{subfigure}[t]{.48\textwidth}
        \resizebox{0.9\textwidth}{!}{ 
\begin{tikzpicture}

  \definecolor{crimson2143940}{RGB}{214,39,40}
  \definecolor{darkgray176}{RGB}{176,176,176}
  \definecolor{darkorange25512714}{RGB}{255,127,14}
  \definecolor{forestgreen4416044}{RGB}{44,160,44}
  \definecolor{lightgray204}{RGB}{204,204,204}
  \definecolor{steelblue31119180}{RGB}{31,119,180}

  \begin{axis}[
      log basis y={10},
      tick align=outside,
      tick pos=left,
      x grid style={darkgray176},
      xlabel={\(\displaystyle k\)},
      xmajorgrids,
      xmin=1, xmax=10,
      xtick style={color=black},
      y grid style={darkgray176},
      ymajorgrids,
      ymin=1e-05, ymax=1,
      ymode=log,
      ytick style={color=black}
    ]
    \addplot [very thick, steelblue31119180, mark=*, mark size=1.5, mark options={solid}]
    table {%
        1 0.0905264259763907
        2 0.0946669622905976
        3 0.103998770142933
        4 0.095192960274342
        5 0.0951635652872056
        6 0.0952551556368201
        7 0.0952480461554244
        8 0.0951933647204304
        9 0.0952028331881995
        10 0.0952065023268951
        11 0.0951844125804985
        12 0.214457959554125
        13 0.161638929201367
        14 0.0952152119475302
        15 0.0952111124811141
        16 0.0952201711600552
        17 0.265235218946444
        18 0.265240288944666
        19 0.0952091029611898
        20 0.0952166742344835
      };
    \addplot [very thick, darkorange25512714, mark=*, mark size=1.5, mark options={solid}]
    table {%
        1 0.00670505244390698
        2 0.000312915212020938
        3 0.668176775313934
        4 0.0013178033145587
        5 0.00134332650744336
        6 0.00144899792276502
        7 0.00139788570774546
        8 0.00140801075632559
        9 0.00140905625981945
        10 0.00136923318293089
        11 0.00188524824119824
        12 0.00141940717704177
        13 0.0014186237906629
        14 0.00142460298906884
        15 0.00142956320388123
        16 0.001430063208599
        17 0.00143210994291539
        18 0.00142384201210487
        19 0.199512134384697
        20 0.201601970999472
      };
    \addplot [very thick, forestgreen4416044, mark=*, mark size=1.5, mark options={solid}]
    table {%
        1 0.0077232833943352
        2 0.000591591482071531
        3 0.608984295173511
        4 0.000396329030416772
        5 0.000420056488069102
        6 0.000517268293529755
        7 0.00047769468149737
        8 0.000473432969138534
        9 0.000474534802923479
        10 0.000485830434477218
        11 0.000487823612296457
        12 0.000500911044078499
        13 0.000490671838536083
        14 0.599919611255767
        15 0.000496220615985604
        16 0.000490495973160227
        17 0.000505866415070486
        18 0.000496448275413344
        19 0.000500961831285905
        20 0.000501898696396647
      };
    \addplot [very thick, crimson2143940, mark=*, mark size=1.5, mark options={solid}]
    table {%
        1 0.00790319733239587
        2 0.00075084547049829
        3 0.0210537560896897
        4 0.00024189267977287
        5 0.000266483510745253
        6 0.000363428257751369
        7 0.000322981127706014
        8 0.000319282855233089
        9 0.000321384744200524
        10 0.00033762671219525
        11 0.000336782672115129
        12 0.000343934662115544
        13 0.000336204465547091
        14 0.000347935569326959
        15 0.000342207228321746
        16 0.000339480190310397
        17 0.000343399044713085
        18 0.000342484618216186
        19 0.00034484196765711
        20 0.000343689156775153
      };
  \end{axis}

\end{tikzpicture}}
        \setlength{\abovecaptionskip}{0pt}
        \caption{4804 elements.}
    \end{subfigure}
    \caption{Approximation error for the elastoplastic case for the two different mesh discretizations.}
    \label{fig:plateplastirelerr}
\end{figure}

To assess the quality of the reduced models, we compare them with the unreduced solution by examining the displacement at point A (see Figure \ref{fig:Plate_quarter}). We calculate the relative error between the unreduced and reduced solutions as follows:

\be
\varepsilon_{u_{A}}=\frac{|u_{A\:FOM}-u_{A\:DEIM}|}{u_{A\:prec}}.
\ee

where $u_{A\:FOM}$ represents the displacement at point A in the full order model (FOM) solution, and $u_{A\:DEIM}$ represents the displacement at point A in the reduced solution obtained through the DEIM method.
In Figure~\ref{fig:plateplastirelerr}, this error is plotted on the y-axis, while the number of DEIM modes used is shown on the x-axis. Since the number of POD modes was also varied, these will be represented by different curves (as indicated in the legend).

It can be seen that as the number of POD modes increases, the error decreases. While the simulations are getting more stable with increasing number of DEIM modes. With just 20 POD modes and 4 or more DEIM modes, an error of less than one percent can be achieved. However, it should be noted that the method becomes unstable with 3 DEIM modes. The cause of this instability is currently under investigation and cannot be reliably explained at present.

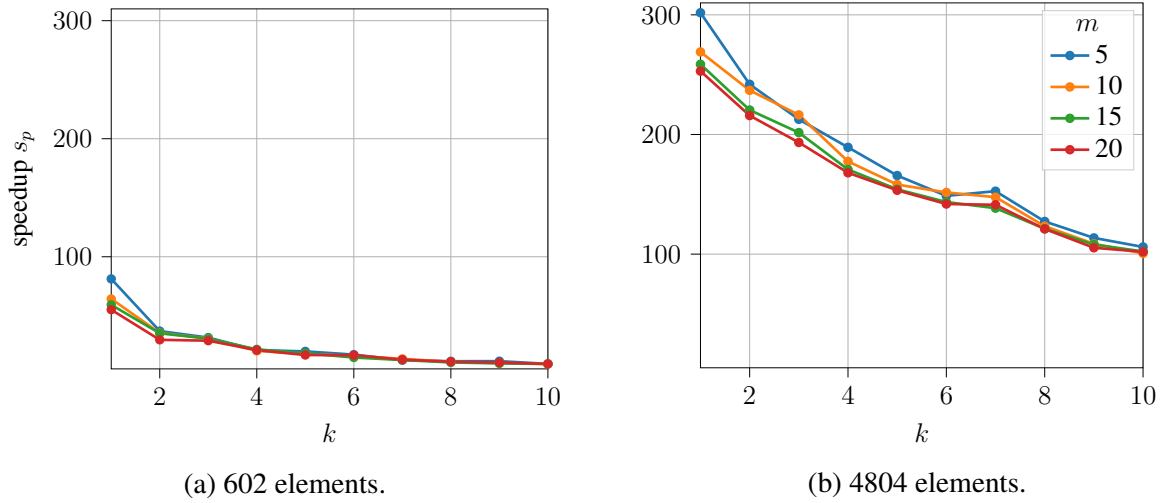
\begin{figure}[!ht]
    \centering
    \begin{subfigure}[t]{.48\textwidth}
        \resizebox{1\textwidth}{!}{ 
\begin{tikzpicture}

    \definecolor{crimson2143940}{RGB}{214,39,40}
    \definecolor{darkgray176}{RGB}{176,176,176}
    \definecolor{darkorange25512714}{RGB}{255,127,14}
    \definecolor{forestgreen4416044}{RGB}{44,160,44}
    \definecolor{lightgray204}{RGB}{204,204,204}
    \definecolor{steelblue31119180}{RGB}{31,119,180}

    \begin{axis}[
            legend cell align={left},
            legend style={fill opacity=0.8, draw opacity=1, text opacity=1, draw=lightgray204},
            tick align=outside,
            tick pos=left,
            x grid style={darkgray176},
            xlabel={\(\displaystyle k\)},
            xmajorgrids,
            xmin=1, xmax=10,
            xtick style={color=black},
            y grid style={darkgray176},
            ylabel={speedup $s_p$},
            ymajorgrids,
            ymin=5, ymax=310,
            ytick style={color=black}
        ]
        \addplot [very thick, steelblue31119180, mark=*, mark size=1.5, mark options={solid}]
        table {%
                1 81.2410770365472
                2 37.0213054966654
                3 31.5382153667479
                4 20.9181605811729
                5 19.8483625500716
                6 17.1005288817299
                7 12.4269522149997
                8 11.4924539676928
                9 11.5511457394378
                10 9.26457082959891
                11 9.18015327286765
                12 8.12862926720597
                13 8.10014113264352
                14 7.92774131288264
                15 7.57661507779642
                16 7.32019954848916
                17 6.62491551554665
                18 6.34325881791677
                19 6.04496095778965
                20 6.24107093662828
            };
        \addplot [very thick, darkorange25512714, mark=*, mark size=1.5, mark options={solid}]
        table {%
                1 64.2989541003397
                2 35.4943994281957
                3 30.8071654454317
                4 20.2759319764568
                5 17.7355476306413
                6 16.0373752190349
                7 13.4859568721756
                8 10.9795893170876
                9 9.87550062665555
                10 9.18917782387808
                11 8.66836988830859
                12 7.83626466058754
                13 7.64437338167139
                14 7.57202211820597
                15 7.15752783577635
                16 6.81514335654279
                17 6.32741567051886
                18 6.44897337299244
                19 5.92834398490636
                20 6.3442370543112
            };
        \addplot [very thick, forestgreen4416044, mark=*, mark size=1.5, mark options={solid}]
        table {%
                1 59.2597811787189
                2 35.188210983903
                3 30.5911715646991
                4 21.5391894784545
                5 18.2582133484869
                6 14.6240906436084
                7 12.5028566829058
                8 10.4481519055699
                9 9.66118139516095
                10 9.17964603914519
                11 8.21747808659511
                12 7.45770997044376
                13 7.459516102257
                14 7.30827017770058
                15 6.98732611433523
                16 6.87217961661484
                17 6.11764277795286
                18 5.98914737956064
                19 5.80173640975115
                20 6.05476357984324
            };
        \addplot [very thick, crimson2143940, mark=*, mark size=1.5, mark options={solid}]
        table {%
                1 55.1362284401325
                2 29.6278106149733
                3 28.891249807817
                4 20.8740447597813
                5 16.7461393894561
                6 16.5994395088918
                7 12.8787533346946
                8 11.0913512953583
                9 10.412778767372
                10 9.17250503324245
                11 8.56173543092038
                12 7.68492667395062
                13 7.5570044861922
                14 7.26251669154161
                15 6.91632910279093
                16 6.73807597874472
                17 6.09217996566534
                18 6.21242360088797
                19 5.94864498943521
                20 5.93032740045513
            };
    \end{axis}

\end{tikzpicture}}
        \setlength{\abovecaptionskip}{-8pt}
        \caption{602 elements.}
    \end{subfigure}
    \hfill
    \begin{subfigure}[t]{.48\textwidth}
        \resizebox{0.937\textwidth}{!}{ 
\begin{tikzpicture}

  \definecolor{crimson2143940}{RGB}{214,39,40}
  \definecolor{darkgray176}{RGB}{176,176,176}
  \definecolor{darkorange25512714}{RGB}{255,127,14}
  \definecolor{forestgreen4416044}{RGB}{44,160,44}
  \definecolor{lightgray204}{RGB}{204,204,204}
  \definecolor{steelblue31119180}{RGB}{31,119,180}

  \begin{axis}[
      legend cell align={left},
      legend style={fill opacity=0.8, draw opacity=1, text opacity=1, draw=lightgray204},
      tick align=outside,
      tick pos=left,
      x grid style={darkgray176},
      xlabel={\(\displaystyle k\)},
      xmajorgrids,
      xmin=1, xmax=10,
      xtick style={color=black},
      y grid style={darkgray176},
      ymajorgrids,
      ymin=5, ymax=310,
      ytick style={color=black}
    ]
    \addlegendimage{empty legend}
    \addlegendentry{\hspace{-.3cm}$m$}
    \addplot [very thick, steelblue31119180, mark=*, mark size=1.5, mark options={solid}]
    table {%
        1 301.777998382892
        2 241.967791754111
        3 212.664708446141
        4 189.283824688065
        5 165.692231672234
        6 148.709667660003
        7 152.62215541315
        8 127.313026599583
        9 113.578664389432
        10 105.985591089271
        11 100.573786956675
        12 96.5324620541834
        13 98.3303357611845
        14 84.2879751128029
        15 78.8095787019648
        16 75.1721470575879
        17 80.5802831101647
        18 77.159773600462
        19 69.566134091581
        20 67.3518827414143
      };
    \addlegendentry{5}
    \addplot [very thick, darkorange25512714, mark=*, mark size=1.5, mark options={solid}]
    table {%
        1 268.991976636253
        2 236.895088116247
        3 216.387502724211
        4 177.490800754997
        5 158.137513332304
        6 151.553949724852
        7 147.617892979422
        8 123.373627660346
        9 108.630879143919
        10 100.595974007195
        11 97.896328156425
        12 86.1521770120292
        13 84.5129686695965
        14 81.4461859566206
        15 74.8923392103608
        16 73.8624826194716
        17 66.9812407451008
        18 67.0684633521561
        19 73.7083253986763
        20 75.9320713689773
      };
    \addlegendentry{10}
    \addplot [very thick, forestgreen4416044, mark=*, mark size=1.5, mark options={solid}]
    table {%
        1 258.694056473821
        2 220.456852082531
        3 201.602613767895
        4 170.765474101758
        5 154.184497392282
        6 143.531935624886
        7 138.406851908404
        8 121.255680816641
        9 107.99926900589
        10 102.222830205355
        11 96.0280510654271
        12 88.4931540497661
        13 83.9381284746797
        14 96.2885638716564
        15 74.3307503720086
        16 75.2868067615381
        17 66.944770947374
        18 68.0945844551485
        19 62.5190801834601
        20 66.1632992892121
      };
    \addlegendentry{15}
    \addplot [very thick, crimson2143940, mark=*, mark size=1.5, mark options={solid}]
    table {%
        1 252.966019184801
        2 215.828539486129
        3 193.329956859895
        4 167.992798005656
        5 153.364483398022
        6 141.895542727604
        7 141.243541153798
        8 121.035083561544
        9 105.307208372836
        10 101.966807546746
        11 94.8477243655022
        12 84.9907634274822
        13 84.4916011948014
        14 80.1463626716341
        15 73.6579995824488
        16 73.3585002343894
        17 67.6235057883795
        18 65.5364339840842
        19 65.2674325531844
        20 65.6993427949703
      };
    \addlegendentry{20}
  \end{axis}

\end{tikzpicture}}
        \caption{4804 elements.}
    \end{subfigure}
    \caption{Speedup for the elastoplastic case for the two different mesh discretizations.}
    \label{fig:plateplastispeedup}
\end{figure}

An essential question that arises in the field of model reduction methods is the issue of speedups and how much time can be saved through these techniques. The speedup $s_p$ is defined as
\be
s_p = \frac{t_{full}}{t_{DEIM}}
\ee
with $t_{full}$ is the time of one Newton step in the unreduced simulation and $t_{DEIM}$ of the DEIM simulation including assembling and solving.
In this publication, only academic examples with very few elements have been examined so far. Therefore, comparing different mesh densities is of interest. The mesh with 602 elements achieved a speedup of approximately 30, while the mesh with 4804 elements already achieved a speedup of about 170 as seen in Figure~\ref{fig:plateplastispeedup}. This means that a calculation only requires around 0.6\% of the time needed for the full calculation, while the error remains well below one percent. With larger examples, the speedups would be even better.

\begin{figure}[!ht]
    \centering
    \input{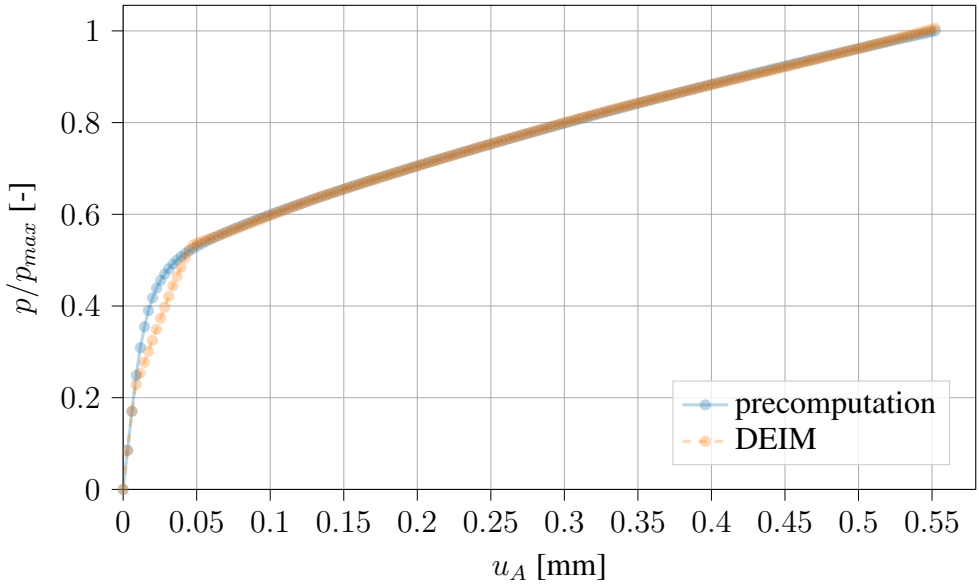}
    \caption{Load-displacement curves of the precomputation and the DEIM-reduced elastoplastic simulation with 20 POD modes and 4 DEIM modes.}
    \label{fig:plate_plasticurveprecdeim}
\end{figure}

To demonstrate that it is not only the displacement at point A that is well approximated, Figure \ref{fig:plate_plasticurveprecdeim} compares the force-displacement curve. For simplicity the force $p$ is devided by $p_{max}$ of the unreduced simulation.  Due to the limited number of modes, there is a larger error in the elastic range, while the plasticity-dominated region is well approximated.

\pgfplotsset{%
    width=.8\textwidth,
    height=.5\textwidth
}

\subsubsection{Damage}

\begin{figure}[!ht]
    \centering
    \pgfplotsset{%
        width=.8\textwidth,
        height=.5\textwidth
    }
    \input{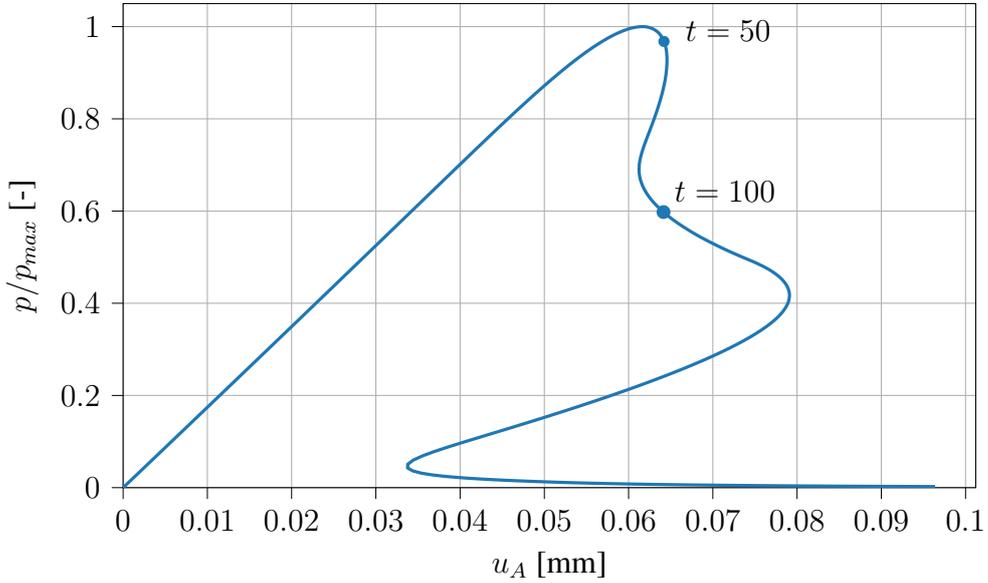}
    \caption{Force-displacement curve for an elasto-damage material behavior.}
    \label{fig:elastodamage}
\end{figure}

In the next step, we will focus solely on damage analysis. To achieve this, we ensure that no plasticity occurs by setting the yield stress of plasticity to a very high value while the remaining parameters are the same as listed in \ref{tab:paraplate}. Furthermore, we will only perform calculations using the mesh with 4804 elements. The force-displacement curve of the unreduced simulation is depicted in Figure \ref{fig:elastodamage}. With the material model used, the plate with a hole can be simulated until complete failure occurs, resulting in a complete crack through the plate. However, performing a complete simulation with model reduction methods poses significant challenges.
These complex phenomena include damage, unloading, and snapback behavior. For each of these phenomena, additional modes would be required. Since the presented method is kept very general, we have limited ourselves and will only simulate up to the 50th time step, which is beyond the limit load. In most practical applications, simulations are typically performed until a certain stress level is exceeded. Hence, we are already simulating well beyond what would be commonly employed in many fields.

In addition to the relative displacement error, we introduce the relative error of the limit load
\be
\varepsilon_{p_{max}}=\frac{|p_{max \, prec}-p_{max \, DEIM}|}{p_{max \, prec}}
\ee
where $p_{max \, prec}$ represents the limit load in the unreduced solution and $p_{max \, DEIM}$ represents the limit load in the reduced solution obtained through the DEIM method. The relative error of the maximum micromorphic damage
\be
\varepsilon_{\Dbar_{B}}=\frac{|D_{B\:prec}-D_{B\:DEIM}|}{D_{B\:prec}}.
\ee
where $D_{B\:prec}$ represents the maximum micromorphic damage in the unreduced solution and $D_{B\:DEIM}$ represents the maximum micromorphic damage in the reduced solution obtained through the DEIM method.

\begin{figure}[!ht]
    \centering
\begin{tikzpicture}

    \definecolor{crimson2143940}{RGB}{214,39,40}
    \definecolor{darkgray176}{RGB}{176,176,176}
    \definecolor{darkorange25512714}{RGB}{255,127,14}
    \definecolor{forestgreen4416044}{RGB}{44,160,44}
    \definecolor{lightgray204}{RGB}{204,204,204}
    \definecolor{mediumpurple148103189}{RGB}{148,103,189}
    \definecolor{steelblue31119180}{RGB}{31,119,180}

    \begin{axis}[
            legend cell align={left},
            legend style={
                    fill opacity=0.8,
                    draw opacity=1,
                    text opacity=1,
                    at={(1.03,1)},
                    anchor=north west,
                    draw=lightgray204
                },
            log basis y={10},
            tick align=outside,
            tick pos=left,
            x grid style={darkgray176},
            xlabel={\(\displaystyle k\)},
            xmajorgrids,
            xmin=0, xmax=50,
            xtick style={color=black},
            y grid style={darkgray176},
            ylabel={\(\displaystyle \varepsilon_{p_{max}}\)},
            ymajorgrids,
            ymin=1.12475503438145e-07, ymax=242.629575192323,
            ymode=log,
            ytick style={color=black}
        ]
        \addlegendimage{empty legend}
        \addlegendentry{\hspace{-.3cm}$m$}
        \addplot [very thick, steelblue31119180, mark=*, mark size=1.5, mark options={solid}]
        table {%
                2 0.00370878620866201
                4 0.00292229578947555
                6 1.94455358392013
                8 0.170543924258387
                10 0.0718904791145451
                12 0.0122157653220153
                14 7.11689260571052
                16 0.00359624771921362
                18 0.0142318143769703
                20 0.0053481525842346
                22 0.0126344506025029
                24 0.00152302672825177
                26 0.291217461259149
                28 0.234390922450151
                30 0.00392434930533792
                32 0.00424104412919252
                34 0.00607460186077299
                36 0.013992941861096
                38 0.00460791865551665
                40 0.00457845200135312
                42 0.00457625193585694
                44 0.00457666047720295
                46 0.00457721328914794
                48 0.0045772830655422
                50 0.00457677335452687
            };
        \addlegendentry{10}
        \addplot [very thick, darkorange25512714, mark=*, mark size=1.5, mark options={solid}]
        table {%
                2 0.00945253749210248
                4 0.00233507076936999
                6 0.0198808905626886
                8 0.00620251869098326
                10 0.00491609323774932
                12 0.000157888776660937
                14 2.06562347098175
                16 0.000606622926118251
                18 91.3431957467226
                20 0.00189297019323444
                22 6.13900916461046e-06
                24 0.00459038011362739
                26 0.043087216662465
                28 0.011928064306219
                30 0.00199480589066868
                32 0.000469258476994273
                34 0.00414561531903681
                36 0.00225822573059904
                38 0.00188515905837042
                40 0.00188351471282948
                42 0.0018832392336725
                44 0.00188330859741916
                46 0.00188391237327305
                48 0.0018833132171369
                50 0.00188409645630384
            };
        \addlegendentry{20}
        \addplot [very thick, forestgreen4416044, mark=*, mark size=1.5, mark options={solid}]
        table {%
                2 0.00855840582998744
                4 0.00238224641271656
                6 0.00155035766286941
                8 0.00659578617632738
                10 0.00457105731853057
                12 0.00222413010926485
                14 6.33802946227559e-05
                16 8.69660511412408e-05
                18 0.000594697140500959
                20 0.000128930416170298
                22 0.000134595859311345
                24 7.03482106809338e-05
                26 0.000853999821632334
                28 0.00033661982240182
                30 0.000179586596475925
                32 0.00445972163582566
                34 9.32907159471739e-05
                36 3.2815035835593e-05
                38 0.000229080451944698
                40 0.000226393519551541
                42 0.000226141194304804
                44 0.000225811504800105
                46 0.000225399166670251
                48 0.000224177900516588
                50 0.000223867454500497
            };
        \addlegendentry{30}
        \addplot [very thick, crimson2143940, mark=*, mark size=1.5, mark options={solid}]
        table {%
                2 0.00854958615670246
                4 0.00238089747530087
                6 0.0015473783760869
                8 0.00652518608839507
                10 0.00457051154829951
                12 0.00220559668989294
                14 5.18384673148983e-05
                16 9.21900084601374e-05
                18 0.000593760541612861
                20 0.000133401937234069
                22 0.00014043411814687
                24 1.03209923178602e-05
                26 1.30942262435808e-05
                28 6.74645597192986e-06
                30 3.13358819052778e-06
                32 1.5654448879343e-05
                34 1.96813474038519e-05
                36 3.27417991489372e-05
                38 1.20744693551924e-06
                40 1.30319331579004e-06
                42 1.55033843587074e-06
                44 1.28665025553634e-06
                46 7.76347678593735e-07
                48 1.30839882966467e-06
                50 3.19864887771127e-07
            };
        \addlegendentry{40}
        \addplot [very thick, mediumpurple148103189, mark=*, mark size=1.5, mark options={solid}]
        table {%
                2 0.00854885919027413
                4 0.00237935156065109
                6 0.00155768867423806
                8 0.0066264368986997
                10 0.00457228302319421
                12 0.00220669933512872
                14 5.17881484695075e-05
                16 9.19976677809966e-05
                18 0.000591105500582523
                20 0.000133265369650631
                22 0.000140252957793306
                24 1.01373539555136e-05
                26 1.24154604583133e-05
                28 6.65370922601928e-06
                30 3.33075368430657e-06
                32 1.5745495815435e-05
                34 1.97161656854114e-05
                36 3.28635070942386e-05
                38 9.32167419776715e-07
                40 1.15223311718628e-06
                42 1.49536700464708e-06
                44 1.07140330223967e-06
                46 6.50579814553113e-07
                48 1.50439205977189e-06
                50 2.98762085075384e-07
            };
        \addlegendentry{50}
    \end{axis}

\end{tikzpicture}
    \caption{Relative error of the limit load with respect to the number of POD basis vectors $m$ and the number of DEIM interpolation points $k$.}
    \label{fig:plate_damage_limit_error}
\end{figure}
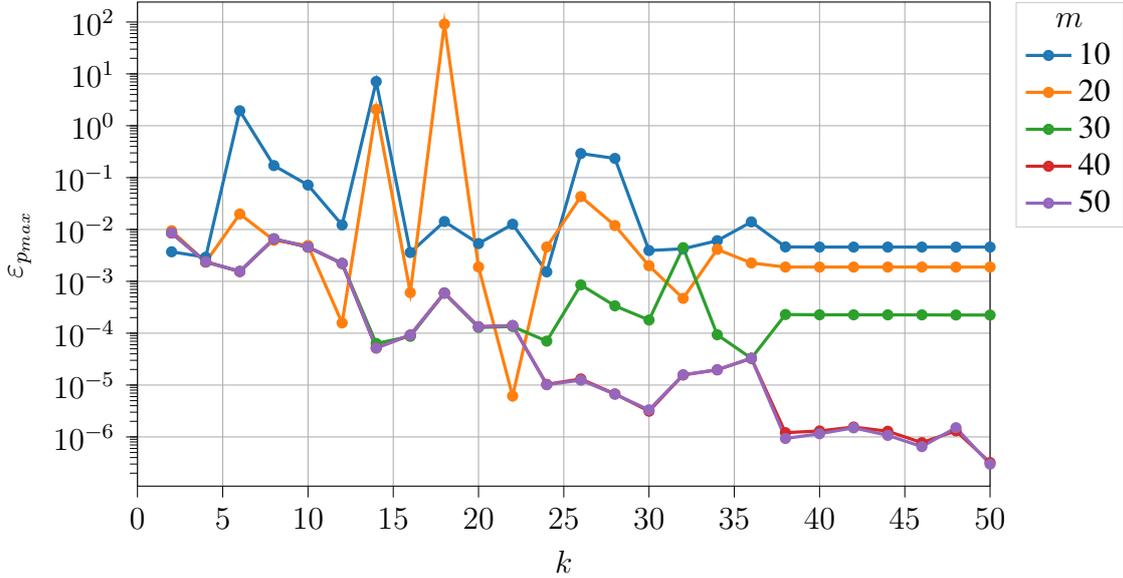

\begin{figure}[!ht]
    \centering
\begin{tikzpicture}

    \definecolor{crimson2143940}{RGB}{214,39,40}
    \definecolor{darkgray176}{RGB}{176,176,176}
    \definecolor{darkorange25512714}{RGB}{255,127,14}
    \definecolor{forestgreen4416044}{RGB}{44,160,44}
    \definecolor{lightgray204}{RGB}{204,204,204}
    \definecolor{mediumpurple148103189}{RGB}{148,103,189}
    \definecolor{steelblue31119180}{RGB}{31,119,180}

    \begin{axis}[
            legend cell align={left},
            legend style={
                    fill opacity=0.8,
                    draw opacity=1,
                    text opacity=1,
                    at={(1.03,1)},
                    anchor=north west,
                    draw=lightgray204
                },
            log basis y={10},
            tick align=outside,
            tick pos=left,
            x grid style={darkgray176},
            xlabel={\(\displaystyle k\)},
            xmajorgrids,
            xmin=0, xmax=50,
            xtick style={color=black},
            y grid style={darkgray176},
            ylabel={\(\displaystyle \varepsilon_{u_{A}}\)},
            ymajorgrids,
            ymin=5.71184610822082e-05, ymax=119.739099424467,
            ymode=log,
            ytick style={color=black}
        ]
        \addlegendimage{empty legend}
        \addlegendentry{\hspace{-.3cm}$m$}
        \addplot [very thick, steelblue31119180, mark=*, mark size=1.5, mark options={solid}]
        table {%
                2 0.0150822079218874
                4 0.00947288020201278
                6 0.211259511458646
                8 0.968848588619464
                10 0.237792282380738
                12 0.672646456483873
                14 1.76783873584464
                16 0.0608551569062446
                18 0.440543950627421
                20 0.246558170664464
                22 0.466817743837336
                24 0.446188779320761
                26 0.352383105151737
                28 0.23491543044696
                30 0.497099526659093
                32 0.497050279027238
                34 0.00216068949238443
                36 0.377602996970357
                38 0.46957757908451
                40 0.469608507245395
                42 0.469603529607037
                44 0.469604692742199
                46 0.469604672670394
                48 0.469603421603999
                50 0.469601905721718
            };
        \addlegendentry{10}
        \addplot [very thick, darkorange25512714, mark=*, mark size=1.5, mark options={solid}]
        table {%
                2 0.0251248583306278
                4 0.352218965422525
                6 0.671420990984495
                8 0.000330049223789365
                10 0.418199778901496
                12 0.00141904230196721
                14 1.35770778957044
                16 0.00263539479739147
                18 61.7869730668133
                20 0.414187844609483
                22 0.00279522398869736
                24 0.495513012487637
                26 0.407117733940228
                28 0.347779771643924
                30 0.468990642885702
                32 0.333135221697982
                34 0.495851116043332
                36 0.468697867749078
                38 0.468791162274367
                40 0.468796316414027
                42 0.468796478340565
                44 0.468796383228696
                46 0.468796338904389
                48 0.468796150990578
                50 0.468796331844051
            };
        \addlegendentry{20}
        \addplot [very thick, forestgreen4416044, mark=*, mark size=1.5, mark options={solid}]
        table {%
                2 0.0139631577824875
                4 0.354097615546369
                6 0.279693912057756
                8 0.190289235682781
                10 0.417287737937466
                12 0.000224925949453884
                14 0.00116243333254698
                16 0.341561554164958
                18 0.412224510180566
                20 0.000190962815483565
                22 0.000242017525329547
                24 0.000472927976755581
                26 0.242154782163052
                28 0.333833300890157
                30 0.277424067584177
                32 0.000179663846058954
                34 0.106885180645233
                36 0.324801477797352
                38 0.26018559346215
                40 0.235923360570915
                42 0.236551150513916
                44 0.236508644517726
                46 0.236437872449626
                48 0.241200846021385
                50 0.240260239880792
            };
        \addlegendentry{30}
        \addplot [very thick, crimson2143940, mark=*, mark size=1.5, mark options={solid}]
        table {%
                2 0.0138876219788035
                4 0.354111589556619
                6 0.279673278537455
                8 0.189770811104737
                10 0.417283820491432
                12 0.000222366569109493
                14 0.00116161423827142
                16 0.341141887451074
                18 0.412228305438775
                20 0.000184451830586673
                22 0.000235259516306492
                24 0.00046426670675037
                26 0.000425631061649513
                28 0.00015352035927329
                30 0.000141193119631832
                32 0.00012945742356537
                34 0.000110691829539851
                36 0.306344954965967
                38 0.38804203144742
                40 0.38791914789867
                42 0.388038346888107
                44 0.388037498037495
                46 0.388029653210129
                48 0.387634581838576
                50 0.387952862508019
            };
        \addlegendentry{40}
        \addplot [very thick, mediumpurple148103189, mark=*, mark size=1.5, mark options={solid}]
        table {%
                2 0.0139764003547895
                4 0.354116009922533
                6 0.279755377260259
                8 0.19010165087057
                10 0.4173031435253
                12 0.00024319131330698
                14 0.00113765795480707
                16 0.341087689415143
                18 0.412247191156888
                20 0.000206021092727919
                22 0.000255873038451923
                24 0.000486613325528144
                26 0.000448598225095313
                28 0.000173149292217235
                30 0.000164367363814432
                32 0.000152674705855198
                34 0.000133965879173902
                36 0.306363067863297
                38 0.388058678488147
                40 0.388053626843781
                42 0.388054467699715
                44 0.388053699107738
                46 0.388053920930228
                48 0.388054065232971
                50 0.388053773360952
            };
        \addlegendentry{50}
    \end{axis}

\end{tikzpicture}
    \caption{Relative error of the displacement in point A with respect to the number of POD basis vectors $m$ and the number of DEIM interpolation points $k$.}
    \label{fig:plate_damage_displacement_error}
\end{figure}
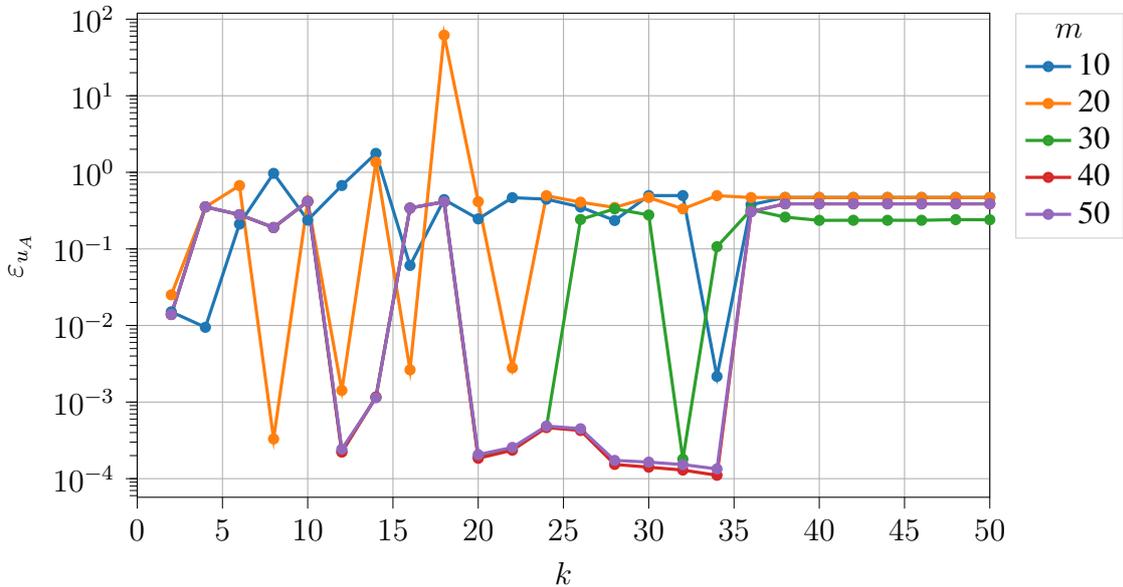

Figure \ref{fig:plate_damage_limit_error} displays the error of the limit load. With 40 and 50 POD modes and a sufficient number of DEIM modes, the error can be reduced to the order of $10^{-6}$.
The displacement error is depicted in Figure \ref{fig:plate_damage_displacement_error}. There are simulations with very good results and errors on the order of $10^{-3}$. However, there is no clear trend where the error consistently decreases with an increasing number of modes.
This is due to an instability in the arc-length method, which needs to be further investigated. The arc-length method follows the unloading path instead of continuing to load, leading to deviations. This is exemplified in Figure \ref{fig:platedmg4040} for 40 POD modes and 40 DEIM modes. This problem of artifical unloading is already known. Pohl et al. investigated different path following schemes for problems with softening \cite{pohl2014adaptive}. Their example of the contrained adaptive arc-length method did not converge with step size adjustment. Without step size adjustment a large number of increments are needed to trace the equilibrium path.
\begin{figure}[!ht]
    \centering
\begin{tikzpicture}

    \definecolor{darkgray176}{RGB}{176,176,176}
    \definecolor{darkorange25512714}{RGB}{255,127,14}
    \definecolor{steelblue31119180}{RGB}{31,119,180}
    \definecolor{lightgray204}{RGB}{204,204,204}

    \begin{axis}[
            legend cell align={left},
            legend style={
                    fill opacity=0.8,
                    draw opacity=1,
                    text opacity=1,
                    at={(0.97,0.05)},
                    anchor=south east,
                    draw=lightgray204
                },
            tick align=outside,
            tick pos=left,
            x grid style={darkgray176},
            xlabel={\(\displaystyle u_A\) [mm]},
            xmajorgrids,
            xmin=0, xmax=0.0677863068482703,
            xtick style={color=black},
            y grid style={darkgray176},
            ylabel={\(\displaystyle p / p_{max}\) [-]},
            ymajorgrids,
            ymin=0, ymax=1.05589161478605,
            ytick style={color=black},
            xticklabel style={
                    /pgf/number format/fixed,
                    /pgf/number format/precision=2
                },
            scaled x ticks=false
        ]
        \addplot [very thick, steelblue31119180, opacity=1]
        table {%
                0 0
                0.00296974457222592 0.0516993732508435
                0.00594015891379931 0.103471221165282
                0.00891121807570502 0.155315159019106
                0.011882898574819 0.207230825333442
                0.0148551782986006 0.259217880354045
                0.0178280364165769 0.311276004638352
                0.0208014532980975 0.363404897742068
                0.0237754104358763 0.415604276997673
                0.0267498903748823 0.467873876377874
                0.0297248766461704 0.520213445437608
                0.0327003537052824 0.572622748328671
                0.0356762993827764 0.625100859831919
                0.0386518965627478 0.677608100823487
                0.0416010458171891 0.729499929124806
                0.0443876613502291 0.778039019021006
                0.0468118681356625 0.819448100646943
                0.0488010038483327 0.852479354235905
                0.0504551401104873 0.879102989197116
                0.0518053600096683 0.899945917836356
                0.05299013170267 0.917573578554906
                0.0539864054727894 0.931622348823799
                0.0548895476977236 0.943827778291797
                0.0556945596597134 0.95412636854962
                0.0564025912112368 0.962568978312566
                0.0570571956359689 0.969932268568213
                0.0576667554366336 0.976379251051819
                0.0582161382832602 0.981665750142129
                0.0587161263447303 0.985987891956528
                0.0591853008722472 0.989670356866429
                0.0596296300380484 0.992823990784904
                0.0600439027344714 0.99536990136186
                0.0604208705074243 0.997197478987712
                0.0607704468111335 0.998475502675921
                0.0611007804842964 0.999345636119345
                0.0614141431621631 0.99984910402631
                0.0617111814695415 1
                0.0619885307394148 0.999750572597278
                0.0622420708665704 0.999043594331068
                0.0624774636947585 0.9979694934043
                0.062697608914555 0.996574250552551
                0.0629061466064984 0.994921571123404
                0.0631024288330209 0.993000237790409
                0.0632873540308292 0.990827914617766
                0.0634576818926892 0.988358459087274
                0.0636121621063368 0.985579320191546
                0.0637520267274608 0.982510376940044
                0.0638796659581687 0.979186287596934
                0.0639961637943732 0.975625050426052
                0.0641018206030761 0.97182995101837
                0.0641980388309257 0.967823438562578
                0.0642845579782385 0.963601302340128
                0.0643595934080081 0.959140187681773
                0.0644232494173472 0.954442049863806
                0.0644728432952225 0.949474318816399
                0.0645103801160941 0.944263825674141
                0.0645369804467912 0.938824095782403
                0.0645531831780822 0.933161315347269
                0.0645583874745431 0.927265702626506
                0.0645534250532233 0.921148301674645
                0.0645378716436684 0.914800955599445
                0.0645115691637861 0.908219961234633
                0.0644740765535127 0.90139966198068
                0.0644242489163166 0.894324635261796
                0.0643620639530994 0.886991672279799
                0.0642857726888685 0.879380443257277
                0.0641966401154408 0.871501966376152
                0.0640940986130234 0.86334544399633
                0.0639781828690559 0.854907145699377
                0.0638490282397947 0.846184162278791
                0.0637062421156551 0.837167167016195
                0.0635493772001141 0.827845612057413
                0.063378096143633 0.818210065033287
                0.0631926749110214 0.808257387653724
                0.062993071911152 0.797981249265802
                0.0627790801462105 0.787378822581891
                0.0625533469421417 0.776475834285114
                0.0623193678207804 0.765315224866201
                0.0620831247771617 0.753979725859921
                0.0618548523452739 0.742605327435656
                0.0616474133972992 0.731377547333481
                0.0614739707940176 0.720507217052823
                0.0613454219810866 0.71018174668677
                0.0612666582411212 0.700513504221718
                0.0612373152665051 0.691530347261188
                0.0612533487662135 0.683195504902805
                0.0613095124128888 0.675441614118612
                0.0613998423771498 0.668190660588456
                0.0615180362609781 0.661365964852529
                0.0616591651273324 0.654900396315301
                0.0618180669187005 0.648734702797047
                0.0619947532277493 0.642835437176059
                0.0621869408602322 0.637167581438499
                0.062393394242813 0.631705173917495
                0.0626127533319585 0.626424655215504
                0.0628440453885387 0.621309264195371
                0.0630872599575271 0.616348738785228
                0.0633409399495801 0.611527760189596
                0.0636029882384772 0.606829179658589
                0.0638717510810119 0.602239932179356
                0.0641448593648129 0.597742894057087
            };
        \addlegendentry{precomputation}
        \addplot [very thick, darkorange25512714, opacity=1, dashed]
        table {%
                0 0
                0.00296983566969531 0.0516995722177976
                0.00594031005335913 0.103471993330853
                0.00891140786576888 0.155316811826091
                0.0118831141790749 0.207233585211018
                0.0148554144095496 0.25922187966511
                0.0178282943068776 0.311281269790846
                0.0208017399438661 0.363411338373586
                0.0237757377065254 0.415611676149919
                0.0267502742845087 0.467881881583587
                0.0297253366619059 0.520221560649359
                0.0327009121083667 0.572630326624128
                0.0356768503036996 0.625110632924133
                0.0386525396121802 0.677619434332409
                0.0415983609701026 0.729452797465097
                0.0444007015138259 0.778264505468662
                0.0468241339404487 0.819655595032825
                0.0488079278706433 0.852593231264445
                0.0504640565243066 0.879241703189879
                0.0518122324832699 0.900050921618273
                0.0529961511592227 0.917659122176276
                0.0539913682958293 0.931691074993582
                0.0548940931481808 0.943888126273167
                0.0556986925316731 0.954177150485128
                0.0564063516601654 0.962612958454995
                0.0570606665199401 0.969969848231524
                0.0576699717060852 0.976411787937094
                0.0582190801090763 0.981692189852476
                0.0587188788405433 0.986010068622818
                0.0591878420347435 0.989688087120249
                0.0596321266807626 0.992840733436776
                0.0600461613797675 0.995381207952982
                0.060422944912826 0.997205297906274
                0.0607723694156429 0.998479998630964
                0.0611026835063322 0.999349530177992
                0.0614157475212415 0.999846585740643
                0.0617130101307211 1.00000130319332
                0.0368296557881168 0.558929227511096
                0.0338534572107029 0.506451845080636
                0.0308777572298236 0.454043154148383
                0.0279025682829908 0.401703534148464
                0.0249279030857564 0.349433371673773
                0.0219537746389286 0.297233060616158
                0.0189801962374423 0.24510300237525
                0.0160071814795691 0.193043606074887
                0.0130347442764684 0.141055288786995
                0.0100628988621083 0.0891384757633931
                0.00714520322497855 0.037293600675835
                0.00431149292415682 -0.0144788941348159
                0.00324475065190775 -0.0661785574010633
                0.00253711874805319 -0.117804928661324
            };
        \addlegendentry{DEIM}
    \end{axis}

\end{tikzpicture}
    \caption{Comparison of the load-displacement curve of the precomputation and the DEIM-reduced simulation with 40 POD modes and 40 DEIM modes. There is an unintended occurrence of artificial unloading, which should be attempted to be avoided.}
    \label{fig:platedmg4040}
\end{figure}
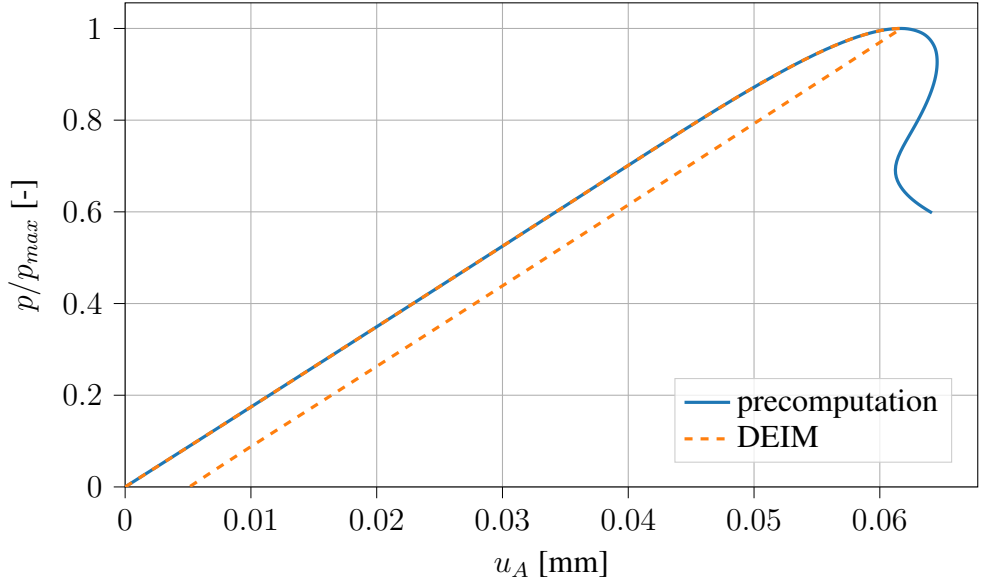

\pgfplotsset{width=.8\textwidth, height=.43\textwidth}
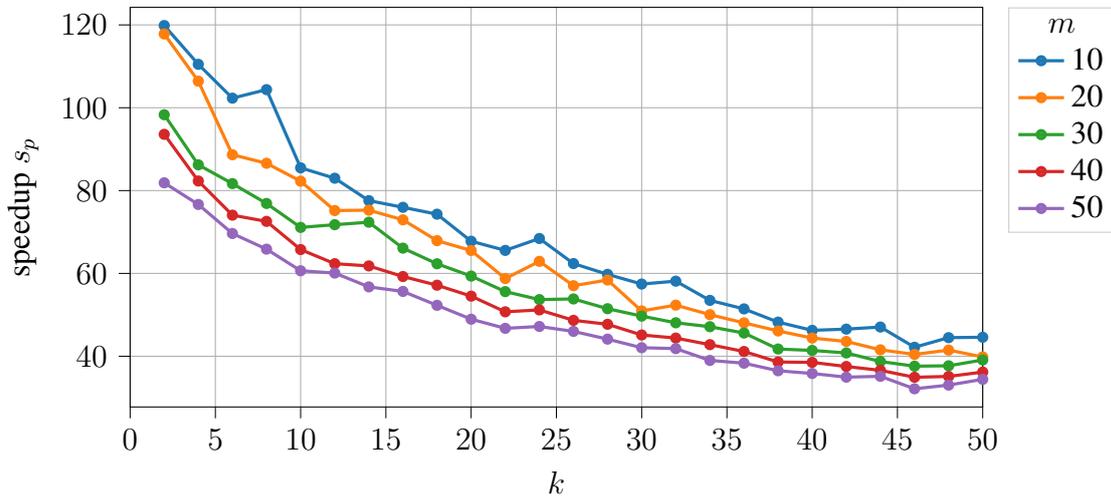
\begin{figure}[!ht]
    \centering
\begin{tikzpicture}

    \definecolor{crimson2143940}{RGB}{214,39,40}
    \definecolor{darkgray176}{RGB}{176,176,176}
    \definecolor{darkorange25512714}{RGB}{255,127,14}
    \definecolor{forestgreen4416044}{RGB}{44,160,44}
    \definecolor{lightgray204}{RGB}{204,204,204}
    \definecolor{mediumpurple148103189}{RGB}{148,103,189}
    \definecolor{steelblue31119180}{RGB}{31,119,180}

    \begin{axis}[
            legend cell align={left},
            legend style={
                    fill opacity=0.8,
                    draw opacity=1,
                    text opacity=1,
                    at={(1.03,1)},
                    anchor=north west,
                    draw=lightgray204
                },,
            tick align=outside,
            tick pos=left,
            x grid style={darkgray176},
            xlabel={\(\displaystyle k\)},
            xmajorgrids,
            xmin=0, xmax=50,
            xtick style={color=black},
            y grid style={darkgray176},
            ylabel={speedup $s_p$},
            ymajorgrids,
            ymin=27.7562034752645, ymax=124.236397007469,
            ytick style={color=black}
        ]
        \addlegendimage{empty legend}
        \addlegendentry{\hspace{-.3cm}$m$}
        \addplot [very thick, steelblue31119180, mark=*, mark size=1.5, mark options={solid}]
        table {%
                2 119.850933665096
                4 110.486005272072
                6 102.3138448121
                8 104.37105642925
                10 85.4876500785631
                12 82.9827399817313
                14 77.5770168763761
                16 75.9592978704431
                18 74.2988918736906
                20 67.7684185932347
                22 65.5544426276402
                24 68.4283642847669
                26 62.3775419908634
                28 59.7750517877502
                30 57.4283496470323
                32 58.128799659274
                34 53.4888624153457
                36 51.4316038606552
                38 48.240422801247
                40 46.2534375192734
                42 46.549978877512
                44 47.0518946141251
                46 42.1724171624798
                48 44.4928066005436
                50 44.5831460841198
            };
        \addlegendentry{10}
        \addplot [very thick, darkorange25512714, mark=*, mark size=1.5, mark options={solid}]
        table {%
                2 117.81645284522
                4 106.419470613758
                6 88.6507811757705
                8 86.5894265513666
                10 82.2680318387894
                12 75.1717153817319
                14 75.2718892746851
                16 72.9348551156772
                18 67.9063109701097
                20 65.522910979875
                22 58.7538232815023
                24 62.908205355041
                26 57.0295674882665
                28 58.4061973844428
                30 50.937228218498
                32 52.3415846841323
                34 50.0406827243024
                36 48.0603956216782
                38 46.1080076675376
                40 44.3920305941411
                42 43.5713308078915
                44 41.5544530572305
                46 40.4415984435956
                48 41.4866535244961
                50 39.8352254987884
            };
        \addlegendentry{20}
        \addplot [very thick, forestgreen4416044, mark=*, mark size=1.5, mark options={solid}]
        table {%
                2 98.3114643079609
                4 86.2160665598025
                6 81.6822690806088
                8 76.8874060861204
                10 71.1037438476863
                12 71.7627898553156
                14 72.3585074610666
                16 66.1012979409934
                18 62.3256743244058
                20 59.3582628421602
                22 55.604602943835
                24 53.6790142789163
                26 53.836893648872
                28 51.477978714183
                30 49.6930215913299
                32 48.0807982086533
                34 47.1320571591777
                36 45.6054659335811
                38 41.7451134238238
                40 41.4039422879735
                42 40.7951549889624
                44 38.7275531098217
                46 37.6023613369217
                48 37.6908435574839
                50 39.1070853181024
            };
        \addlegendentry{30}
        \addplot [very thick, crimson2143940, mark=*, mark size=1.5, mark options={solid}]
        table {%
                2 93.5814716312623
                4 82.311022421711
                6 74.066613317708
                8 72.5590166899708
                10 65.7606403281945
                12 62.3452432286627
                14 61.807711927864
                16 59.2470869667752
                18 57.1525379739556
                20 54.5113382284413
                22 50.7279094698539
                24 51.1848002841684
                26 48.6714236539253
                28 47.7216724704137
                30 45.1320709599235
                32 44.3894921891046
                34 42.8002289812251
                36 41.1500719009053
                38 38.612775487132
                40 38.5196045500358
                42 37.5270322645078
                44 36.5995416478635
                46 34.9241672299685
                48 35.1233287329268
                50 36.1819198797575
            };
        \addlegendentry{40}
        \addplot [very thick, mediumpurple148103189, mark=*, mark size=1.5, mark options={solid}]
        table {%
                2 81.8717341403509
                4 76.645273072728
                6 69.6339572270164
                8 65.8600920707468
                10 60.6389437090442
                12 60.0854881725507
                14 56.7494899838476
                16 55.6499643041886
                18 52.2969550923896
                20 48.9460655275882
                22 46.7367433902724
                24 47.1888783610893
                26 46.0004650604027
                28 44.1356973215786
                30 42.0607308367242
                32 41.8466454548889
                34 38.9923529108508
                36 38.3262748155376
                38 36.4939701286009
                40 35.838786200467
                42 34.9516422796755
                44 35.1551215054555
                46 32.1416668176374
                48 33.0370623613347
                50 34.4264920530537
            };
        \addlegendentry{50}
    \end{axis}

\end{tikzpicture}
    \caption{Speedup of the reduced simulation with respect to the number of POD basis vectors $m$ and the number of DEIM interpolation points $k$ for damage.}
    \label{fig:plate_damage_speedup}
\end{figure}
For completeness, a comparison of the speedups for the selected number of modes is presented. Even for this complex material behavior, speedups on the order of 40-80 can be achieved, as shown in Figure \ref{fig:plate_damage_speedup}.

\begin{figure}[!ht]
    \centering
    \begin{subfigure}{.8\textwidth}
        \centering
        \begin{subfigure}{.5\textwidth}
            \centering
            \includegraphics[width=\textwidth]{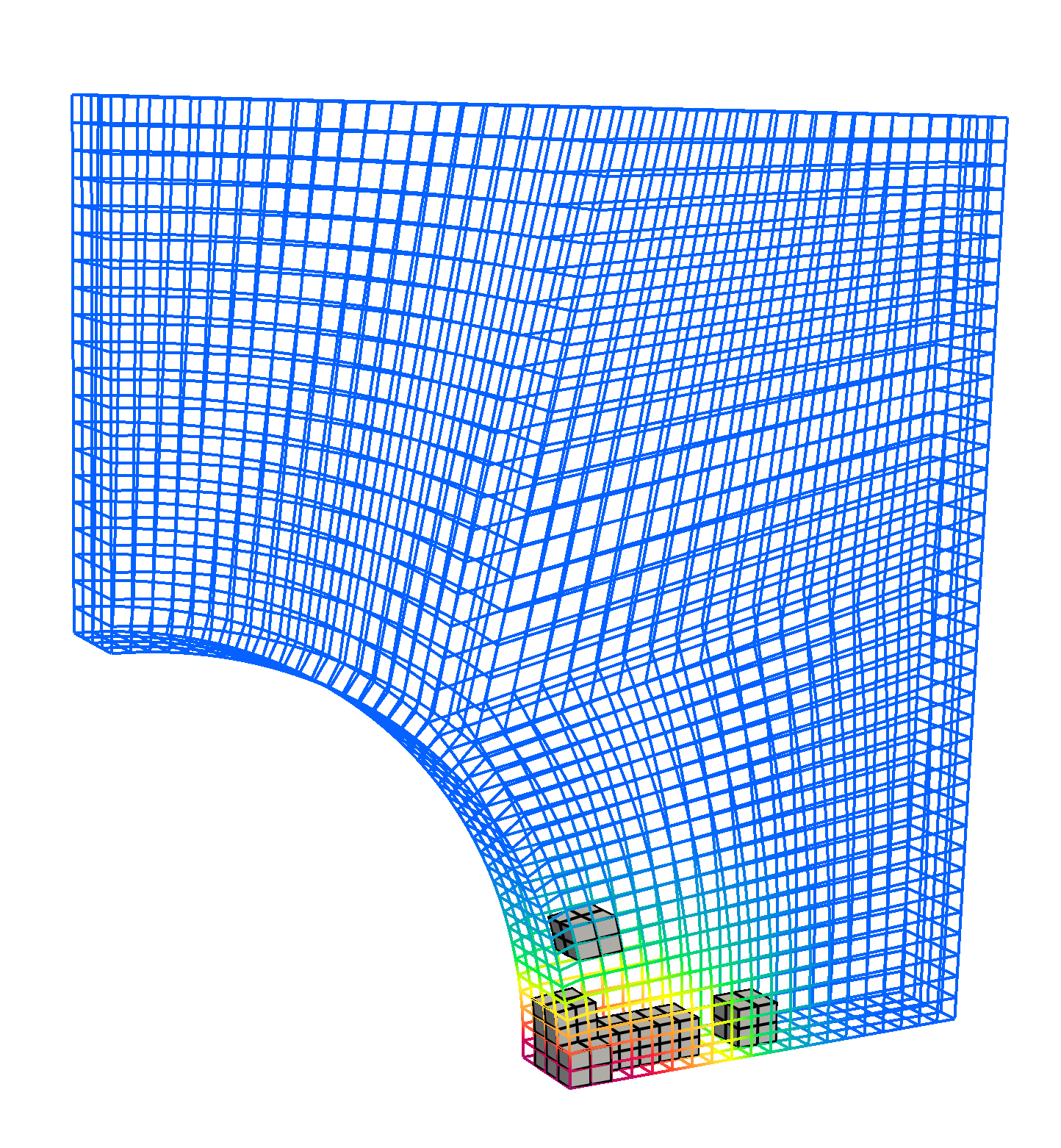}
            \caption{10 interpolation points}
        \end{subfigure}%
        \begin{subfigure}{.5\textwidth}
            \centering
            \includegraphics[width=\textwidth]{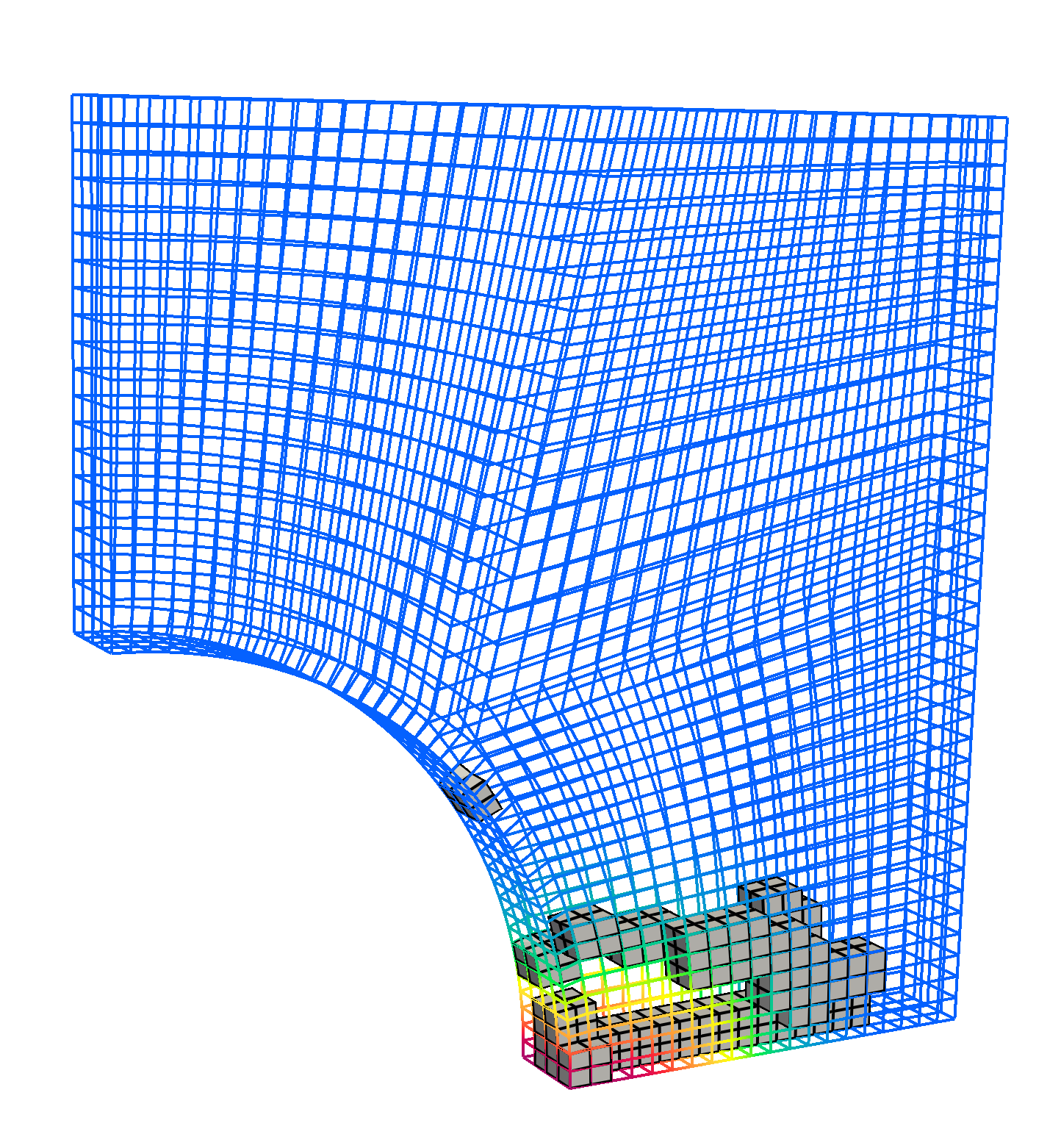}
            \caption{40 interpolation points}
        \end{subfigure}
        \begin{subfigure}{.5\textwidth}
            \centering
            \includegraphics[width=\textwidth]{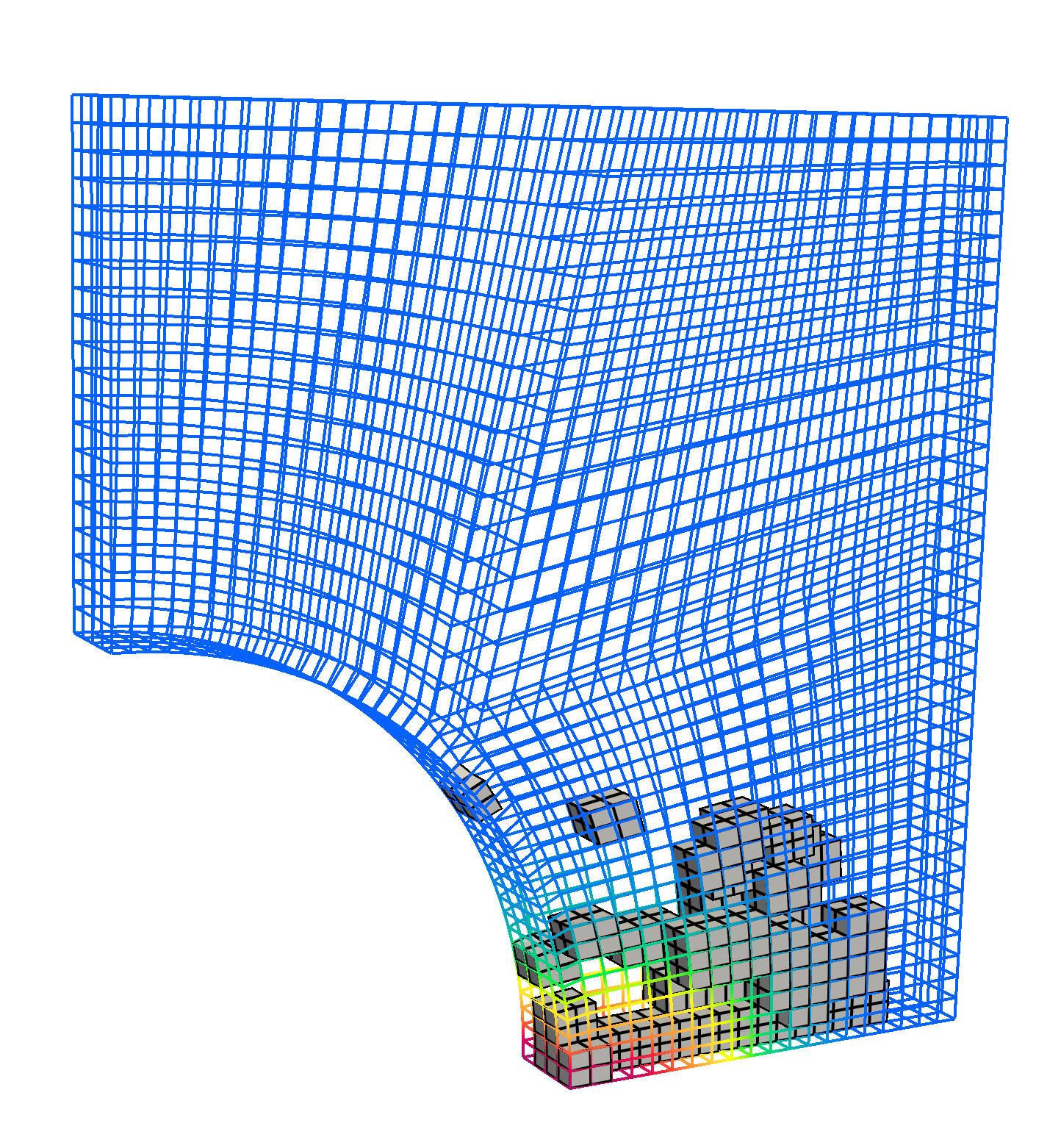}
            \caption{70 interpolation points}
        \end{subfigure}%
        \begin{subfigure}{.5\textwidth}
            \centering
            \includegraphics[width=\textwidth]{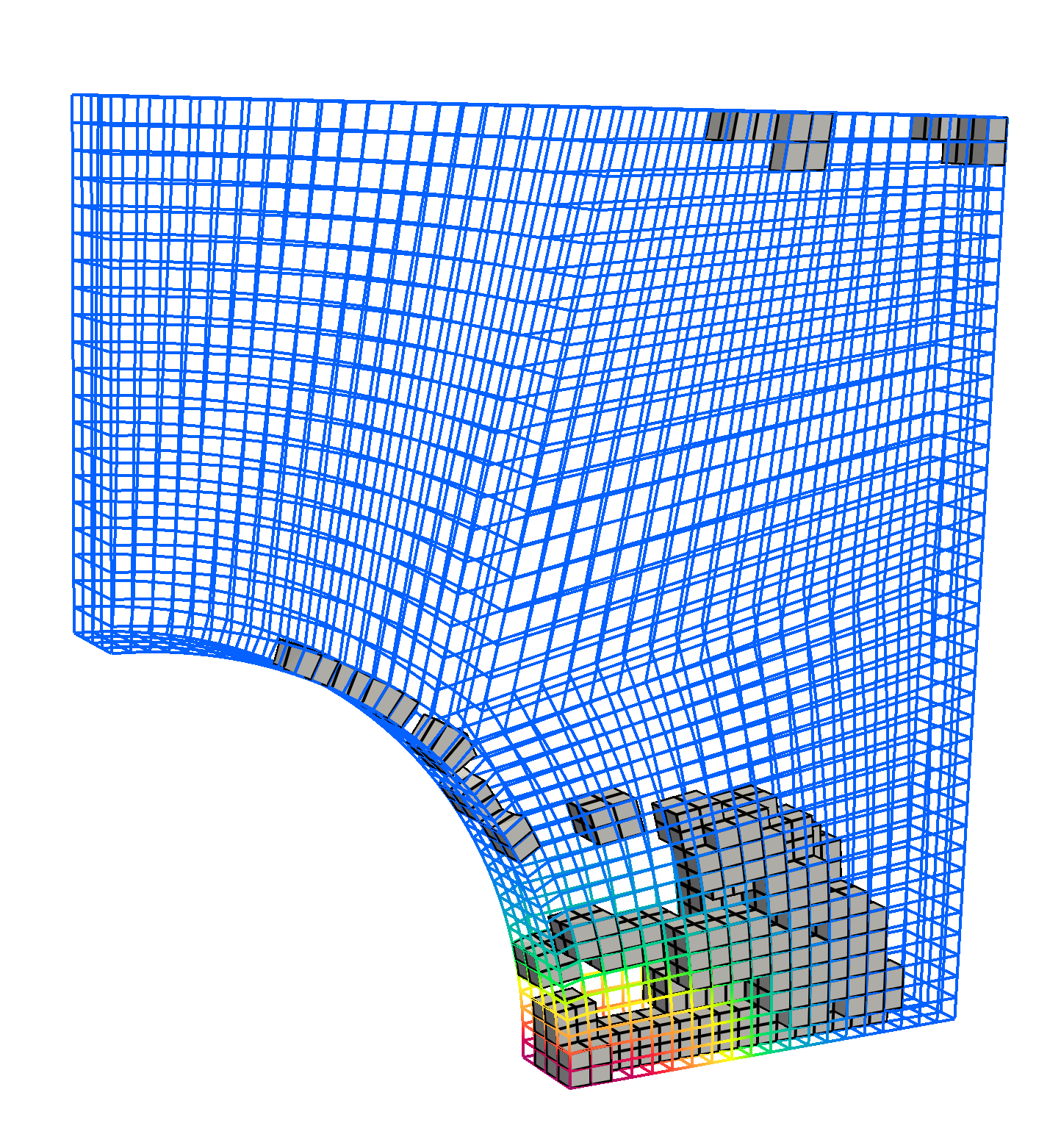}
            \caption{100 interpolation points}
        \end{subfigure}
    \end{subfigure}
    \begin{subfigure}{.12\textwidth}
        \includegraphics[width=\textwidth]{figures/legend.png}
    \end{subfigure}
    \caption{Elements that need to be evaluated in the reduced order model for different numbers of DEIM interpolation points.}
    \label{fig:dmginterpolationpoints}
\end{figure}

In Figure \ref{fig:dmginterpolationpoints}, the DEIM interpolation points selected by the Greedy algorithm are shown. Only the highlighted elements need to be evaluated which is really remarkable. The algorithm primarily selects elements from the region where damage occurs. Only after a sufficient number of elements from this region have been selected, the algorithm chooses elements from other regions.

\FloatBarrier
\subsubsection{Damage and plasticity}

In this chapter, we examine the combination of plasticity and damage, using the material parameters from Table \ref{tab:paraplate}. The corresponding load-displacement curves have already been shown in Figure \ref{fig:mesh_conv_plate}. Consequently, Figure \ref{fig:plate_damage_plasticity_limit_error} directly presents the relative error of the limit load. While the error steadily decreases during the first 10 DEIM modes and is already below one percent, numerical instabilities and artifical unloading occur once again. However, for 50 and 70 POD modes, the error remains predominantly below one percent.

\pgfplotsset{width=.8\textwidth, height=.43\textwidth}
\begin{figure}[!ht]
    \centering
\begin{tikzpicture}

    \definecolor{crimson2143940}{RGB}{214,39,40}
    \definecolor{darkgray176}{RGB}{176,176,176}
    \definecolor{darkorange25512714}{RGB}{255,127,14}
    \definecolor{forestgreen4416044}{RGB}{44,160,44}
    \definecolor{lightgray204}{RGB}{204,204,204}
    \definecolor{steelblue31119180}{RGB}{31,119,180}

    \begin{axis}[
            legend cell align={left},
            legend style={
                    fill opacity=0.8,
                    draw opacity=1,
                    text opacity=1,
                    at={(1.03,1)},
                    anchor=north west,
                    draw=lightgray204
                },
            log basis y={10},
            tick align=outside,
            tick pos=left,
            x grid style={darkgray176},
            xlabel={\(\displaystyle k\)},
            xmajorgrids,
            xmin=1, xmax=50,
            xtick style={color=black},
            y grid style={darkgray176},
            ylabel={\(\displaystyle \varepsilon_{p_{max}}\)},
            ymajorgrids,
            ymin=1e-05, ymax=10,
            ymode=log,
            ytick style={color=black},
            ytick={1e-06,1e-05,0.0001,0.001,0.01,0.1,1,10,100},
            yticklabels={
                    \(\displaystyle {10^{-6}}\),
                    \(\displaystyle {10^{-5}}\),
                    \(\displaystyle {10^{-4}}\),
                    \(\displaystyle {10^{-3}}\),
                    \(\displaystyle {10^{-2}}\),
                    \(\displaystyle {10^{-1}}\),
                    \(\displaystyle {10^{0}}\),
                    \(\displaystyle {10^{1}}\),
                    \(\displaystyle {10^{2}}\)
                }
        ]
        \addlegendimage{empty legend}
        \addlegendentry{\hspace{-.3cm}$m$}
        \addplot [thick, steelblue31119180, mark=*, mark size=1.5, mark options={solid}]
        table {%
                2 0.0682299844868119
                4 0.0227358064054878
                6 0.0125809448420111
                8 0.00793879725608674
                10 0.0688405484420737
                12 0.00277251144097103
                14 0.0322101776637043
                16 2.15215535232163
                18 0.054691544112248
                20 0.00187005212199052
                22 0.0668515743740051
                24 0.00499039845516552
                26 0.00356249079072743
                28 0.000857252860034189
                30 0.0558258037369838
                32 0.0385224995698235
                34 0.0234095752529874
                36 0.0544453243711679
                38 0.0511579367332605
                40 0.0518775650185782
                42 0.296842762385783
                44 0.287631455443242
                46 0.12093273500741
                48 0.201262734081016
                50 0.0606910566590514
            };
        \addlegendentry{20}
        \addplot [thick, darkorange25512714, mark=*, mark size=1.5, mark options={solid}]
        table {%
                2 0.0754640490016576
                4 0.0220004900601723
                6 0.00808414911309063
                8 0.00670660417798433
                10 0.00344367294911366
                12 0.00175942303358875
                14 0.0673150910846558
                16 0.353741420092755
                18 0.000913035182365048
                20 0.00191035580465823
                22 0.000801041555373898
                24 0.000643095037991946
                26 0.000246876363904948
                28 0.00138090188920953
                30 0.000606171616133312
                32 0.00218731547230473
                34 0.00787056110713832
                36 0.000507760545033857
                38 0.00732712450221966
                40 0.00829601235378927
                42 0.079058480036668
                44 0.323254828709181
                46 0.271503285889067
                48 0.00647889805090745
                50 0.000889452845388535
            };
        \addlegendentry{40}
        \addplot [thick, forestgreen4416044, mark=*, mark size=1.5, mark options={solid}]
        table {%
                2 0.0741329528065932
                4 0.0223416816546378
                6 0.0129279219683065
                8 0.00675171514121806
                10 0.00293517103245916
                12 0.0019600939832993
                14 0.0747480633997276
                16 0.00110534647604369
                18 0.000741737188662474
                20 0.00142331015103291
                22 0.000303480216040233
                24 0.000448618009460996
                26 0.00573635031783741
                28 0.000816448144465594
                30 0.000407089789834156
                32 0.00225396270000448
                34 0.000238213999707178
                36 4.40814644543473e-05
                38 0.00880092886922592
                40 0.000382310764887949
                42 0.0470139603599503
                44 0.0469298957903447
                46 0.00271600540119932
                48 0.0288323511129501
                50 0.0130222847622232
            };
        \addlegendentry{50}
        \addplot [thick, crimson2143940, mark=*, mark size=1.5, mark options={solid}]
        table {%
                2 0.0740907484565517
                4 0.0224293668332438
                6 0.0129277854089393
                8 0.00676073657515588
                10 0.00816162162470156
                12 0.00191757332599481
                14 0.745200607027763
                16 0.00543071604825801
                18 0.322559983667542
                20 0.00144670155555923
                22 0.00682587879723153
                24 0.000146295585239023
                26 0.00674443586670469
                28 0.000674660413623722
                30 0.000566150649524498
                32 0.00666999100912622
                34 0.000316083571647328
                36 0.000142613643169066
                38 0.00774309423502432
                40 0.00381107291259377
                42 0.0258930097369749
                44 0.0308159624111518
                46 0.0246118195089566
                48 0.0285704222101934
                50 0.0330160340088024
            };
        \addlegendentry{70}
    \end{axis}

\end{tikzpicture}
    \caption{Relative error of the limit load with respect to the number of POD basis vectors $m$ and the number of DEIM interpolation points $k$.}
    \label{fig:plate_damage_plasticity_limit_error}
\end{figure}
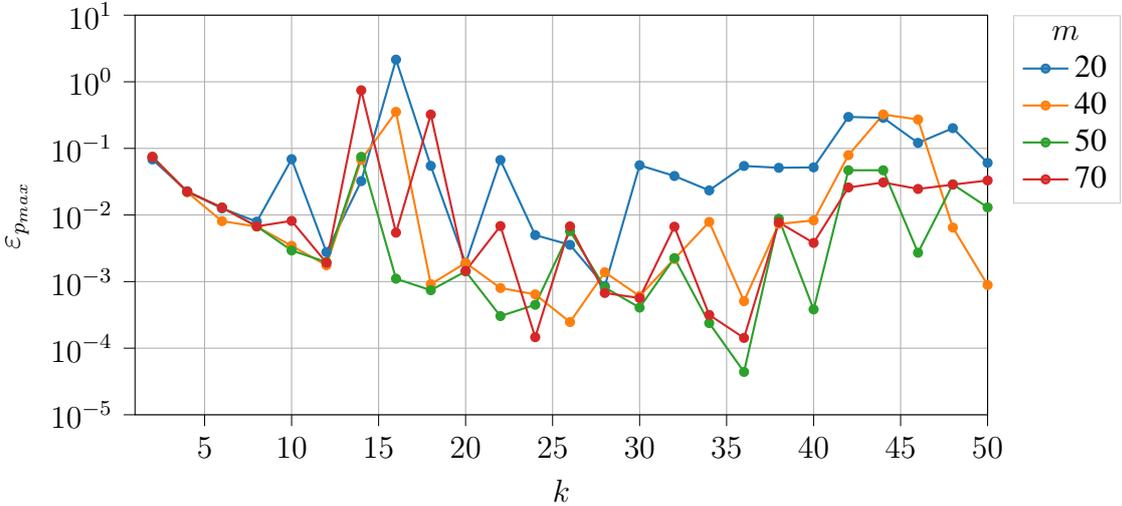

\begin{figure}[!ht]
    \centering
\begin{tikzpicture}

    \definecolor{crimson2143940}{RGB}{214,39,40}
    \definecolor{darkgray176}{RGB}{176,176,176}
    \definecolor{darkorange25512714}{RGB}{255,127,14}
    \definecolor{forestgreen4416044}{RGB}{44,160,44}
    \definecolor{lightgray204}{RGB}{204,204,204}
    \definecolor{steelblue31119180}{RGB}{31,119,180}

    \begin{axis}[
            legend cell align={left},
            legend style={
                    fill opacity=0.8,
                    draw opacity=1,
                    text opacity=1,
                    at={(1.03,1)},
                    anchor=north west,
                    draw=lightgray204
                },
            log basis y={10},
            tick align=outside,
            tick pos=left,
            x grid style={darkgray176},
            xlabel={\(\displaystyle k\)},
            xmajorgrids,
            xmin=1, xmax=50,
            xtick style={color=black},
            y grid style={darkgray176},
            ylabel={\(\displaystyle \varepsilon_{u_{A}}\)},
            ymajorgrids,
            ymin=1e-05, ymax=10,
            ymode=log,
            ytick style={color=black},
            ytick={1e-06,1e-05,0.0001,0.001,0.01,0.1,1,10,100},
            yticklabels={
                    \(\displaystyle {10^{-6}}\),
                    \(\displaystyle {10^{-5}}\),
                    \(\displaystyle {10^{-4}}\),
                    \(\displaystyle {10^{-3}}\),
                    \(\displaystyle {10^{-2}}\),
                    \(\displaystyle {10^{-1}}\),
                    \(\displaystyle {10^{0}}\),
                    \(\displaystyle {10^{1}}\),
                    \(\displaystyle {10^{2}}\)
                }
        ]
        \addlegendimage{empty legend}
        \addlegendentry{\hspace{-.3cm}$m$}
        \addplot [thick, steelblue31119180, mark=*, mark size=1.5, mark options={solid}]
        table {%
                2 0.00155336481678711
                4 0.00964493004075724
                6 0.131650445499123
                8 0.000479844049417917
                10 0.0251426241581524
                12 0.102643295870976
                14 0.0248534667030697
                16 0.693188348321637
                18 0.283146011986559
                20 0.000261491976019537
                22 0.306023676969002
                24 0.00594977486638429
                26 0.0596122964261695
                28 4.5862602981909e-05
                30 0.169630417010838
                32 0.228841352601069
                34 0.194708764702557
                36 0.0443563145384848
                38 0.0319331402353558
                40 0.262747013508319
                42 0.738297283613156
                44 0.784395837781388
                46 0.122387997496353
                48 0.782167735656077
                50 0.00202587392168881
            };
        \addlegendentry{20}
        \addplot [thick, darkorange25512714, mark=*, mark size=1.5, mark options={solid}]
        table {%
                2 0.0149033536826648
                4 0.0102473104707314
                6 0.00290776548535477
                8 0.00214502000841119
                10 0.0132186669545736
                12 0.00195781038734397
                14 0.357074488525539
                16 0.888070421918862
                18 0.000374917791631649
                20 0.0427281848859916
                22 9.91894956906035e-05
                24 0.000333985336958983
                26 0.000769768600875378
                28 0.0219483269191083
                30 0.000304139024678476
                32 0.0345113594569251
                34 0.100831305612874
                36 0.0103155943913121
                38 0.10814462982877
                40 0.114844403712157
                42 0.229679771322353
                44 0.728397004525471
                46 0.527277942263641
                48 0.0177370848316226
                50 0.0624190150007514
            };
        \addlegendentry{40}
        \addplot [thick, forestgreen4416044, mark=*, mark size=1.5, mark options={solid}]
        table {%
                2 0.0150476134852989
                4 0.0101702429939864
                6 0.131578912910531
                8 0.00204982910285346
                10 0.0136702702339943
                12 0.00210416910307965
                14 0.357015753969469
                16 0.10251384741738
                18 0.000225930808226413
                20 0.0423464677032971
                22 0.000247228177325976
                24 0.000739623794485693
                26 0.0922684329328754
                28 0.00678922381250603
                30 0.000509892105763906
                32 0.000467156441315764
                34 0.000310922887092085
                36 0.000534850739166785
                38 0.100648510759796
                40 0.00107016428793654
                42 0.207122222171715
                44 0.232380812573861
                46 0.0959742125986049
                48 0.0207961161604425
                50 0.0181209739830768
            };
        \addlegendentry{50}
        \addplot [thick, crimson2143940, mark=*, mark size=1.5, mark options={solid}]
        table {%
                2 0.0151563914548546
                4 0.0102088353785796
                6 0.131534231511474
                8 0.00209759890909932
                10 0.0809113554729257
                12 0.00210354272755335
                14 1.40281752513989
                16 0.114164469862046
                18 0.570810523830016
                20 0.0284200282629581
                22 0.092902888094349
                24 7.32109460135546e-05
                26 0.0926727998830147
                28 0.000267368001316175
                30 0.000423122726397663
                32 0.101571979785951
                34 0.000379414452976637
                36 0.000509388430715392
                38 0.132107352235385
                40 0.0406553029701697
                42 0.0218322043471633
                44 0.020847280870308
                46 0.0179902737784082
                48 0.0206025934403314
                50 0.0216891869510534
            };
        \addlegendentry{70}
    \end{axis}

\end{tikzpicture}
    \caption{Relative error of the displacement in point A with respect to the number of POD basis vectors $m$ and the number of DEIM interpolation points $k$.}
    \label{fig:plate_damage_plasticity_displacement_error}
\end{figure}
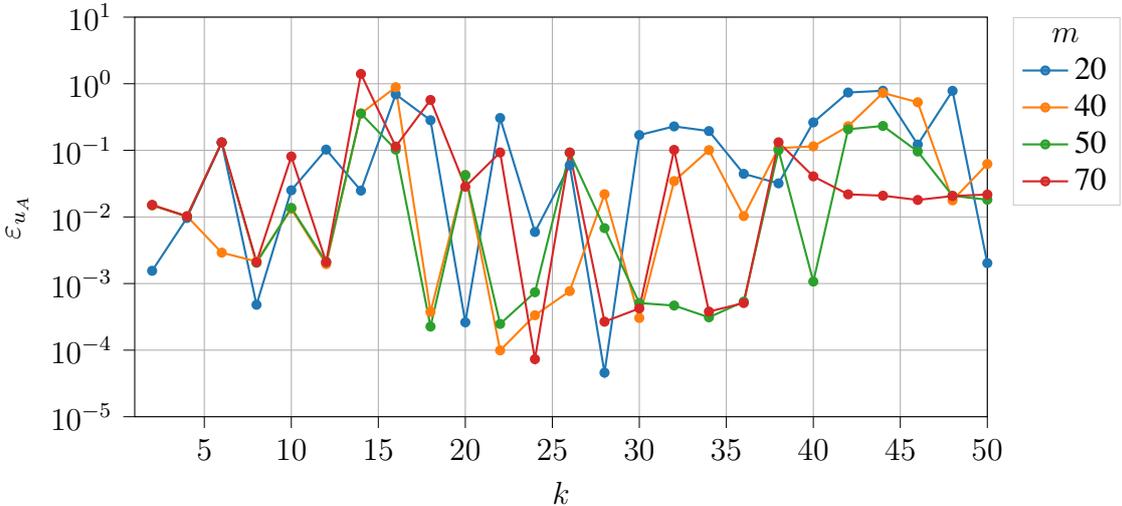

For the displacement error (shown in Figure \ref{fig:plate_damage_plasticity_displacement_error}), there are also many results with very good accuracy.

\begin{figure}[!ht]
    \centering
    \input{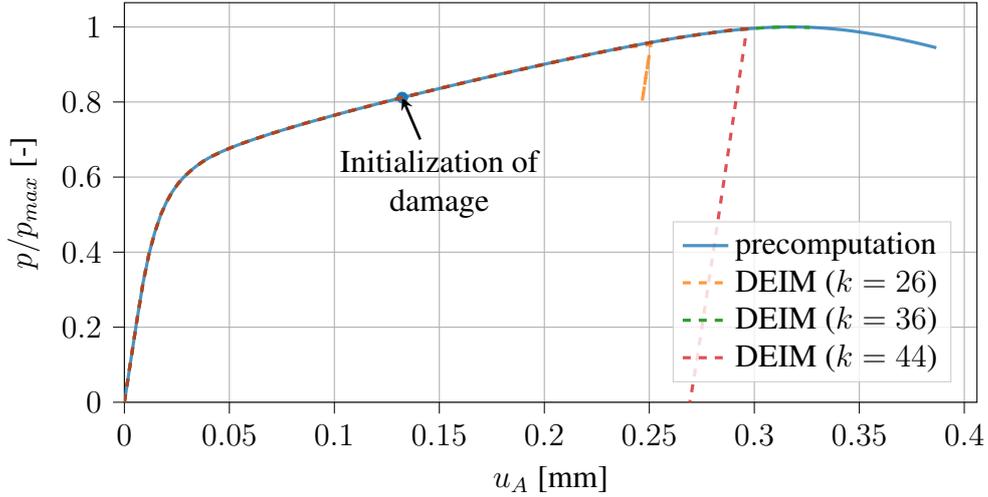}
    \caption{Comparison of the load-displacement curve of the precomputation and the DEIM-reduced simulation with 50 POD modes and 26, 36, and 44 DEIM modes.}
    \label{fig:platedmg502644}
\end{figure}

However, the issue with the arc-length method arises once again, where it sometimes follows the unloading path instead of continuing to load further. As a result, the displacement at the end of the simulation is small and incorrect, even though the majority of the load-displacement curve can be approximated very well. This is exemplified in Figure \ref{fig:platedmg502644}.

\begin{figure}[!ht]
    \centering
\begin{tikzpicture}

\definecolor{crimson2143940}{RGB}{214,39,40}
\definecolor{darkgray176}{RGB}{176,176,176}
\definecolor{darkorange25512714}{RGB}{255,127,14}
\definecolor{forestgreen4416044}{RGB}{44,160,44}
\definecolor{lightgray204}{RGB}{204,204,204}
\definecolor{steelblue31119180}{RGB}{31,119,180}

\begin{axis}[
        legend cell align={left},
        legend style={
                fill opacity=0.8,
                draw opacity=1,
                text opacity=1,
                at={(1.03,1)},
                anchor=north west,
                draw=lightgray204
            },,
        tick align=outside,
        tick pos=left,
        x grid style={darkgray176},
        xlabel={\(\displaystyle k\)},
        xmajorgrids,
        xmin=0, xmax=50,
        xtick style={color=black},
        y grid style={darkgray176},
        ylabel={speedup $s_p$},
        ymajorgrids,
        ymin=0, ymax=150,
        ytick style={color=black}
    ]
\addlegendimage{empty legend}
\addlegendentry{\hspace{-.3cm}$m$}
\addplot [very thick, steelblue31119180, mark=*, mark size=1.5, mark options={solid}]
table {%
2 148.100119332211
4 111.870200584764
6 95.4766848535707
8 72.6614670925455
10 59.0144849615968
12 54.1957962128735
14 49.1839995437298
16 50.1130842100328
18 43.462137263073
20 35.1314930781886
22 37.3370320286627
24 33.1922863022317
26 32.2311813862518
28 27.7669999137576
30 25.9544706849964
32 25.5436035189774
34 23.4694175548215
36 22.8646073530229
38 23.4821143923371
40 21.513577825857
42 28.2268331865462
44 25.4525075902526
46 20.3634815280199
48 23.5386546885987
50 17.5386529270572
};
\addlegendentry{20}
\addplot [very thick, darkorange25512714, mark=*, mark size=1.5, mark options={solid}]
table {%
2 124.23362623721
4 99.0427928119048
6 79.347295703046
8 66.0392317732579
10 55.2224932339301
12 47.769325302968
14 48.6416168911857
16 56.3810631028615
18 36.159296442221
20 34.2909899938016
22 32.1176679299093
24 31.4209606854948
26 27.5336491812248
28 26.350085320006
30 26.7141751990073
32 25.7364218143936
34 26.0661413511367
36 23.1845176483319
38 22.685061068372
40 21.2843384571018
42 21.1247544314384
44 24.5498492733274
46 21.7907509173792
48 18.505418727882
50 16.0806399183944
};
\addlegendentry{40}
\addplot [very thick, forestgreen4416044, mark=*, mark size=1.5, mark options={solid}]
table {%
2 116.847829438542
4 93.8777127912806
6 79.4077725654668
8 62.8454508457887
10 52.9419006743298
12 46.5404688442308
14 49.9197970615481
16 38.6328879708949
18 35.0031058222173
20 32.8598123097778
22 31.5277355112977
24 29.896433822997
26 28.5011995413634
28 24.6577345923005
30 25.7934260690529
32 25.1561620001999
34 22.6472632843039
36 23.3175360673071
38 23.7236922310419
40 20.8313798825179
42 19.7303109238605
44 18.8762515236532
46 17.5651733391203
48 17.1224057450036
50 16.2903056828741
};
\addlegendentry{50}
\addplot [very thick, crimson2143940, mark=*, mark size=1.5, mark options={solid}]
table {%
2 101.693207580934
4 81.2142075261194
6 74.882652233972
8 58.6694193427414
10 49.5963178323621
12 43.5122050212088
14 41.6504294719172
16 38.0449460049248
18 40.1382200673671
20 30.3435787988648
22 31.0328305768756
24 28.2441073668742
26 26.777540540032
28 23.7777096222742
30 23.8287310071633
32 24.7211120628274
34 21.2705964264877
36 21.746042435395
38 20.9830742378903
40 20.0632173598388
42 18.7536146731173
44 16.2492909797552
46 17.1266517065492
48 16.4915402921554
50 15.5189086711494
};
\addlegendentry{70}
\end{axis}

\end{tikzpicture}
    \caption{Speedup of the reduced simulation with respect to the number of POD basis vectors $m$ and the number of DEIM interpolation points $k$ for damage and plasticity.}
    \label{fig:plate_damage_plasticity_speedup}
\end{figure}
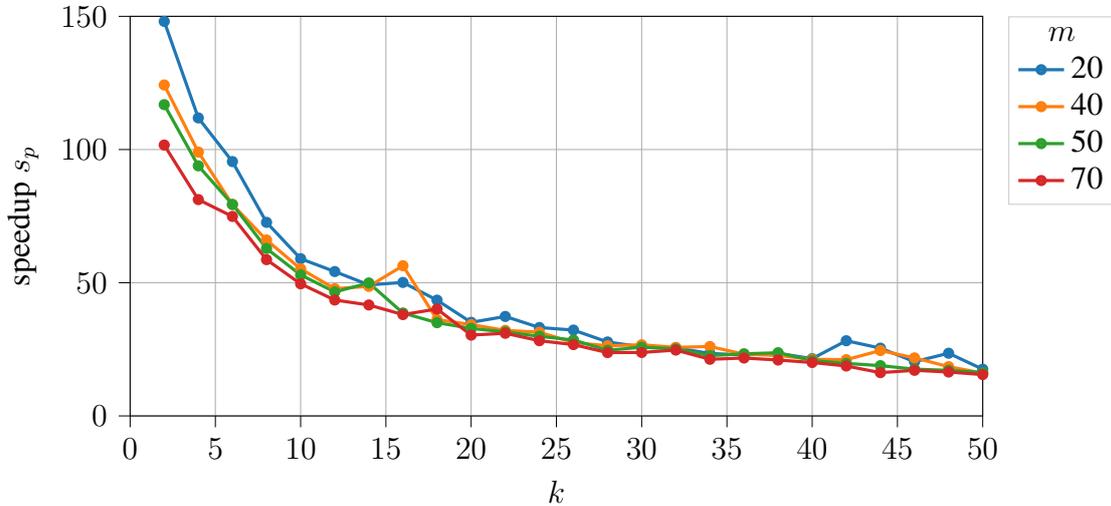

Even for this complex material behaviour decent speedups can be achieved. The speedups $s_p$ are shown in Figure \ref{fig:plate_damage_plasticity_speedup}. With 50 DEIM modes and and 70 POD modes there is still a speedup of factor 15.52. This is because as the complexity of the material increases, the computation time for each individual element goes up, but at the same time, more time can be saved for each element that does not have to be calculated.

\FloatBarrier
\subsection{Asymmetrically notched specimen}

\begin{figure}[ht]
    \centering
    \begin{subfigure}[b]{.6\textwidth}
        \centering
        \def\svgwidth{\textwidth}
        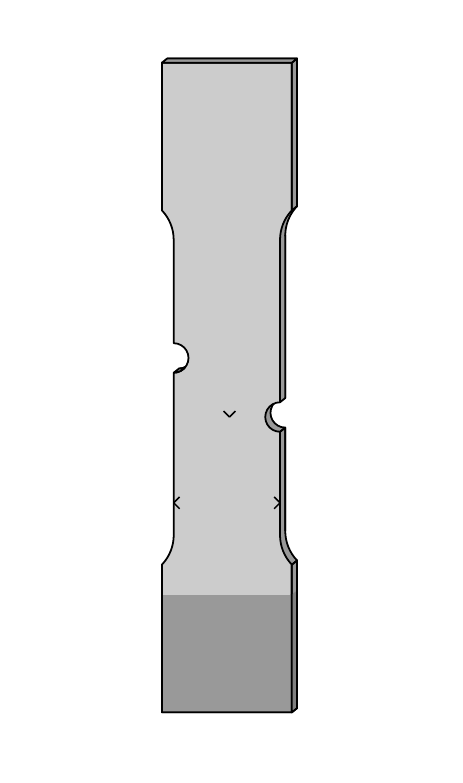
        \caption{Geometry and boundary conditions of the asymmetrically notched specimen (dimensions in mm).}
        \label{fig:asymnotch}
    \end{subfigure}
    \hspace{2cm}
    \begin{subfigure}[b]{.2\textwidth}
        \centering
        \includegraphics[width=\textwidth]{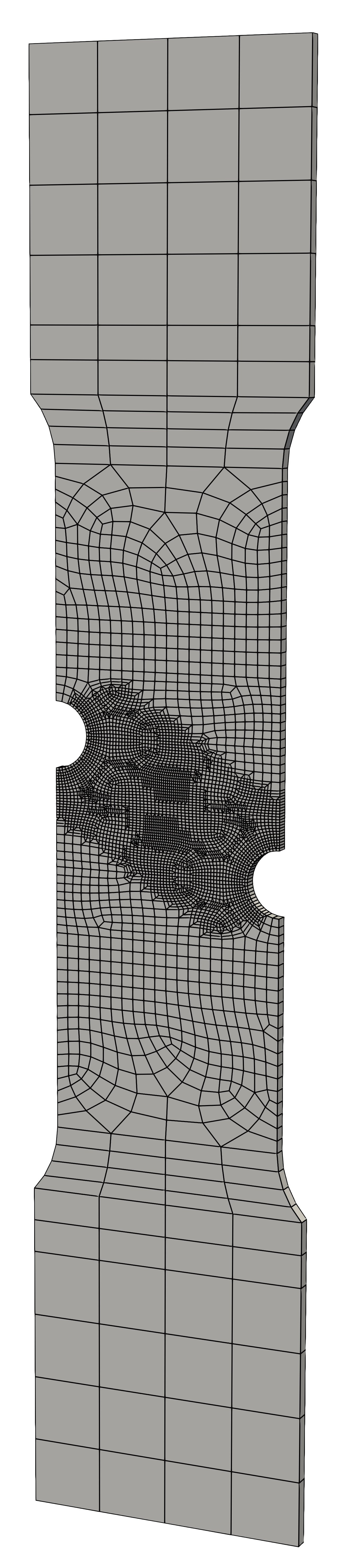}
        \caption{Mesh discretization of the specimen.}
        \label{fig:asymmesh}
    \end{subfigure}
    \caption{Asymetrically notched specimen.}
\end{figure}

The second example is the asymmetrically notched specimen. In this case, no symmetry can be exploited. The corresponding dimensions can be seen in the drawing in Figure \ref{fig:asymnotch}.
The meshing is shown in Figure \ref{fig:asymmesh}. The mesh is refined in the area where damage and plasticity are expected to occur due to the geometric weakening. Since the structure is slightly more complex than the previous example, the material will be examined in three parts again. First, we will consider purely elastoplastic behavior before investigating damage without plasticity. In the final step, we will examine the behavior of plasticity and damage for this example as well.
The general material parameters are listed in Table \ref{tab:asympara}, based on the work of Brepols et al. \cite{brepols2020}.

\begin{spacing}{1.0}
    \begin{table}[ht]
        \centering
        \begin{tabular}{l | l | l | l}
            \hline
            Symbol     & Material parameter                   & Value \hspace{1.5cm} & Unit                     \\
            \hline
            $\Lambda$  & first Lamé parameter                 & 25000                & MPa                      \\
            $\mu$      & second Lamé parameter                & 55000                & MPa                      \\
            $\sigma_0$ & yield stress                         & 100                  & MPa                      \\
            $a$        & first kinematic hardening parameter  & 62.5                 & MPa                      \\
            $b$        & second kinematic hardening parameter & 2.5                  & [-]                      \\
            $e$        & first isotropic hardening parameter  & 125                  & MPa                      \\
            $f$        & second isotropic hardening parameter & 5                    & [-]                      \\
            $Y_0$      & damage threshold                     & 2.5                  & MPa                      \\
            $r$        & first damage hardening parameter     & 5                    & MPa                      \\
            $s$        & second damage hardening parameter    & 100                  & MPa                      \\
            $A$        & internal length scale parameter      & 75                   & MPa $\text{mm}^\text{2}$ \\
            $H$        & penalty parameter                    & $\text{10}^\text{5}$ & MPa
        \end{tabular}
        \caption{Parameter set for the asymmetrically notched specimen.}
        \label{tab:asympara}
    \end{table}
\end{spacing}

\FloatBarrier
\subsubsection{Plasticity}

\begin{figure}[ht]
    \centering
    \begin{subfigure}[t]{.62\textwidth}
        \pgfplotsset{%
            width=1.0\textwidth,
            height=.85\textwidth
        }
        \centering
        \begin{tikzpicture}

    \definecolor{darkgray176}{RGB}{176,176,176}
    \definecolor{steelblue31119180}{RGB}{31,119,180}

    \begin{axis}[
            tick align=outside,
            tick pos=left,
            x grid style={darkgray176},
            xlabel={\(\displaystyle u_A\) [mm]},
            xmajorgrids,
            xmin=0, xmax=0.528967326307322,
            xtick style={color=black},
            y grid style={darkgray176},
            ylabel={\(\displaystyle p / p_{max}\) [-]},
            ymajorgrids,
            ymin=0, ymax=1.05,
            ytick style={color=black},
            clip marker paths
        ]
        \addplot[steelblue31119180, mark=*, mark size=2]
        table {
                0.503778406006973 1
            };
        \addplot [very thick, steelblue31119180, opacity=1]
        table {%
                0 0
                0.00412391286723319 0.0488328579096794
                0.00824784448452829 0.0977433658763376
                0.012371724124637 0.146729969183535
                0.0164954829100232 0.19579113619572
                0.0206190538171993 0.244925358393154
                0.0247425238837412 0.294129476590044
                0.0288767717227959 0.343283161579652
                0.0330357226623155 0.392228311007089
                0.0372288410958955 0.440838823242128
                0.0414628697078556 0.488976106262693
                0.0457475989220842 0.536442208709082
                0.0500974438445083 0.582918557112397
                0.0545332468683603 0.627878978293442
                0.0592256831267338 0.670368370521083
                0.064300441556639 0.708626924808088
                0.0696605473911738 0.740342387165297
                0.0748990812557999 0.764771987169926
                0.0805236203917809 0.782951198814
                0.0854756075341051 0.794951939489267
                0.0893971521042151 0.801996178219255
                0.0929594177300001 0.807379276684806
                0.0963396070544407 0.81194792082041
                0.0996178661554879 0.816014313736951
                0.102835610176106 0.819736999950984
                0.106006046108058 0.823196155281559
                0.109136136704718 0.826451122165428
                0.112246507416316 0.82954958171427
                0.115350796006819 0.832512671806658
                0.11843188366118 0.83535080956805
                0.121496707405803 0.838083712581521
                0.124549665034012 0.840731228287806
                0.12758938509874 0.843299086211125
                0.130622184296134 0.845797070773662
                0.133649586406222 0.848227156661797
                0.136667523581377 0.850599629720423
                0.139674537415712 0.852915989648873
                0.142675133842483 0.855186954509624
                0.14567009156844 0.857411843388634
                0.148662936159097 0.859593952070643
                0.151658074447996 0.861734724121817
                0.154717736803061 0.863834238215827
                0.15778907912276 0.865897708253565
                0.160853093007664 0.867928833227603
                0.163915462772996 0.869931328876043
                0.166979943017852 0.871903842893432
                0.17004587309921 0.873846804135209
                0.173113531158325 0.875762677877787
                0.176177041501463 0.877653575311261
                0.17927672681033 0.879520020585608
                0.182386934543239 0.881363269358867
                0.185492298071331 0.883182089769544
                0.188600038152102 0.884981043582242
                0.191708449588508 0.886757635632008
                0.194813160809162 0.888512134892112
                0.197918618232766 0.890245688469912
                0.201031496010823 0.891958842302854
                0.204138203092577 0.893653190811014
                0.207245103865764 0.895329425294806
                0.210357940923213 0.896988253879588
                0.213479512471931 0.898628426062487
                0.216604786004865 0.900252201555308
                0.219738524744878 0.901860856201648
                0.222880699908859 0.90345320040434
                0.226027032539656 0.9050291759247
                0.229184104522652 0.906588414335959
                0.232352443870989 0.90813212845928
                0.235534334213399 0.909662309121407
                0.238727193938191 0.911176758366878
                0.241923023284335 0.912676255472091
                0.24512605383047 0.914161563653444
                0.248334481182572 0.915632792118271
                0.251560092125028 0.917091101260275
                0.254832203585531 0.918536685498609
                0.258103664402399 0.919968404422293
                0.261380287219442 0.92138819463506
                0.264665284765107 0.922796471362616
                0.267959208781738 0.92419311360414
                0.271263070987886 0.92557936347153
                0.274567466156826 0.926954884256461
                0.277866911012895 0.928317971141152
                0.281163345403582 0.92966781447306
                0.284456276192895 0.931006918609829
                0.287747748412082 0.932335737566654
                0.291040271049161 0.9336536014343
                0.294333499972666 0.934960917838377
                0.297626558796146 0.936258594307913
                0.300918445072941 0.937545687197747
                0.304212062138666 0.938822111922711
                0.30750581588829 0.940088543340888
                0.310797709329002 0.941345650486277
                0.314089926568077 0.942593940689532
                0.317374814557526 0.943832324454367
                0.320658554831857 0.945061567666791
                0.3239496357623 0.946281459574721
                0.327241303768865 0.947493330003355
                0.330530302410378 0.94869560892204
                0.333811559957457 0.949887356614444
                0.33709220256561 0.951071452038286
                0.340372754376567 0.952246641640576
                0.343686539585375 0.953413684166721
                0.346999835089798 0.954571480393117
                0.350307946643317 0.955718662459083
                0.353610628467449 0.956856004360363
                0.356910344964138 0.957982249608791
                0.36020643317523 0.959091085074583
                0.363499509732883 0.960187282321663
                0.366793189713209 0.961270355655746
                0.370078327021758 0.962337817328322
                0.373356965067245 0.963392358389868
                0.376634273385385 0.96443709192849
                0.379905430430279 0.9654720909685
                0.383167763916358 0.966494783843908
                0.3864308407868 0.967508276646163
                0.389688058085777 0.968511891700123
                0.392939527638492 0.969505889243563
                0.396188062975078 0.970491486461151
                0.39943014462364 0.971465602300444
                0.402663384119149 0.97243098182514
                0.405891288995266 0.973389233043719
                0.409113204846444 0.974340865547836
                0.412329567472605 0.975284223423557
                0.415545528194029 0.976219515434221
                0.418752411683739 0.977145920483667
                0.421952312794805 0.978065513013617
                0.425145150075351 0.97897820642036
                0.428331181435781 0.97988375266186
                0.43151201217464 0.980783356132742
                0.4346878903862 0.98167700266746
                0.437857724756458 0.982563568192298
                0.441026681614979 0.983445197951379
                0.444192800103446 0.984319852188114
                0.447351570052201 0.985186892690254
                0.450507578697902 0.986049310319739
                0.45366276647738 0.986906761387684
                0.456816414960513 0.987760331656254
                0.459969392492272 0.988609707572602
                0.463117014750187 0.989453148940342
                0.466259015197603 0.990290913093619
                0.469394483681313 0.99112289287857
                0.472526178407012 0.991950406463283
                0.475656025525901 0.992774199272109
                0.478785683563689 0.993593479872932
                0.481910891534656 0.994406435074196
                0.485036454501369 0.9952158000149
                0.488167708129983 0.996021691495858
                0.491295220251594 0.996823968850091
                0.494420825188833 0.997623186829441
                0.497542618383671 0.99841888667309
                0.500661986371481 0.999211246981507
                0.503778406006973 1
            };
        \node [anchor=north] (time) at (0.48,0.93) {$t=150$};
        \draw [-stealth, line width=1pt, black] (time) -- (0.5038, 1);
    \end{axis}

\end{tikzpicture}
        \caption{Force-displacement curve of the asymmetrically notched specimen for elasto-plastic material behavior.}
        \label{fig:asym_loaddispplasti}
    \end{subfigure}
    \hfill
    \begin{subfigure}[t]{.3\textwidth}
        \includegraphics[width=\textwidth]{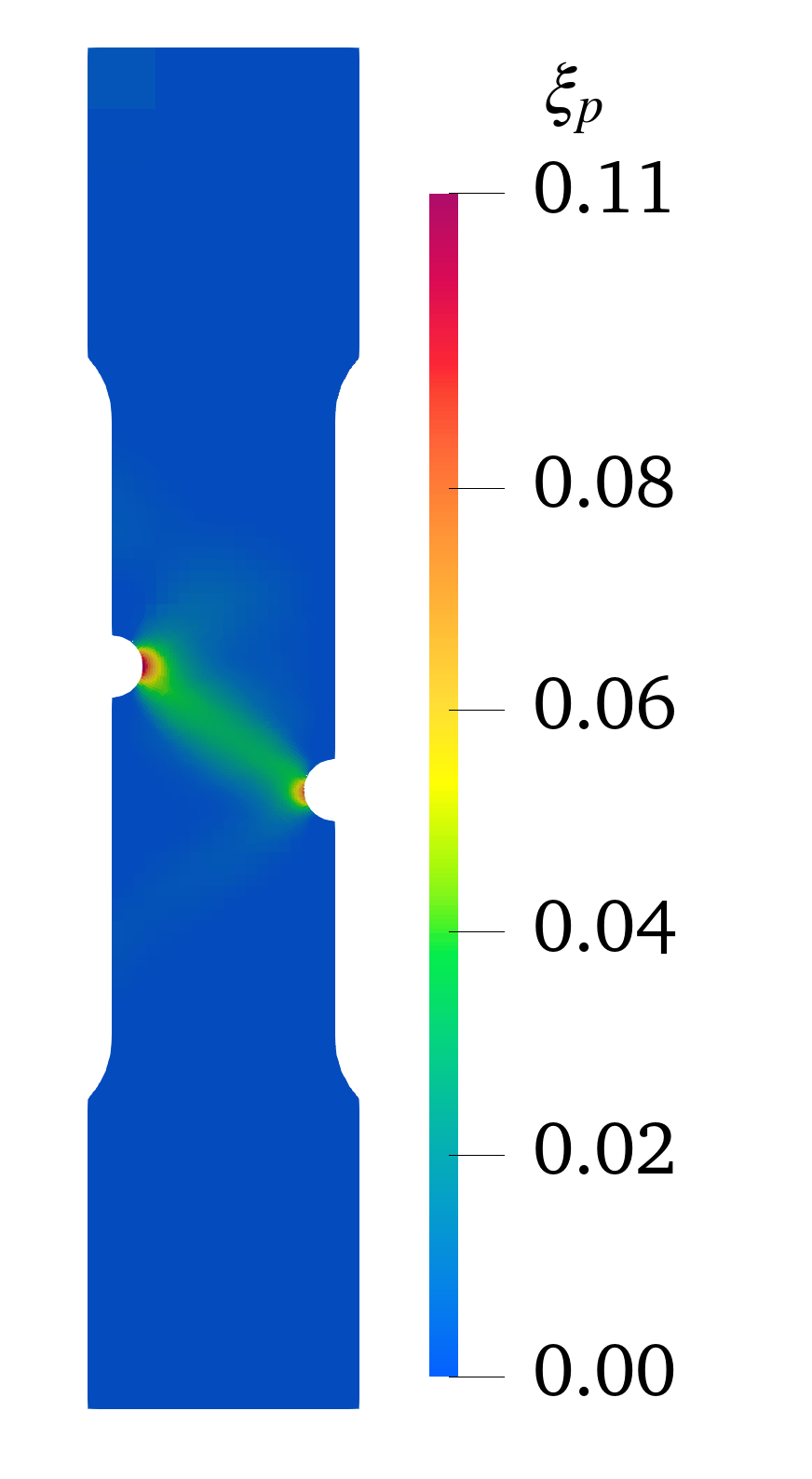}
        \caption{Accumulated plastic strain $\xi_p$ after 150 pseudo-timesteps.}
        \label{fig:asym_plasti}
    \end{subfigure}
    \caption{Simulation results of the asymmetrically notched specimen for elasto-plastic material behavior.}
\end{figure}
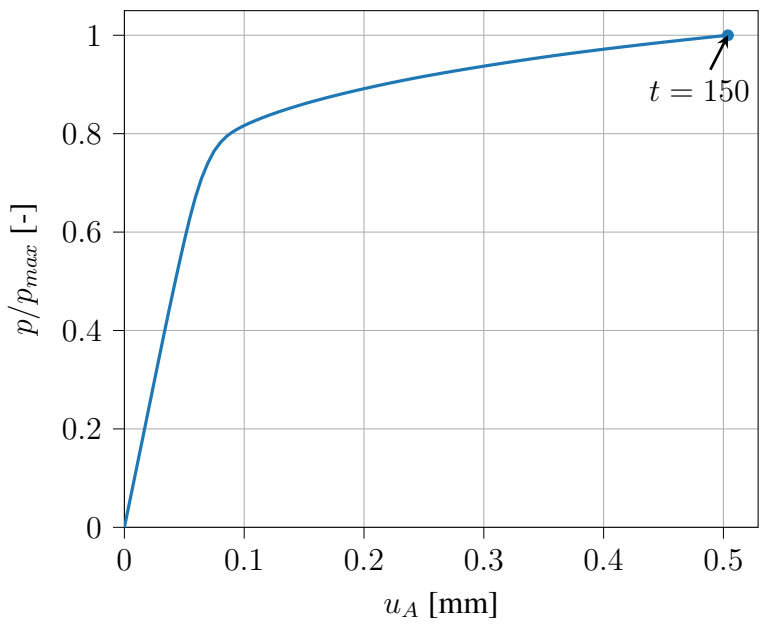
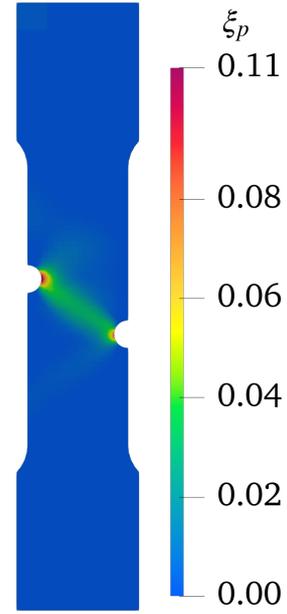

To examine only plasticity, the damage threshold $\sigma_0$ is again set to a high value to prevent any damage. The evolution of accumulated plastic strain is shown in Figure \ref{fig:asym_plasti}. As expected, plastic deformation occurs between the two notches. The calculation was performed with 150 pseudo-timesteps, which were also used as snapshots for the reduced calculations. The load-displacement curve is depicted in Figure \ref{fig:asym_loaddispplasti}.

\begin{figure}[!ht]
    \pgfplotsset{%
        width=.8\textwidth,
        height=.5\textwidth
    }
    \centering
    \input{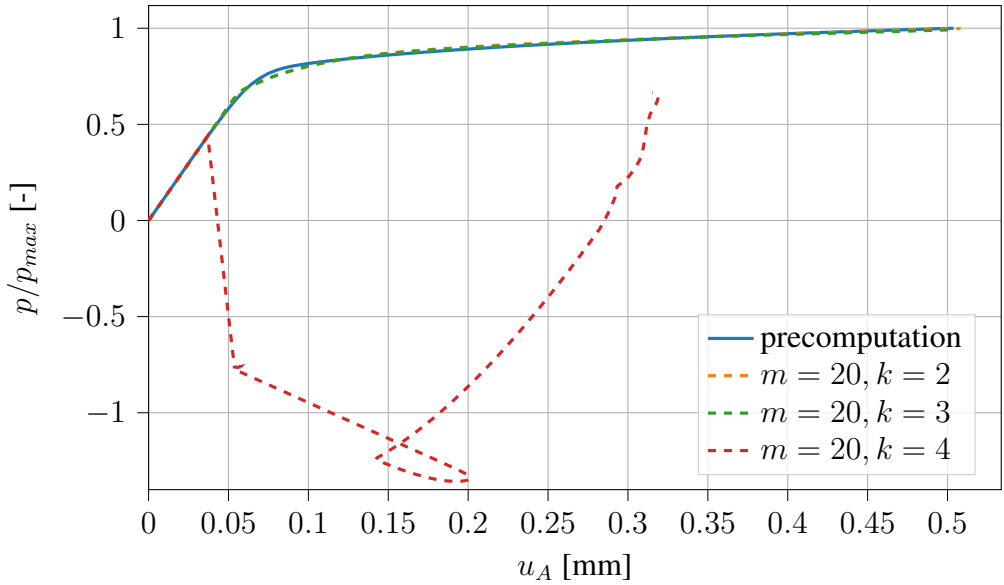}
    \caption{Load-displacement curves of the asymmetrically notched specimen for 20 POD modes and 2,3, and 4 DEIM modes compared to the precomputation.}
    \label{fig:20234}
\end{figure}

Figure \ref{fig:20234} shows the load-displacement curve for several reduced DEIM solutions along with the reference solution. It can be observed that with 20 POD bases and only a few DEIM bases, the behavior can be approximated very well. However, it should be noted that there are instabilities with DEIM in such complex behavior, as evident in the case of 20 POD bases and 4 DEIM bases. While it is usually expected that more DEIM bases would improve the solution, instabilities can also occur in this case.

\begin{figure}
    \centering
    \begin{subfigure}[t]{.1\textwidth}
        \centering
        \def\svgwidth{\textwidth}
\begingroup%
  \makeatletter%
  \providecommand\color[2][]{%
    \errmessage{(Inkscape) Color is used for the text in Inkscape, but the package 'color.sty' is not loaded}%
    \renewcommand\color[2][]{}%
  }%
  \providecommand\transparent[1]{%
    \errmessage{(Inkscape) Transparency is used (non-zero) for the text in Inkscape, but the package 'transparent.sty' is not loaded}%
    \renewcommand\transparent[1]{}%
  }%
  \providecommand\rotatebox[2]{#2}%
  \newcommand*\fsize{\dimexpr\f@size pt\relax}%
  \newcommand*\lineheight[1]{\fontsize{\fsize}{#1\fsize}\selectfont}%
  \ifx\svgwidth\undefined%
    \setlength{\unitlength}{253.59775814bp}%
    \ifx\svgscale\undefined%
      \relax%
    \else%
      \setlength{\unitlength}{\unitlength * \real{\svgscale}}%
    \fi%
  \else%
    \setlength{\unitlength}{\svgwidth}%
  \fi%
  \global\let\svgwidth\undefined%
  \global\let\svgscale\undefined%
  \makeatother%
  \begin{picture}(1,1.06245777)%
    \lineheight{1}%
    \setlength\tabcolsep{0pt}%
    \put(0,0){\includegraphics[width=\unitlength,page=1]{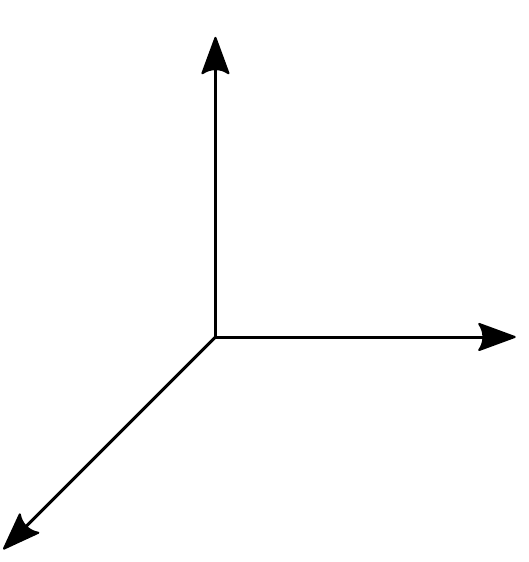}}%
    \put(0.92854113,0.46612945){\color[rgb]{0,0,0}\makebox(0,0)[lt]{\lineheight{1.25}\smash{\begin{tabular}[t]{l}$x$\end{tabular}}}}%
    \put(0.25862288,0.96800355){\color[rgb]{0,0,0}\makebox(0,0)[lt]{\lineheight{1.25}\smash{\begin{tabular}[t]{l}$y$\end{tabular}}}}%
    \put(0.10411092,0.00722451){\color[rgb]{0,0,0}\makebox(0,0)[lt]{\lineheight{1.25}\smash{\begin{tabular}[t]{l}$z$\end{tabular}}}}%
  \end{picture}%
\endgroup%

    \end{subfigure}
    \begin{subfigure}[t]{.49\textwidth}
        \centering
        \includegraphics[width=.99\textwidth]{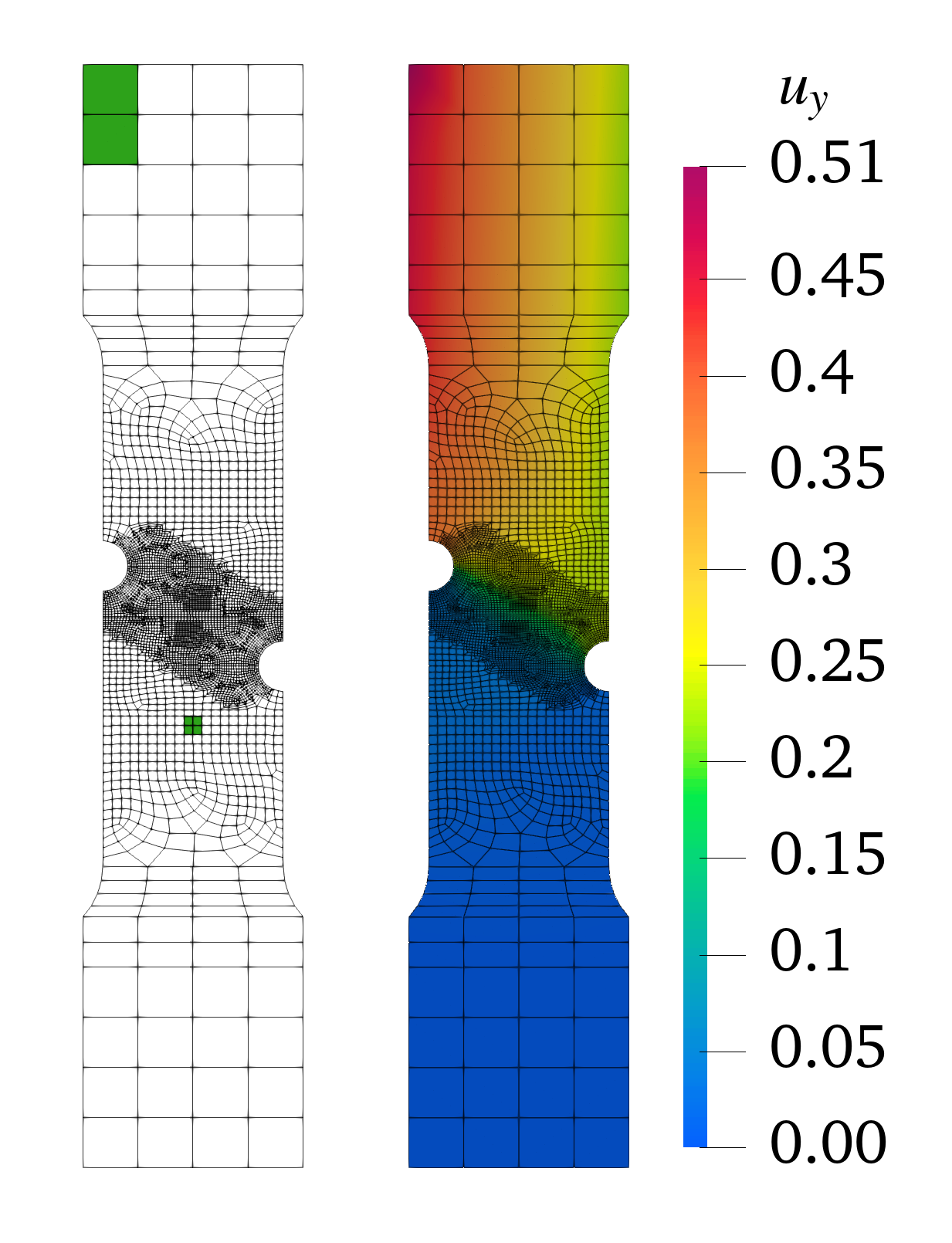}
        \caption{DEIM elements and results with $k=3$.}
        \label{fig:notch_plasti_strain_k3}
    \end{subfigure}
    \begin{subfigure}[t]{.353\textwidth}
        \centering
        \includegraphics[width=.905\textwidth]{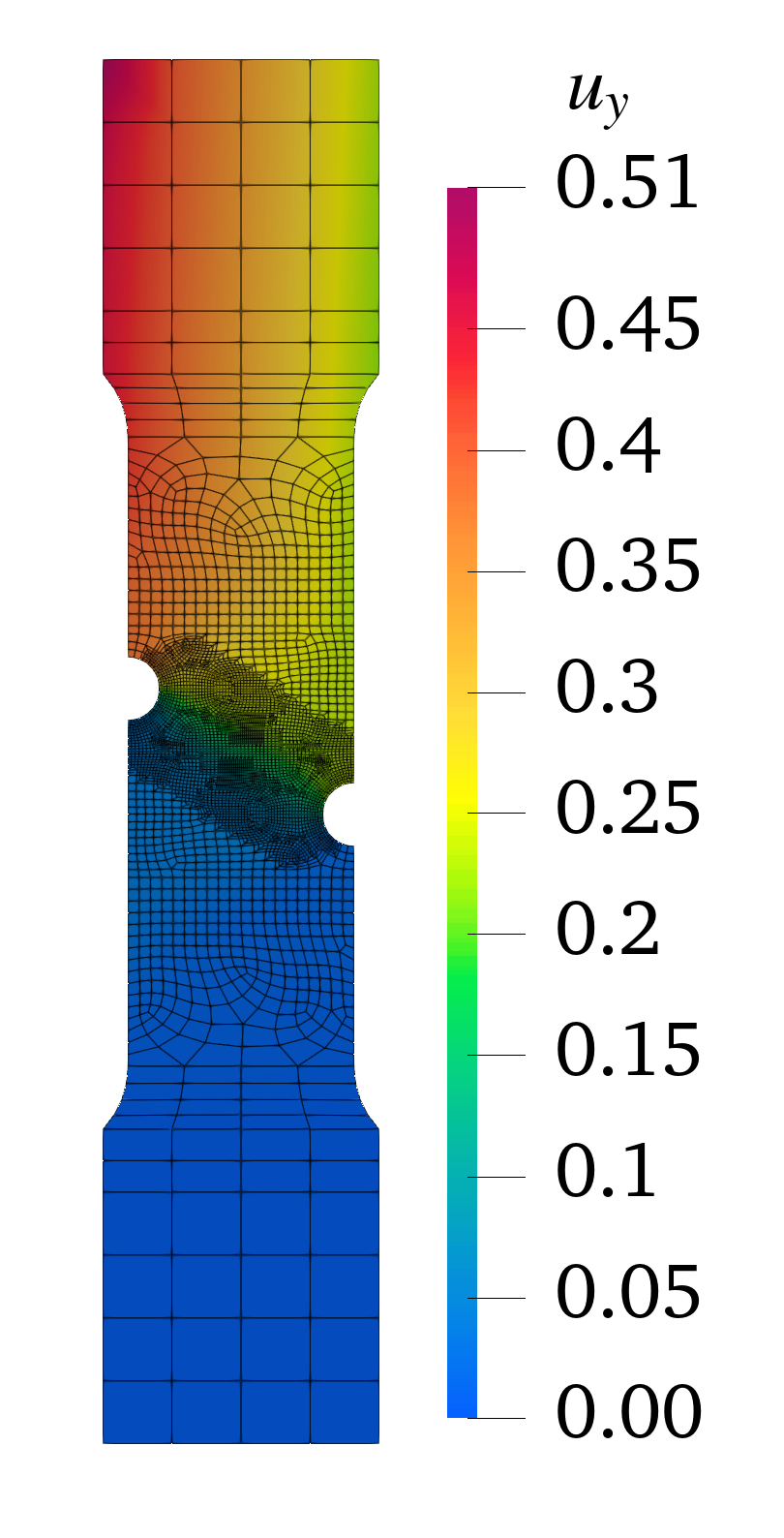}
        \caption{Displacements in y-direction at the end of the precomputation.}
        \label{fig:notch_plasti_strain_prec}
    \end{subfigure}
    \caption{Results and DEIM elements for 3 DEIM interpolation points.}
    \label{fig:asym_deimelem}
\end{figure}

Figure \ref{fig:notch_plasti_strain_k3} shows the displacement field in the y-direction for the DEIM computation with 20 POD modes and 3 DEIM bases. The elements that needed to be evaluated in the reduced computation are highlighted in green, while all other elements did not need to be evaluated. The reference solution from the unreduced precomputation is shown in Figure \ref{fig:notch_plasti_strain_prec}, which looks identical.

\subsubsection{Damage}

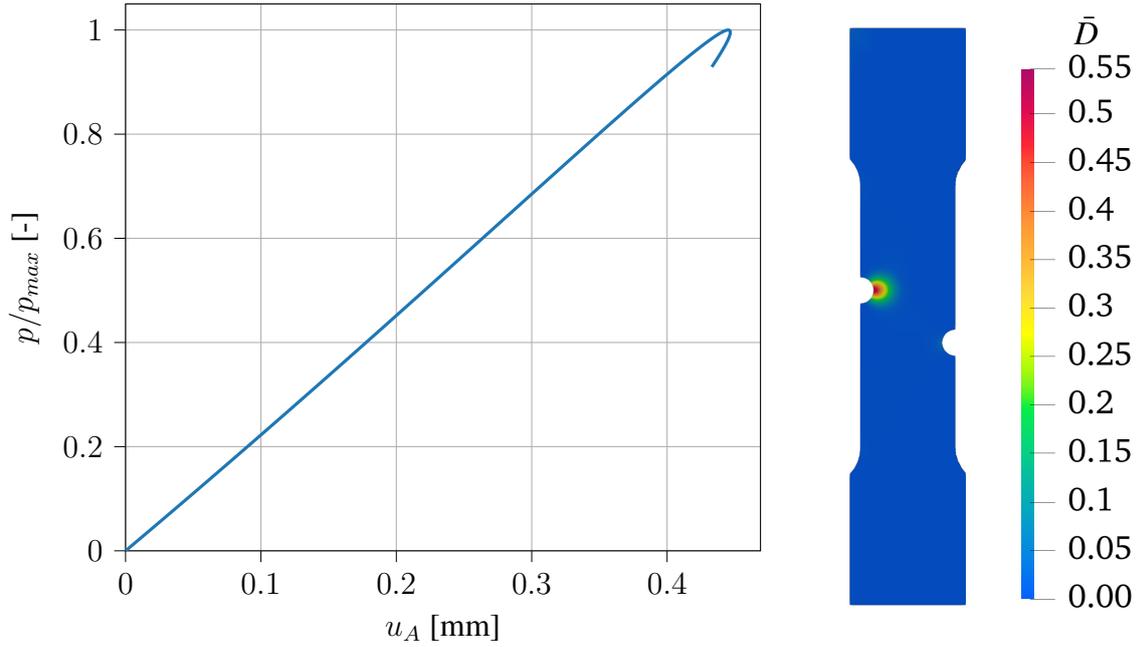
\begin{figure}[ht]
    \centering
    \begin{subfigure}[t]{.69\textwidth}
        \pgfplotsset{width=.9\textwidth, height=.8\textwidth }
        \centering
\begin{tikzpicture}

    \definecolor{darkgray176}{RGB}{176,176,176}
    \definecolor{lightgray204}{RGB}{204,204,204}
    \definecolor{steelblue31119180}{RGB}{31,119,180}

    \begin{axis}[
            legend cell align={left},
            legend style={
                    fill opacity=0.8,
                    draw opacity=1,
                    text opacity=1,
                    at={(0.03,0.97)},
                    anchor=north west,
                    draw=lightgray204
                },
            tick align=outside,
            tick pos=left,
            x grid style={darkgray176},
            xlabel={\(\displaystyle u_A\) [mm]},
            xmajorgrids,
            xmin=0, xmax=0.468982927586658,
            xtick style={color=black},
            y grid style={darkgray176},
            ylabel={\(\displaystyle p / p_{max}\) [-]},
            ymajorgrids,
            ymin=0, ymax=1.05,
            ytick style={color=black}
        ]
        \addplot [very thick, steelblue31119180, opacity=1]
        table {%
                0 0
                0.00412391286723319 0.00902074490549623
                0.00824784448452829 0.0180558338691919
                0.012371724124637 0.0271049796930609
                0.0164954829100232 0.0361678994427392
                0.0206190538171993 0.0452443144539764
                0.0247423716584323 0.0543339503349932
                0.0288653730597416 0.0634365369649345
                0.0329879964374173 0.0725518084885157
                0.0371101819736097 0.0816795033071745
                0.0412318715911789 0.0908193640667979
                0.0453530089278846 0.099971137642267
                0.0494735393100312 0.109134575118957
                0.0535934097256213 0.118309431771388
                0.0577125687970867 0.127495467039173
                0.0618309667536579 0.13669244450045
                0.0659485554034126 0.145900131842916
                0.0700652881050785 0.155118300832654
                0.0741811197396385 0.164346727280893
                0.0782960066817732 0.173585191008843
                0.0824099067711871 0.182833475810729
                0.0865227794538965 0.192091369108091
                0.0906345874329431 0.201358658813513
                0.0947452961276388 0.210635133225651
                0.0988548803833834 0.219920567233592
                0.102963307852097 0.229214748889663
                0.107070553801941 0.23851746693425
                0.111176595679771 0.247828487421196
                0.11528139720719 0.257147638140677
                0.119384941347962 0.266474689934938
                0.123487198963525 0.275809459931313
                0.127588146584533 0.285151753472926
                0.131687762662044 0.294501354605142
                0.135786015488147 0.303858116990237
                0.139882882449072 0.313221815246958
                0.143978342570437 0.322592261795757
                0.148072367485142 0.33196931869423
                0.152164937333934 0.341352753494248
                0.156256032857846 0.350742381368972
                0.160345633386236 0.36013806365384
                0.164433715819286 0.369539581631105
                0.168520261980693 0.378946760833908
                0.172605263427816 0.388359462613008
                0.176688701702529 0.397777455026805
                0.180770551515928 0.407200577141159
                0.184850814179722 0.416628664095052
                0.188929470452317 0.42606149840978
                0.193006499428592 0.435498927332639
                0.197081925179127 0.444940736237615
                0.201155826632031 0.454386774520993
                0.20522817186676 0.463836785610073
                0.209298869161675 0.47329059450476
                0.213368379947587 0.482747994003837
                0.217436880591633 0.492208758924464
                0.2215043864096 0.501672635678896
                0.225570730807777 0.511139437775605
                0.229635514840952 0.520608914114933
                0.233698777200342 0.53008077676291
                0.23776048396196 0.539554786128951
                0.241820642565577 0.549030631871822
                0.245879349186042 0.558507996480176
                0.249936998308289 0.567986396367204
                0.253993279082699 0.577465441649364
                0.258047912895263 0.586943292362639
                0.262100915670408 0.596420082926116
                0.266153028003002 0.60589537092789
                0.270203765953488 0.615368444799417
                0.274252607304766 0.624838410869492
                0.27829935364253 0.634304443174402
                0.282344253364934 0.643766071787211
                0.286387249770232 0.653221931221418
                0.290427913227558 0.662670160328539
                0.294465268498666 0.672107232250319
                0.298499506770034 0.681532933319569
                0.302530864721542 0.690948534996195
                0.306559253295468 0.700354280071328
                0.310584680950579 0.709749933228255
                0.314606834743457 0.719133978541009
                0.318625456309893 0.728504752269388
                0.322640654924967 0.737860419372375
                0.326651807165984 0.747198860419998
                0.330658206302078 0.756518114031739
                0.334658832996261 0.765815536607023
                0.338652706695362 0.775088432296417
                0.342639699025872 0.784334407871462
                0.346618786760015 0.79354931221011
                0.350588270036499 0.802728833729883
                0.354546909509479 0.811867812778261
                0.358492572129335 0.820960276980875
                0.362423192312458 0.830000554092481
                0.366336456900032 0.838981009413142
                0.370229882004269 0.847893708944652
                0.37410016491814 0.856728667354487
                0.377943186523177 0.865474514815981
                0.381754268109138 0.874118255706003
                0.385528155035434 0.882645137423761
                0.389258775175756 0.891038334031128
                0.39293914458792 0.899278809762464
                0.396561462107003 0.907345664297936
                0.400116933115585 0.915215571502006
                0.403595863274016 0.922863059715953
                0.406987711768516 0.93026115804578
                0.41028120800175 0.937381744455917
                0.413464778311946 0.944195953376142
                0.416526830225081 0.950675488637111
                0.419455988609178 0.956793657863682
                0.422241908075792 0.962526498543588
                0.424875478936219 0.96785386094294
                0.427349493165266 0.972760445586027
                0.429658866354044 0.97723648530337
                0.431800718444088 0.981277940282157
                0.433774401539595 0.984886287015951
                0.435581320962364 0.988067960519043
                0.437224715079427 0.990833715110834
                0.438709290189798 0.993197649585567
                0.440040806406466 0.995176140334423
                0.441225889256978 0.996787357891201
                0.44227160295273 0.99805021337407
                0.443185328767414 0.998984076127112
                0.443974490793823 0.999608089483158
                0.444646459602198 0.999940785218269
                0.445208842342478 1
                0.445669791119699 0.99980251086648
                0.446035400215571 0.999363359688413
                0.446311261842651 0.998696779811862
                0.446502575756933 0.997815901305806
                0.446614146786736 0.996732813564853
                0.446650407225389 0.995458649278846
                0.446615439222932 0.994003642301279
                0.446512990817413 0.992377169353222
                0.446346519691817 0.990587862036632
                0.446119195565532 0.988643618714418
                0.445833922732408 0.986551661076682
                0.445493389117024 0.984318669037641
                0.445100055517966 0.981950747348426
                0.444656176986073 0.979453478044393
                0.444163815760056 0.976831959119297
                0.443624899252707 0.974090939580008
                0.443041181732303 0.971234741502435
                0.44241424034022 0.968267253404637
                0.441745594473905 0.965192194940201
                0.441036610469375 0.962012915283747
                0.440288541128973 0.958732474592568
                0.439502536588064 0.955353679740677
                0.438679657725675 0.951879117033213
                0.437820879003054 0.94831115986143
                0.436927095004068 0.944651982763016
                0.435999127419591 0.940903579804643
                0.435037729926413 0.937067776907905
                0.434043593901963 0.933146246619119
                0.433017350376193 0.929140512522674
            };
    \end{axis}

\end{tikzpicture}
        \caption{Force-displacement curve of the asymmetrically notched specimen for micromorphic damage material behavior.}
        \label{fig:asym_loaddispdmg}
    \end{subfigure}
    \begin{subfigure}[t]{.3\textwidth}
        \includegraphics[width=\textwidth]{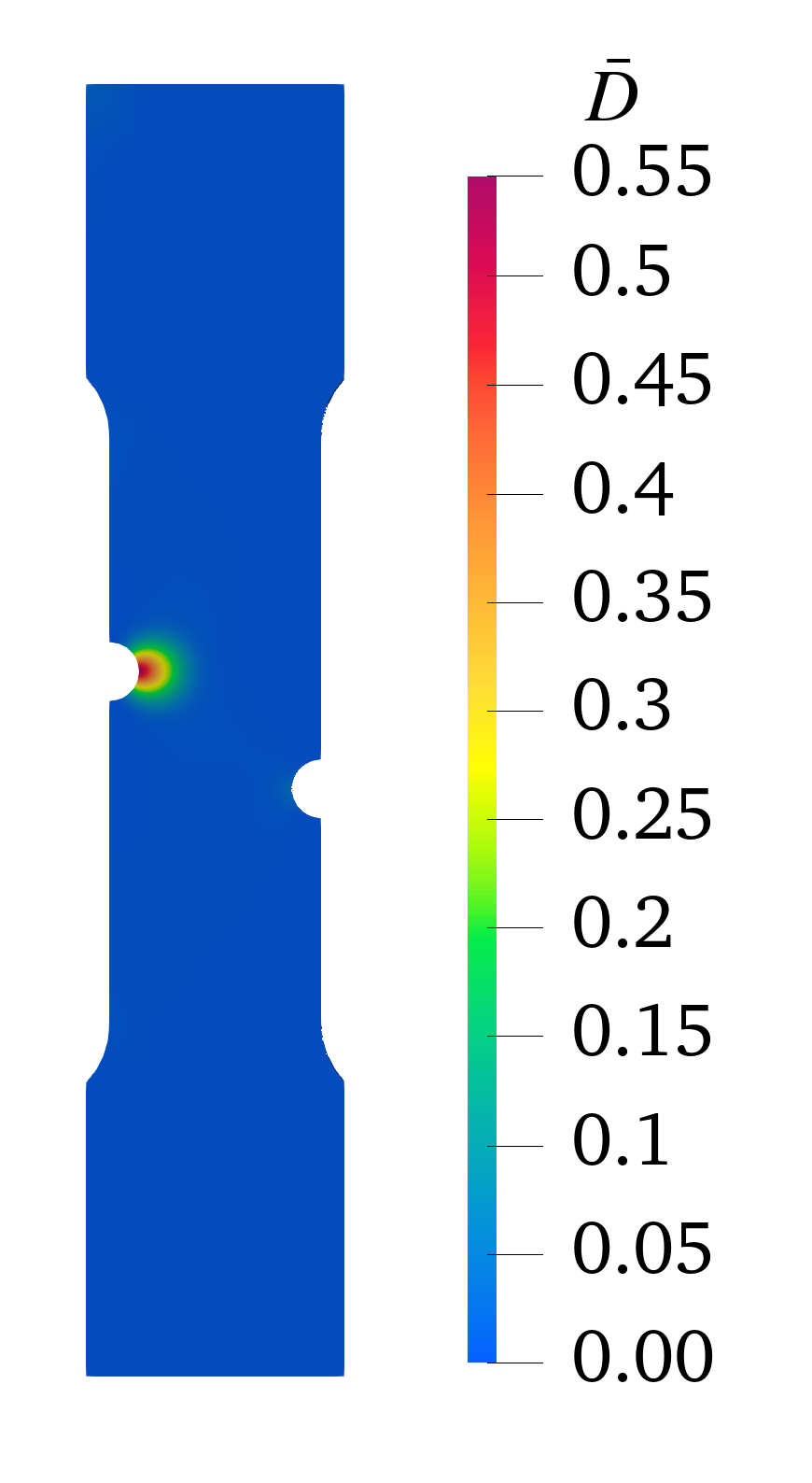}
        \caption{Micromorphic damage $\Dbar$ after 150 pseudo-timesteps.}
        \label{fig:asym_dmg}
    \end{subfigure}
    \caption{Simulation results of the asymmetrically notched specimen for micromorphic damage material behavior.}
\end{figure}

Figure \ref{fig:asym_loaddispdmg} shows the load-displacement curve for the case of damage without plasticity. The yield stress of plasticity has been set to a sufficiently high value to ensure that no plastic deformation occurs. The limit load is clearly exceeded in this case. The field of microstructural damage $\bar{D}$ is shown in Figure \ref{fig:asym_dmg}, where the initiation of damage can be observed in one of the notches, as expected.

Figure \ref{fig:notch_dmg_pmax_error} shows the error in the limit load, while Figure \ref{fig:notch_dmg_disp_error} shows the error in the displacement at point A. In both plots, it can be observed that with approximately 40 POD modes, the reduction methods are capable of reducing the error to below one percent. However, there are still some instabilities present. These instabilities did not improve even with a higher number of modes. Nevertheless, it can be seen from the load-displacement curves that the simulation follows the unloading path, allowing good and bad solutions to be distinguished even without a reference solution.

\pgfplotsset{width=.8\textwidth, height=.43\textwidth}
\begin{figure}[!ht]
    \centering
\begin{tikzpicture}

    \definecolor{crimson2143940}{RGB}{214,39,40}
    \definecolor{darkgray176}{RGB}{176,176,176}
    \definecolor{darkorange25512714}{RGB}{255,127,14}
    \definecolor{forestgreen4416044}{RGB}{44,160,44}
    \definecolor{lightgray204}{RGB}{204,204,204}
    \definecolor{steelblue31119180}{RGB}{31,119,180}

    \begin{axis}[
            legend cell align={left},
            legend style={
                    fill opacity=0.8,
                    draw opacity=1,
                    text opacity=1,
                    at={(1.03,1)},
                    anchor=north west,
                    draw=lightgray204
                },
            log basis y={10},
            tick align=outside,
            tick pos=left,
            x grid style={darkgray176},
            xlabel={\(\displaystyle k\)},
            xmajorgrids,
            xmin=1, xmax=20,
            xtick style={color=black},
            y grid style={darkgray176},
            ylabel={\(\displaystyle \varepsilon_{p_{max}}\)},
            ymajorgrids,
            ymin=0.0001, ymax=10,
            ymode=log,
            ytick style={color=black}
        ]
        \addlegendimage{empty legend}
        \addlegendentry{\hspace{-.3cm}$m$}
        \addplot [very thick, steelblue31119180, mark=*, mark size=1.5, mark options={solid}]
        table {%
                2 0.0144001511659116
                3 0.102041382390119
                4 0.0113749525290889
                5 0.0863242512141104
                6 0.00917194410875376
                7 0.304736089211757
                8 0.149384059659742
                9 0.782150975354247
                10 0.772525720856395
                11 0.772174262950429
                12 0.77206014566886
                13 0.771763927979459
                14 0.760359936592988
                15 0.685851281049244
                16 0.237996623622971
                17 0.701004161753639
                18 0.000397469035991909
                19 0.0152458083660133
                20 0.0845859074793025
            };
        \addlegendentry{20}
        \addplot [very thick, darkorange25512714, mark=*, mark size=1.5, mark options={solid}]
        table {%
                2 0.229992279241693
                3 0.194175468792591
                4 0.142142836215959
                5 0.115090611305203
                6 0.10289378201392
                7 0.490343215089833
                8 0.000543993082276269
                9 0.0497303098041828
                10 0.125398549220973
                11 0.112237630321669
                12 0.257278686823769
                13 0.168301947895163
                14 0.0188767802156054
                15 0.0450041257896154
                16 0.00141127192578696
                17 0.54828929942392
                18 0.200600302470085
                19 0.160501443285088
                20 0.276551386752259
            };
        \addlegendentry{30}
        \addplot [very thick, forestgreen4416044, mark=*, mark size=1.5, mark options={solid}]
        table {%
                2 0.0416968792964234
                3 0.226530850129833
                4 0.301200738349179
                5 0.127605463224574
                6 0.173785465372821
                7 0.0622257194755762
                8 0.000919339153423072
                9 0.00122546516060505
                10 0.00153407107802805
                11 0.147941556032087
                12 0.181589327661393
                13 0.00162608462855253
                14 0.00146115032461363
                15 0.00184264641458133
                16 9.3595166894696e-05
                17 0.895044855900148
                18 0.234238974360009
                19 3.15645411911669
                20 0.0298682575197874
            };
        \addlegendentry{40}
        \addplot [very thick, crimson2143940, mark=*, mark size=1.5, mark options={solid}]
        table {%
                2 0.329050855236241
                3 0.224223138474167
                4 0.252101062014577
                5 0.115472756441403
                6 0.00410064467549094
                7 0.228863117893771
                8 0.00107053425208747
                9 0.000340200595933967
                10 0.00086776179632838
                11 0.0890215016431539
                12 0.180269142048689
                13 0.000954389409115768
                14 0.000613893642582415
                15 0.000107772396471227
                16 0.131551945011649
                17 0.112937389689447
                18 0.0955191579653113
                19 0.0436350000154326
                20 0.253729656505954
            };
        \addlegendentry{50}
    \end{axis}

\end{tikzpicture}
    \caption{Relative error of the limit load with respect to the number of POD basis vectors $m$ and the number of DEIM interpolation points $k$.}
    \label{fig:notch_dmg_pmax_error}
\end{figure}
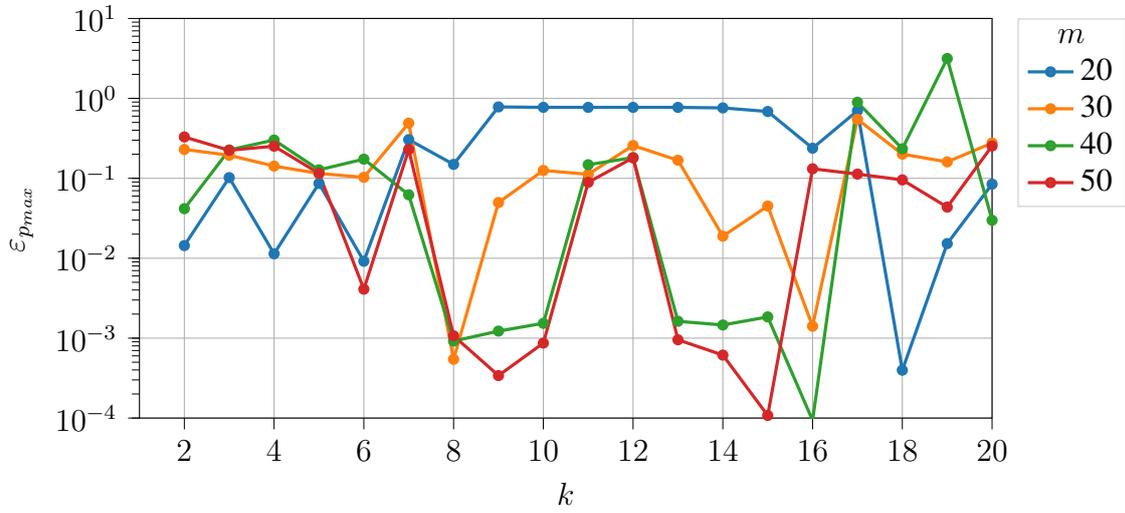
\begin{figure}[!ht]
    \centering
\begin{tikzpicture}

    \definecolor{crimson2143940}{RGB}{214,39,40}
    \definecolor{darkgray176}{RGB}{176,176,176}
    \definecolor{darkorange25512714}{RGB}{255,127,14}
    \definecolor{forestgreen4416044}{RGB}{44,160,44}
    \definecolor{lightgray204}{RGB}{204,204,204}
    \definecolor{steelblue31119180}{RGB}{31,119,180}

    \begin{axis}[
            legend cell align={left},
            legend style={
                    fill opacity=0.8,
                    draw opacity=1,
                    text opacity=1,
                    at={(1.03,1)},
                    anchor=north west,
                    draw=lightgray204
                },
            log basis y={10},
            tick align=outside,
            tick pos=left,
            x grid style={darkgray176},
            xlabel={\(\displaystyle k\)},
            xmajorgrids,
            xmin=1, xmax=20,
            xtick style={color=black},
            y grid style={darkgray176},
            ylabel={\(\displaystyle \varepsilon_{u_{A}}\)},
            ymajorgrids,
            ymin=0.0001, ymax=10,
            ymode=log,
            ytick style={color=black}
        ]
        \addlegendimage{empty legend}
        \addlegendentry{\hspace{-.3cm}$m$}
        \addplot [very thick, steelblue31119180, mark=*, mark size=1.5, mark options={solid}]
        table {%
                2 0.0409769075613766
                3 0.10842441588901
                4 0.0270864714843033
                5 0.0468125327388763
                6 0.0361560190720409
                7 0.300958073683581
                8 0.019188472464867
                9 0.374361086062329
                10 5.1395551388022
                11 0.524364388879247
                12 0.039077765835743
                13 0.0392250674381297
                14 0.291280411463654
                15 0.257012663774895
                16 0.237297127898558
                17 0.0570568219464738
                18 0.0281494953046878
                19 0.0419259814735943
                20 0.0654763815044667
            };
        \addlegendentry{20}
        \addplot [very thick, darkorange25512714, mark=*, mark size=1.5, mark options={solid}]
        table {%
                2 0.224503598962251
                3 0.176823027118285
                4 0.0352884695139064
                5 0.0391936752775291
                6 0.0446003723436091
                7 0.194701716212338
                8 0.00253940299413547
                9 0.0581764806986679
                10 0.148086688619269
                11 0.117390424712145
                12 0.278033640363904
                13 0.177434326145657
                14 0.10189977982764
                15 0.0420547383915119
                16 0.00388884771057534
                17 0.847016745661436
                18 0.167688830490216
                19 0.179615747487288
                20 0.931349372105028
            };
        \addlegendentry{30}
        \addplot [very thick, forestgreen4416044, mark=*, mark size=1.5, mark options={solid}]
        table {%
                2 0.0629988751650543
                3 0.149337546054624
                4 0.306441287766487
                5 0.032125445295688
                6 0.0144012317474887
                7 0.128007786658201
                8 0.00516985726173263
                9 0.000456163240626303
                10 0.000349345577628089
                11 0.166983090786616
                12 0.199301090057016
                13 0.000661737841765483
                14 0.000620616556630873
                15 0.000268081945593197
                16 0.000863718616562882
                17 0.924826866809037
                18 0.250705986270884
                19 2.96894997679085
                20 0.0489201046951334
            };
        \addlegendentry{40}
        \addplot [very thick, crimson2143940, mark=*, mark size=1.5, mark options={solid}]
        table {%
                2 0.188535993555927
                3 0.147555942835197
                4 0.268531992416192
                5 0.0377220957574452
                6 0.00653559892100786
                7 0.238459612107746
                8 0.00549158637978857
                9 0.000985319940903761
                10 0.000784468296195141
                11 0.107794552682864
                12 0.198230266051546
                13 0.000661544752641118
                14 0.000861467787380393
                15 0.00136898330666116
                16 0.146613330252566
                17 0.126842773027864
                18 0.115307946155338
                19 0.0737075985419691
                20 0.212849963469065
            };
        \addlegendentry{50}
    \end{axis}

\end{tikzpicture}
    \caption{Relative error of the displacement in point A with respect to the number of POD basis vectors $m$ and the number of DEIM interpolation points $k$.}
    \label{fig:notch_dmg_disp_error}
\end{figure}
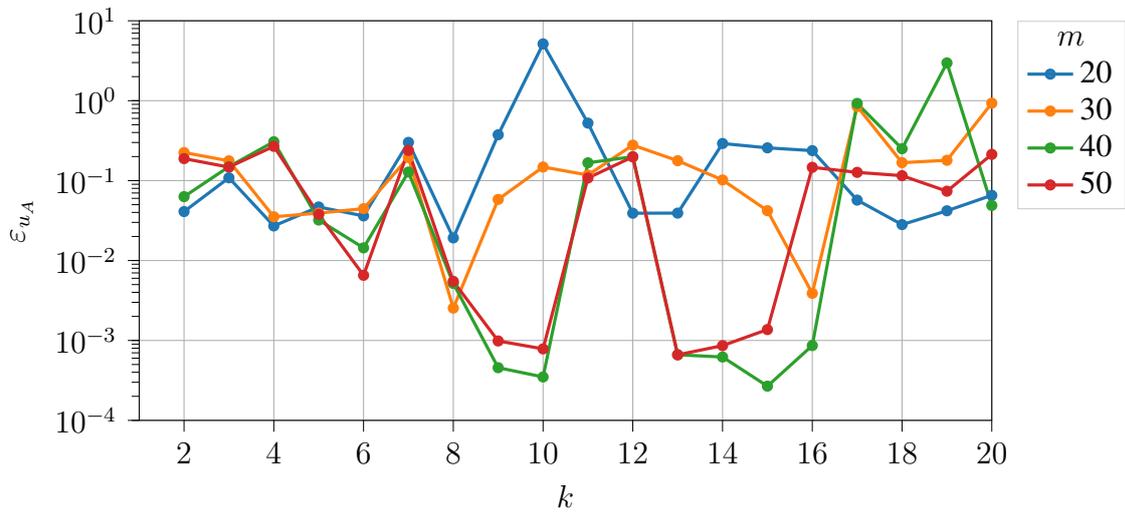
In this example, speedups of over 100 can be achieved. However, the actual speedup depends on the number of selected POD and DEIM modes. The specific speedups are shown in Figure \ref{fig:notch_dmg_speedup}. As a comparison, all the reduced calculations depicted here are still faster than a full-scale calculation.
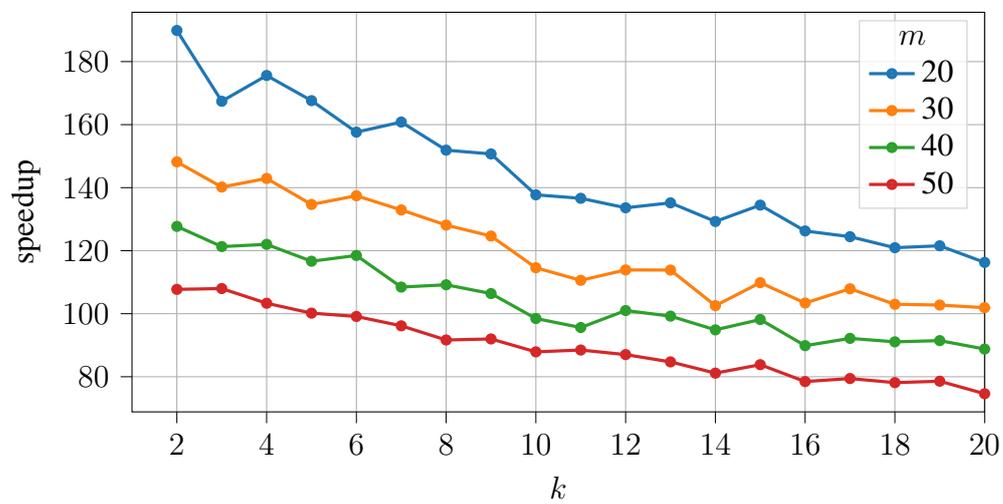
\begin{figure}[!ht]
    \centering
\begin{tikzpicture}

    \definecolor{crimson2143940}{RGB}{214,39,40}
    \definecolor{darkgray176}{RGB}{176,176,176}
    \definecolor{darkorange25512714}{RGB}{255,127,14}
    \definecolor{forestgreen4416044}{RGB}{44,160,44}
    \definecolor{lightgray204}{RGB}{204,204,204}
    \definecolor{steelblue31119180}{RGB}{31,119,180}

    \begin{axis}[
            legend cell align={left},
            legend style={fill opacity=0.8, draw opacity=1, text opacity=1, draw=lightgray204},
            tick align=outside,
            tick pos=left,
            x grid style={darkgray176},
            xlabel={\(\displaystyle k\)},
            xmajorgrids,
            xmin=1, xmax=20,
            xtick style={color=black},
            y grid style={darkgray176},
            ylabel={speedup},
            ymajorgrids,
            ymin=68.850778302974, ymax=195.651557713764,
            ytick style={color=black}
        ]
        \addlegendimage{empty legend}
        \addlegendentry{\hspace{-.3cm}$m$}
        \addplot [very thick, steelblue31119180, mark=*, mark size=1.5, mark options={solid}]
        table {%
                2 189.887885922364
                3 167.418151168045
                4 175.58820024244
                5 167.613803986868
                6 157.619963872141
                7 160.814526033436
                8 151.897339486506
                9 150.698989692531
                10 137.723946953283
                11 136.624228468688
                12 133.610624348969
                13 135.17643990294
                14 129.265571870567
                15 134.453595261284
                16 126.259601088294
                17 124.441826224624
                18 120.94385392408
                19 121.552068476769
                20 116.288650776242
            };
        \addlegendentry{20}
        \addplot [very thick, darkorange25512714, mark=*, mark size=1.5, mark options={solid}]
        table {%
                2 148.183023866524
                3 140.192534083624
                4 142.919347374381
                5 134.66900533824
                6 137.436251269792
                7 132.930486389753
                8 128.124192201574
                9 124.643452649247
                10 114.587144237335
                11 110.594980572876
                12 113.896641422133
                13 113.851693282573
                14 102.555269496158
                15 109.854109765015
                16 103.398304208789
                17 107.918730079141
                18 102.993683803847
                19 102.748554608632
                20 101.918148282132
            };
        \addlegendentry{30}
        \addplot [very thick, forestgreen4416044, mark=*, mark size=1.5, mark options={solid}]
        table {%
                2 127.708590552035
                3 121.301800282612
                4 122.001471615572
                5 116.64913503181
                6 118.481568707041
                7 108.452337969328
                8 109.177755109079
                9 106.39866937419
                10 98.4765176522264
                11 95.6161612589746
                12 101.012531737018
                13 99.2332847780286
                14 94.8775910320137
                15 98.1613330136409
                16 89.8587849982516
                17 92.1780111395092
                18 91.0784100337145
                19 91.460933646515
                20 88.7873919194435
            };
        \addlegendentry{40}
        \addplot [very thick, crimson2143940, mark=*, mark size=1.5, mark options={solid}]
        table {%
                2 107.727712312732
                3 107.980029476156
                4 103.321449800878
                5 100.183787953992
                6 99.1276553867962
                7 96.1622055196782
                8 91.6609361833038
                9 91.9942753010087
                10 87.899975854013
                11 88.4810895521894
                12 87.026261043307
                13 84.7129918834745
                14 81.1476877654156
                15 83.8294978367279
                16 78.4710196312753
                17 79.4331890869483
                18 78.1320467907254
                19 78.5846933516998
                20 74.6144500943735
            };
        \addlegendentry{50}
    \end{axis}

\end{tikzpicture}
    \caption{Speedup of the reduced simulation with respect to the number of POD basis vectors $m$ and the number of DEIM interpolation points $k$.}
    \label{fig:notch_dmg_speedup}
\end{figure}

\FloatBarrier
\subsubsection{Damage and plasticity}

\begin{figure}[!ht]
    \centering
    \input{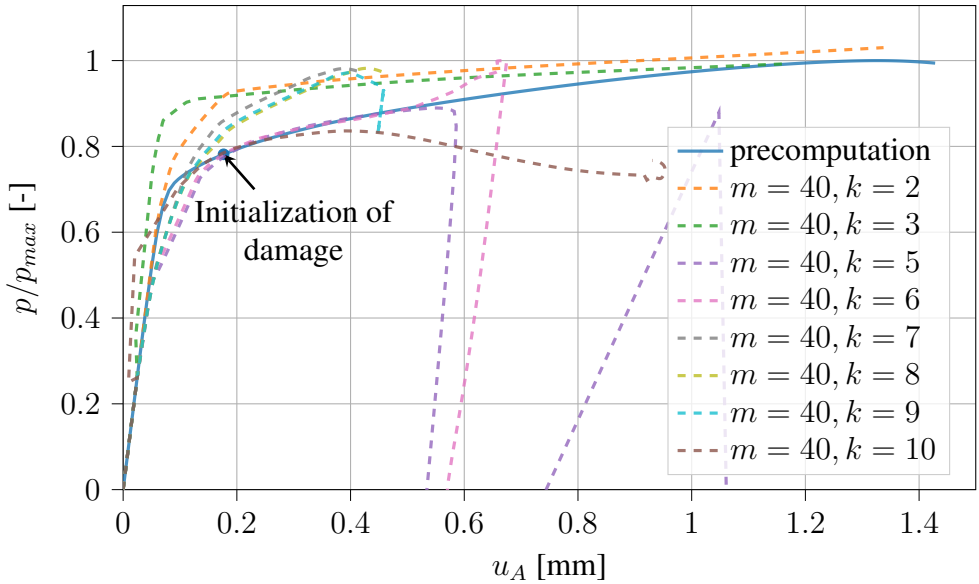}
    \caption{Comparison of the load-displacement curves of the precomputation with the curves of the ROM with 40 POD modes and an increasing number of DEIM modes.}
    \label{fig:asym_compare_pd_40k}
\end{figure}

In the case of combining plasticity and damage, as expected, the difficulties and instabilities observed in the individual cases persist. In order to present a comprehensive picture, we include these challenges in this paper. The parameters used for this case are the same as those in Table \ref{tab:asympara}. In Figure \ref{fig:asym_compare_pd_40k}, we compare the load-displacement curves between the reference solution and reduced calculations using 40 POD modes with varying numbers of DEIM modes. While all the reduced calculations provide reasonably accurate results in the initial range, only a few modes are able to capture the plasticity region effectively. Overall, the methods were unable to produce satisfactory results beyond the limit load for such a complex problem.

\pgfplotsset{width=.8\textwidth, height=.5\textwidth}

This highlights the challenges in accurately capturing the combined behavior of plasticity and damage using reduced-order modeling techniques. Further investigation and development are required to address these instabilities and improve the reliability of the methods for such complex problems.
\FloatBarrier
\section{Conclusion}

In conclusion, we have investigated the application of reduced-order modeling techniques for the simulation of plasticity and damage in structural problems. We have examined different scenarios, including the separate analysis of plasticity and damage, as well as their combination.

For plasticity analysis, we observed that with a sufficient number of POD modes and a small number of DEIM modes, the behavior can be accurately approximated with errors below 1\%. However, we also encountered numerical instabilities in the calculations, which require further investigation and refinement of the methods.

In the case of damage analysis without plasticity, we achieved good results with errors for the displacements in the range of $10^{-3}$. The reduction in computational time was significant, with speedups ranging from 40 to 80 compared to full calculations.

When considering the combined effects of plasticity and damage, the challenges became more pronounced. Instabilities persisted even with higher numbers of POD and DEIM modes, making it difficult to obtain reliable and accurate results beyond the limit load.

Overall, the results demonstrate the great potential of reduced-order modeling for analyses involving highly nonlinear and complex material behaviors such as plasticity and damage, with significant computational speedups and accurate approximations achievable in many cases. Nevertheless, further research and development is still needed to address the observed instabilities in the study and to improve the robustness of the methods.
\section*{Acknowledgments}
S. Kastian and S. Reese gratefully acknowledge the financial support of the project RE 1057/40 (project number 312911604) within SPP 1886 by the German Research Foundation (DFG).
Furthermore, J. Kehls, T. Brepols and S. Reese acknowledge the financial support of subproject B05 within SFB/TRR 339(project number 453596084) and S. Reese thankfully acknowledges the funding of subproject A01 within SFB/TRR 280 (project number 417002380) by the DFG. In addition, S. Reese and T. Brepols gratefully acknowledge the funding of the project RE 1057/51 (project number 453715964).

\newpage


\bibliographystyle{agsm}
\bibliography{Bibliography}

@article{pohl2014adaptive,
  title={Adaptive path following schemes for problems with softening},
  author={Pohl, T. and Ramm, E. and Bischoff, M.},
  journal={Finite elements in analysis and design},
  volume={86},
  pages={12--22},
  year={2014},
  publisher={Elsevier}
}

@article{mooreslaw,
  title     = {Fifty years of Moore's law},
  author    = {Mack, C. A},
  journal   = {IEEE Transactions on semiconductor manufacturing},
  volume    = {24},
  number    = {2},
  pages     = {202--207},
  year      = {2011},
  publisher = {IEEE}
}

@book{MOR1,
  title     = {System-and Data-Driven Methods and Algorithms},
  author    = {Benner, P. and Grivet-Talocia, S. and Quarteroni, A. and Rozza, G. and Schilders, W. and Silveira, L. M.},
  year      = {2021},
  publisher = {De Gruyter}
}

@article{forest2009,
  title     = {Micromorphic approach for gradient elasticity, viscoplasticity, and damage},
  author    = {Forest, S.},
  journal   = {Journal of Engineering Mechanics},
  volume    = {135},
  number    = {3},
  pages     = {117--131},
  year      = {2009},
  publisher = {American Society of Civil Engineers}
}

@article{brepols2020,
  title     = {A gradient-extended two-surface damage-plasticity model for large deformations},
  author    = {Brepols, T. and Wulfinghoff, S. and Reese, S.},
  journal   = {International Journal of Plasticity},
  volume    = {129},
  pages     = {102635},
  year      = {2020},
  publisher = {Elsevier}
}

@article{chatsor,
  title     = {Nonlinear model reduction via discrete empirical interpolation},
  author    = {Chaturantabut, S. and Sorensen, D. C.},
  journal   = {SIAM Journal on Scientific Computing},
  volume    = {32},
  number    = {5},
  pages     = {2737--2764},
  year      = {2010},
  publisher = {SIAM}
}

@article{riks1979,
  title     = {An incremental approach to the solution of snapping and buckling problems},
  author    = {Riks, E.},
  journal   = {International journal of solids and structures},
  volume    = {15},
  number    = {7},
  pages     = {529--551},
  year      = {1979},
  publisher = {Elsevier}
}

@article{brepols2017,
  title     = {Gradient-extended two-surface damage-plasticity: micromorphic formulation and numerical aspects},
  author    = {Brepols, T. and Wulfinghoff, S. and Reese, S.},
  journal   = {International Journal of Plasticity},
  volume    = {97},
  pages     = {64--106},
  year      = {2017},
  publisher = {Elsevier}
}

@article{barfusz2021,
  title     = {A single Gauss point continuum finite element formulation for gradient-extended damage at large deformations},
  author    = {Barfusz, O. and Brepols, T. and van der Velden, T. and Frischkorn, J. and Reese, S.},
  journal   = {Computer Methods in Applied Mechanics and Engineering},
  volume    = {373},
  pages     = {113440},
  year      = {2021},
  publisher = {Elsevier}
}

@article{radermacher2016,
  title     = {POD-based model reduction with empirical interpolation applied to nonlinear elasticity},
  author    = {Radermacher, A. and Reese, S.},
  journal   = {International Journal for Numerical Methods in Engineering},
  volume    = {107},
  number    = {6},
  pages     = {477--495},
  year      = {2016},
  publisher = {Wiley Online Library}
}

@article{ritz1,
  title    = {A reduction method for nonlinear structural dynamic analysis},
  journal  = {Computer Methods in Applied Mechanics and Engineering},
  volume   = {49},
  number   = {3},
  pages    = {253-279},
  year     = {1985},
  author   = {Idelsohn, S. R. and Cardona, A.}
}

@article{ritz2,
  title     = {An efficient crankshaft dynamic analysis using substructuring with Ritz vectors},
  author    = {Mourelatos, Z. P.},
  journal   = {Journal of Sound and vibration},
  volume    = {238},
  number    = {3},
  pages     = {495--527},
  year      = {2000},
  publisher = {Elsevier}
}

@article{ritz3,
  title   = {Reduced basis technique for nonlinear analysis of structures},
  author  = {Noor, A. K. and Peters, J. M.},
  journal = {Aiaa journal},
  volume  = {18},
  number  = {4},
  pages   = {455--462},
  year    = {1980}
}

@article{riks1972,
  author   = {Riks, E.},
  title    = {{The Application of Newton’s Method to the Problem of Elastic Stability}},
  journal  = {Journal of Applied Mechanics},
  volume   = {39},
  number   = {4},
  pages    = {1060-1065},
  year     = {1972},
  month    = {12},
  eprint   = {https://asmedigitalcollection.asme.org/appliedmechanics/article-pdf/39/4/1060/5452567/1060\_1.pdf}
}

@article{radermacher2013,
  title     = {A comparison of projection-based model reduction concepts in the context of nonlinear biomechanics},
  author    = {Radermacher, A. and Reese, S.},
  journal   = {Archive of Applied Mechanics},
  volume    = {83},
  number    = {8},
  pages     = {1193--1213},
  year      = {2013},
  publisher = {Springer}
}

@article{hyper,
  title     = {Hyper-reduced predictions for lifetime assessment of elasto-plastic structures},
  author    = {Ryckelynck, D. and Lampoh, K. and Quilicy, S.},
  journal   = {Meccanica},
  volume    = {51},
  pages     = {309--317},
  year      = {2016},
  publisher = {Springer}
}

@article{GPOD1,
  title     = {Unsteady flow sensing and estimation via the gappy proper orthogonal decomposition},
  author    = {Willcox, K.},
  journal   = {Computers \& fluids},
  volume    = {35},
  number    = {2},
  pages     = {208--226},
  year      = {2006},
  publisher = {Elsevier}
}

@article{GPOD2,
  title     = {Stability of discrete empirical interpolation and gappy proper orthogonal decomposition with randomized and deterministic sampling points},
  author    = {Peherstorfer, B. and Drmac, Z. and Gugercin, S.},
  journal   = {SIAM Journal on Scientific Computing},
  volume    = {42},
  number    = {5},
  pages     = {A2837--A2864},
  year      = {2020},
  publisher = {SIAM}
}

@article{DEIM2,
  title     = {Application of the discrete empirical interpolation method to reduced order modeling of nonlinear and parametric systems},
  author    = {Antil, H. and Heinkenschloss, M. and Sorensen, D. C.},
  journal   = {Reduced order methods for modeling and computational reduction},
  pages     = {101--136},
  year      = {2014},
  publisher = {Springer}
}

@article{DEIM3,
  title     = {A new selection operator for the discrete empirical interpolation method---improved a priori error bound and extensions},
  author    = {Drmac, Z. and Gugercin, S.},
  journal   = {SIAM Journal on Scientific Computing},
  volume    = {38},
  number    = {2},
  pages     = {A631--A648},
  year      = {2016},
  publisher = {SIAM}
}

@article{EC1,
  title     = {Dimensional hyper-reduction of nonlinear finite element models via empirical cubature},
  author    = {Hernandez, J. A. and Caicedo, M. A. and Ferrer, A.},
  journal   = {Computer methods in applied mechanics and engineering},
  volume    = {313},
  pages     = {687--722},
  year      = {2017},
  publisher = {Elsevier}
}

@article{EC2,
  title     = {An adaptive domain-based POD/ECM hyper-reduced modeling framework without offline training},
  author    = {Rocha, I. B. C. M. and van der Meer, F. P. and Sluys, L. J.},
  journal   = {Computer Methods in Applied Mechanics and Engineering},
  volume    = {358},
  pages     = {112650},
  year      = {2020},
  publisher = {Elsevier}
}

@article{NI1,
  title   = {Investigation of nonlinear model order reduction of the quasigeostrophic equations through a physics-informed convolutional autoencoder},
  author  = {Cooper, R. and Popov, A. A. and Sandu, A.},
  journal = {arXiv preprint arXiv:2108.12344},
  year    = {2021}
}

@article{NI2,
  title     = {Neural network closures for nonlinear model order reduction},
  author    = {San, O. and Maulik, R.},
  journal   = {Advances in Computational Mathematics},
  volume    = {44},
  pages     = {1717--1750},
  year      = {2018},
  publisher = {Springer}
}

@article{NI3,
  title     = {Deep neural networks for nonlinear model order reduction of unsteady flows},
  author    = {Eivazi, H. and Veisi, H. and Naderi, M. H. and Esfahanian, V.},
  journal   = {Physics of Fluids},
  volume    = {32},
  number    = {10},
  pages     = {105104},
  year      = {2020},
  publisher = {AIP Publishing LLC}
}

@incollection{crisfield1981,
  title     = {A fast incremental/iterative solution procedure that handles “snap-through”},
  author    = {Crisfield, M. A.},
  booktitle = {Computational methods in nonlinear structural and solid mechanics},
  pages     = {55--62},
  year      = {1981},
  publisher = {Elsevier}
}

@article{hrarclen,
  title    = {Hyper-reduced arc-length algorithm for stability analysis in elastoplasticity},
  journal  = {International Journal of Solids and Structures},
  volume   = {208-209},
  pages    = {167-180},
  year     = {2021},
  author   = {Launay, H. and Besson, J. and Ryckelynck, D. and Willot, F.}
}

@inproceedings{ramm1981,
  author    = {Ramm, E.},
  editor    = {Wunderlich, W.
               and Stein, E.
               and Bathe, K.-J.},
  title     = {Strategies for Tracing the Nonlinear Response Near Limit Points},
  booktitle = {Nonlinear Finite Element Analysis in Structural Mechanics},
  year      = {1981},
  publisher = {Springer Berlin Heidelberg},
  address   = {Berlin, Heidelberg},
  pages     = {63--89},
  isbn      = {978-3-642-81589-8}
}

@article{wempner1971,
  title   = {Discrete approximations related to nonlinear theories of solids},
  journal = {International Journal of Solids and Structures},
  volume  = {7},
  number  = {11},
  pages   = {1581-1599},
  year    = {1971},
  author  = {Wempner, G. A.}
}

@article{PGD1,
  title     = {Recent advances and new challenges in the use of the proper generalized decomposition for solving multidimensional models},
  author    = {Chinesta, F. and Ammar, A. and Cueto, E.},
  journal   = {Archives of Computational methods in Engineering},
  volume    = {17},
  number    = {4},
  pages     = {327--350},
  year      = {2010},
  publisher = {Springer}
}

@article{PGD2,
  title     = {A priori model reduction through proper generalized decomposition for solving time-dependent partial differential equations},
  author    = {Nouy, Anthony},
  journal   = {Computer Methods in Applied Mechanics and Engineering},
  volume    = {199},
  number    = {23-24},
  pages     = {1603--1626},
  year      = {2010},
  publisher = {Elsevier}
}

@article{RB1,
  title     = {On the stability of the reduced basis method for Stokes equations in parametrized domains},
  author    = {Rozza, Gianluigi and Veroy, Karen},
  journal   = {Computer methods in applied mechanics and engineering},
  volume    = {196},
  number    = {7},
  pages     = {1244--1260},
  year      = {2007},
  publisher = {Elsevier}
}

@article{RB2,
  title     = {The reduced basis method for the electric field integral equation},
  author    = {Fares, M and Hesthaven, Jan S and Maday, Yvon and Stamm, Benjamin},
  journal   = {Journal of Computational Physics},
  volume    = {230},
  number    = {14},
  pages     = {5532--5555},
  year      = {2011},
  publisher = {Elsevier}
}

@proceedings{MDEIM,
  author = {Tiso, P. and Dedden, R. and Rixen, D.},
  title  = {{A Modified Discrete Empirical Interpolation Method for Reducing Non-Linear Structural Finite Element Models}},
  volume = {Volume 7B: 9th International Conference on Multibody Systems, Nonlinear Dynamics, and Control},
  series = {International Design Engineering Technical Conferences and Computers and Information in Engineering Conference},
  year   = {2013},
  month  = {08}
}

\end{document}